\documentclass[twocolumn]{aastex631}

\usepackage{amsmath}
\usepackage{multirow}
\usepackage{enumitem}

\newcommand{\Mone}{$\rm{\frac{M_{1kpc}}{M_*}}$}
\newcommand{\Se}{$\Sigma_e$}
\newcommand{\Sone}{$\Sigma_{\rm{1kpc}}$}
\newcommand{\UVc}{(U $-$ V)$_{\rm{corr}}$}
\newcommand{\VJc}{(V $-$ J)$_{\rm{corr}}$}

\begin{document}

\title{AGN Selection Methods Have Profound Impacts on the Distributions of Host Galaxy Properties} 

\correspondingauthor{Zhiyuan Ji}
\email{zhiyuanji@astro.umass.edu}

\author[0000-0001-7673-2257]{Zhiyuan Ji}
\affiliation{University of Massachusetts Amherst, 710 North Pleasant Street, Amherst, MA 01003-9305, USA}
\author[0000-0002-7831-8751]{Mauro Giavalisco}
\affiliation{University of Massachusetts Amherst, 710 North Pleasant Street, Amherst, MA 01003-9305, USA}
\author{Allison Kirkpatrick}
\affiliation{Department of Physics \& Astronomy, University of Kansas, Lawrence, KS 66045, USA}
\author[0000-0001-9201-4706]{Dale Kocevski}
\affiliation{Department of Physics and Astronomy, Colby College, Waterville, ME 04961, USA}
\author[0000-0002-3331-9590]{Emanuele Daddi}
\affiliation{CEA, Irfu, DAp, AIM, Universit\'{e} Paris-Saclay, Universit\'{e} de Paris, CNRS, F-91191 Gif-sur-Yvette, France}
\author[0000-0001-8706-2252]{Ivan Delvecchio}
\affiliation{INAF - Osservatorio Astronomico di Brera, via Brera 28, I-20121, Milano, Italy}
\author{Cassandra Hatcher}
\affiliation{Department of Physics \& Astronomy, University of Kansas, Lawrence, KS 66045, USA}

\begin{abstract}
 We present a comparative study of X-ray and IR AGNs at $z\approx2$ to highlight the important AGN selection effects on the distributions of host galaxy properties. Compared with non-AGN star-forming galaxies (SFGs) on the main sequence, X-ray AGNs have similar median star formation (SF) properties, but their incidence (q$_{\rm{AGN}}$) is higher among galaxies with either enhanced or suppressed SF, and among galaxies with larger stellar mass surface density, regardless if it is measured within half-light radius (\Se) or central 1kpc (\Sone). Unlike X-ray AGNs, IR AGNs are less massive, and have enhanced SF and similar distributions of colors, \Se\ and \Sone\ relative to non-AGN SFGs. Given that \Se\ and \Sone\ strongly correlate with M$_*$, we introduce the fractional mass within central 1kpc (\Mone), which only {\it weakly} depends on M$_*$, to quantify galaxy compactness. Both AGN populations have similar \Mone\ distributions compared to non-AGN SFGs'. While q$_{\rm{AGN}}$ increases with \Se\ and \Sone, it remains constant with \Mone, indicating that the trend of increasing q$_{\rm{AGN}}$ with $\rm{\Sigma}$ is driven by M$_*$ more than morphology. While our findings are not in conflict with the scenario of AGN quenching, they do not imply it either, because the incidence of AGNs hosted in transitional galaxies depends crucially on AGN selections. Additionally, despite the relatively large uncertainty of AGN bolometric luminosities, their very weak correlation, if any, with SF activities, regardless of AGN selections, also argues against a direct causal link between the presences of AGNs and the quenching of massive galaxies at $z\sim2$. 

\end{abstract}


\section{Introduction} \label{sec:intro}

Modern observational cosmology, primarily the observations of cosmic large-scale structures such as the Baryon Acoustic Oscillations (e.g. \citealt{Eisenstein2005}) and the polarization of the Cosmic Microwave Background (e.g. \citealt{Planck2016}), are, for the most part, in good quantitative agreement with the predictions of the Lambda Cold Dark Matter ($\Lambda$CDM) paradigm. However, a key prediction of the theory -- the mass function of dark matter halos -- significantly differs from the observed galaxy stellar mass function at both the low-mass and high-mass ends (see \citealt{Wechsler2018} and references therein), which reflects the complex, and still poorly-understood, dependence of the physics of star formation on the halo mass and the environment. 

In order to reproduce the observations at the high-mass end, a crucial ingredient required by most theoretical models (see \citealt{Somerville2015} and references therein) is the so-called AGN feedback, which refers to the effects produced by the active nucleus activities (winds, jets, radiation) of a massive galaxy on the surrounding interstellar medium (ISM) and circum-galactic medium (CGM). The concept of AGN feedback was initially introduced by \citet{Silk1998} and \citet{Haehnelt1998} to explain the observed tight correlations among black hole mass (M$\rm{_{BH}}$), bulge mass/luminosity and velocity dispersion. Recently, depending on the nature of energy output, two major modes of AGN feedback are being considered: radiative and kinetic feedback (see the review of \citealt{Fabian2012} and references therein). Kinetic mode, sometimes also known as radio mode, refers to the feedback effects generated by the mechanical energy of radio jets which are often observed when AGN radiative activities are operating at low levels. In contrast, radiative mode refers to the feedback effects occurring when AGNs are very luminous. In this work, we will specifically focus on the radiative AGNs.

In the absence of AGN feedback, cosmological simulations under the $\Lambda$CDM paradigm produce too many massive galaxies compared to the observations \citep[e.g.][]{Oppenheimer2010,Kaviraj2017} and, the simulated massive galaxies also are too blue \citep[e.g.][]{Hatton2003} and too compact \citep[e.g.][]{Peirani2017}. 
For the simulations, one resolution to those discrepancies is to add the sub-grid AGN feedback models to suppress star formation in massive galaxies, a process generically referred to as AGN quenching. While including such models has become increasingly popular in modern cosmological simulations, a big concern is the large uncertainty on how to properly implement AGN physics and couple the feedback effects to the ISM \citep[e.g.][]{DiMatteo2005,Booth2009,Weinberger2017}. It is therefore of great importance to observationally investigate the effects of AGNs on the host galaxies. 

Taking advantage of deep and high-angular resolution X-ray observations, significant progress has been recently made in understanding the relationship between X-ray AGNs \citep[e.g.][]{Xue2016, Luo2017,Fornasini2018,Brown2019} and the properties of their host galaxies \citep[e.g.][]{Xue2010,Yang2017,Yang2018,Kocevski2017}. Yet, observational evidence of the feedback effects from X-ray AGNs is far from conclusive. For example, at $z\approx2$, where both the quasar activities \citep{Hasinger2005} and cosmic star formation rate density \citep{Madau2014} peak, AGN feedback (if any) is expected to be strong. A number of studies have been carried out to investigate the star formation properties for the host galaxies of X-ray AGNs out to z$\sim$3 \citep[e.g.][]{Lutz2010,Santini2012,Rosario2012,Rovilos2012,Page2012,Harrison2012,Barger2015, Hatziminaoglou2010,Harrison2012,Stanley2015,Barger2019}. While many of these studies have consistently shown that the median star formation intensity in galaxies hosting moderate luminous X-ray AGNs ($42<LogL_X<44$) is similar to that in normal SFGs, diverging conclusions emerge in luminous ($LogL_X>44$) X-ray AGN hosts. For example, using far-infrared (FIR) luminosity as the star formation rate estimator, some groups \citep[e.g.][]{Page2012,Barger2015} reported suppressed star formation in luminous X-ray AGN hosting galaxies, while others \citep[e.g.][]{Lutz2010,Santini2012,Rovilos2012} reached the opposite conclusion that their samples of luminous X-ray AGNs show enhanced star formation. Yet, other investigators \citep{Harrison2012, Stanley2015} reported no dependence of star formation activity on the X-ray AGN luminosity.

One general issue for the observational studies of the effects of the AGN presences on hosting galaxy properties is the interpretation of the data. Empirically speaking, compared with non-AGNs, any distinct distribution of physical properties of AGN hosts can be attributed to the presences of AGNs. However, such attribution does not necessarily imply a causal relationship in the sense that the real cause(s) behind might be some other mechanisms which are also likely to trigger AGN activities, even if the latter is only weakly related, if any, to the properties of the host. One example is galaxy major merger, where
strong gravitational torques induced by the merging galaxy/galaxies can drive gas to the center which as a result can simultaneously (1) make the gas distribution more nucleated; (2) trigger a central starburst and increase galactic wide star formation rate and (3) trigger a bright AGN \citep[e.g.][]{Mihos1996,Sanders1988,Hopkins2006}.

The other issue comes from the AGN selection, which is the focus of this work. While selecting AGNs in X-ray has been shown to be one of the most efficient ways to study them, it is by no mean complete. Since X-ray photons (soft ones in particular) heavily suffer from the line-of-sight obscuration, X-ray selection itself can miss a significant fraction of obscured AGNs \citep[e.g.][]{Gilli2007}, which become increasingly important at higher redshifts where the fraction of obscured AGNs becomes larger \citep[e.g.][]{Liu2017}. To get a comprehensive observational picture of AGN feedback, the missing population of AGNs must be taken into account.

Observations at mid-IR (MIR) are efficient to identify those highly-obscured AGNs missed by the X-ray selection (e.g. \citealt{Daddi2007,Donley2008}), because MIR directly probes the re-processed radiation from the absorbed X-ray, UV and optical photons. The primary issue of studying AGNs in MIR is the confusion with light from the host galaxies. Unless the AGNs are powerful enough, their spectral energy distribution (SED) in MIR is always a comparable mixture of the reprocessed emission from AGNs and the emission from star formations. Despite that the shape of AGN MIR spectra remains to be characterized in details by future studies, e.g. with JWST \citep[e.g.][]{Kirkpatrick2017}, substantial progress has been recently made in identifying IR AGNs using the broad band photometry in MIR, including the selection methods based on {\it Spitzer} IRAC colors  \citep[e.g.][]{Lacy2004,Stern2005,Donley2008,Donley2012,Kirkpatrick2013}, {\it WISE} colors \citep[e.g.][]{Eisenhardt2012,Stern2012}) and SED decomposition techniques \citep[e.g.][]{Armus2007,Pope2008,Kirkpatrick2012,Berta2013}. Finally, some progress has also been made in understanding the MIR spectroscopic properties of AGNs at high redshifts using the observations from {\it Spitzer}/IRS \citep{Kirkpatrick2013}, although such studies are only limited to the luminous AGNs given the MIR sensitivities of current instruments.

In this work, we present a comparative study of the properties of the host galaxies of X-ray- and IR-selected AGNs. Specifically, we will compare the star-formation and morphological properties of the AGN and the non-AGN hosting galaxies, focusing on the effects, if any, of the presences of AGNs on their host galaxies. Throughout this paper, we adopt a $\Lambda$CDM cosmology with $\Omega_m = 0.3$, $\Omega_\Lambda = 0.7$ and $\rm{h = H_0/(100 kms^{-1}Mpc^{-1}) = 0.7}$.

\section{Sample selection}
In this Section, we describe in details about the sample selections in this work. Table \ref{tab:info} lists the number of galaxies in each sample.

\begin{table}
\begin{flushleft}
\begin{tabular}{ c|c|c|c } 
\hline
Field & Parent sample & X-ray AGN & IR AGN  \\
 \hline
 GOODS-S & 2500 & 164 & 69 \\ 
 GOODS-N & 2309 & 74 & 69 \\ 
    All & 4809 & 238 & 138 \\ 
 \hline
\end{tabular}
 \caption{The number of galaxies in each sample.}
\label{tab:info}
\end{flushleft}
\end{table}

\subsection{Parent Sample}
\begin{figure}
\gridline{\fig{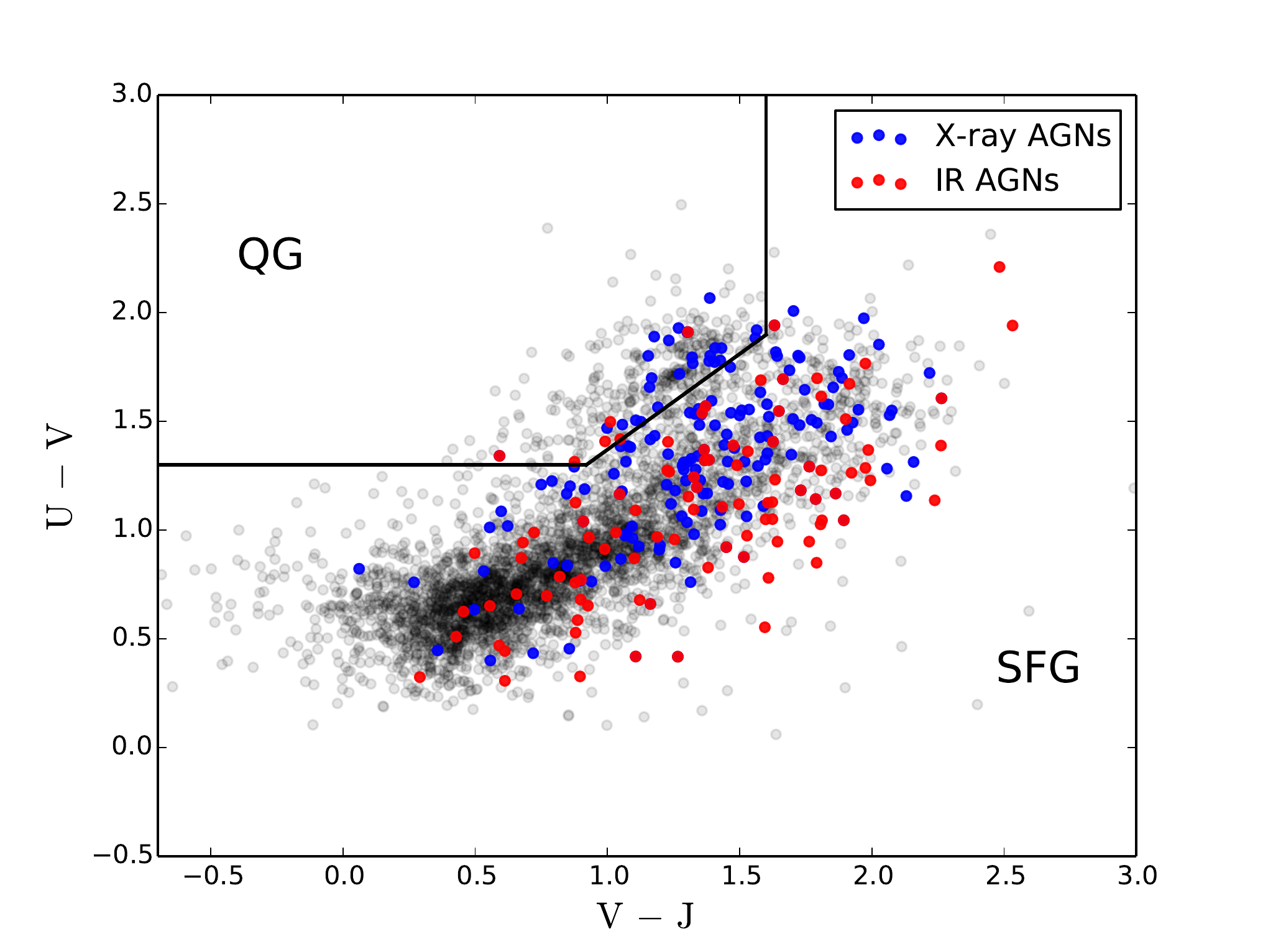}{0.5\textwidth}{}
        }
\caption{Rest-frame UVJ-color diagram. Black dots are galaxies in the parent sample. Black solid lines mark the boundaries used to separate SFGs and QGs, where QGs are in the region with $\rm{U-V>0.88\, (V-J) + 0.49,\,U-V>1.3}$ and $\rm{V-J <1.6}$. Overplotted blue and red dots are galaxies identified as X-ray AGNs and IR AGNs respectively (see Section \ref{sec:X-ray AGN} and \ref{sec:IR AGN} for details).} \label{fig:uvj}
\end{figure}

Our parent sample is the same as that of \citet{Lee2018}, which is drawn from {\it Hubble Space Telescope} ({\it HST}) H$_{160}$-band selected 4809 galaxies in the GOODS-S (2500 galaxies) and GOODS-N (2309 galaxies) fields. Both fields have the deep {\it HST}/ACS data acquired during the GOODS survey \citep{Giavalisco2004} and the deep {\it HST}/WFC3 data acquired during the CANDELS survey \citep{Grogin2011, Koekemoer2011}. The sample galaxies are selected to be in the redshift range of 1.2 $<$ z $<$ 4 with $\rm{M_* >10^{9.5}\,M_\odot}$ and their isophotal H$_{160}$ signal-to-noise ratios (SNRs) are required to be SNR $>$ 10 in order to get good photometry and hence high-quality photometric redshifts (photo-z) and spectral energy distribution (SED) fitting measures. The full sample is divided into two subsamples according to star formation properties of the galaxies (Figure \ref{fig:uvj}). Star-forming galaxies (SFGs) and quiescent galaxies (QGs) are distinguished using the rest-frame UVJ-color diagram (see Section \ref{sec:SED} for the measurements of rest-frame colors) that was initially proposed by \citet{Williams2009}. In this work, we adopt the SFG-QG separation boundary from \citet{Schreiber2015} which is built upon CANDELS galaxies and has been demonstrated to be valid up to $z = 4$. \citet{Ji2018} used the simulation done by \citet{Guo2013} to show that the parent sample is $\rm{\approx 80\%}$ complete down to $\rm{10^9\,M_\sun}$. In this work, we decide to ignore galaxies with stellar mass less than $\rm{10^{9.5}\,M_\sun}$ because (1) a lower-mass galaxy statistically tends to have a lower metallicity \citep[e.g.][]{Tremonti2004} such that the AGN selection based on IR colors can mimic an AGN when really there is none \citep{Satyapal2014,Hainline2016,Marleau2017,Kaviraj2019} and (2) it is hard for a $\rm{<10^{9.5}\,M_\sun}$ galaxy's black hole to accrete actively enough to become an AGN from a theoretical point of view \citep[e.g.][]{Fontanot2011}. 

\subsection{X-ray AGNs}\label{sec:X-ray AGN}
The identifications of X-ray AGNs are done by spatially cross-matching the CANDELS catalog of the parent sample with the AGN catalogs of the 7Ms {\it Chandra} Deep Field South \citep[CDF-S,][]{Luo2017} and the 2Ms {\it Chandra} Deep Field North \citep[CDF-N,][]{Xue2016}. Details of AGN classifications in both fields can be found in \citet{Xue2016} and \citet{Luo2017}. In short, an X-ray source is classified as an AGN if it meets the criteria built upon intrinsic X-ray luminosity threshold and spectral shape (hardness ratio), as well as the flux ratio between X-ray and other bands (optical, IR and radio). We cross-match the coordinates of the parent sample (H$_{160}$ coordinates from the CANDELS catalog) with X-ray AGNs using a 0.5'' radius, the same matching radius has also been used in other works \citep[e.g.][]{Yang2017,Yang2018}. We have checked that our results do not change if we use a smaller (0.3'', 0.4'') or larger (0.6'', 0.7'') radius. To further secure the cross-matching, we request the redshifts of sample galaxies ($\rm{z_{CANDELS}}$) and those assigned to the matched X-ray counterparts ($\rm{z_{Xray}}$) are either the same if spectroscopic redshifts are available or within 10\% difference (i.e. $\rm{|z_{Xray}-z_{CANDELS}|/z_{CANDELS} \le 10\%}$) if photo-z are used. The 10\% tolerance of photo-z difference is because of the different photo-z catalogs used in \citet{Luo2017} and \citet{Lee2018}. We have checked that our results do not change if we set the tolerance to be 5\% or 15\%.
With a 0.5'' matching radius and 10\% tolerance of the photo-z difference, we find that 238 galaxies in the parent sample have X-ray AGNs (no duplicated match). 

It is worth pointing out that the approach of searching for counterpart within a small radius is not ideal for faint galaxies, given both the centroid errors of X-ray sources and sometimes high background optical/NIR source density. An alternative approach is to use the likelihood-ratio method which has been carried out in both fields (see Section 2.3.3 in \citealt{Xue2016} and Section 4.2 in \citealt{Luo2017} for details). We have checked, by comparing the 238 cross-matched X-ray AGNs with the counterparts identified using the likelihood-ratio technique, the two matching results are the same, which is not surprising given that the parent sample are relatively bright (recall that we require all galaxies have SNR $>$ 10 in H$_{160}$) and the addition redshift difference tolerance can further secure our cross matching.

\subsection{IR AGNs}\label{sec:IR AGN}

Because of the availability of deep {\it Spitzer}/IRAC photometry in the GOODS fields, IR AGNs are selected using the IRAC color-color diagram from \citet{Donley2012}, which was built on a large sample of galaxies in the COSMOS field. This selection is able to effectively identify IR AGNs at high redshifts, which has been demonstrated by many other surveys where IRAC photometry is available \citep[e.g.][]{Mendez2016,Delvecchio2017,Leung2017,Donley2018}.

Two IRAC colors are used to select IR AGNs, namely $\rm{x = Log(S_{5.8}/S_{3.6})}$ and $\rm{y = Log(S_{8.0}/S_{4.5})}$. A galaxy is classified to be an IR AGN host if it meets the following criteria:
\begin{align} \label{eqn:C1}
&\begin{cases}
	1.21x-0.27\le y\le 1.21x+0.27
	\\
	x\ge0.08
	\\
	y\ge0.15
	\\
	S_8>S_{5.8}>S_{4.5}>S_{3.6}
\end{cases}\\ \label{eqn:C2}
&\begin{cases}
	z\ge2.7
	\\
	x/y\le0.95
	\\
	\rm{Log\,S_8/S_{3.6}\ge}\begin{cases}
		0.39z-0.69\quad if\,\,z=2.7-3.1
		\\
		0.18z-0.04\quad if\,\,z=3.1-4.0
	\end{cases}
\end{cases}
\end{align}

\begin{figure}
\gridline{\fig{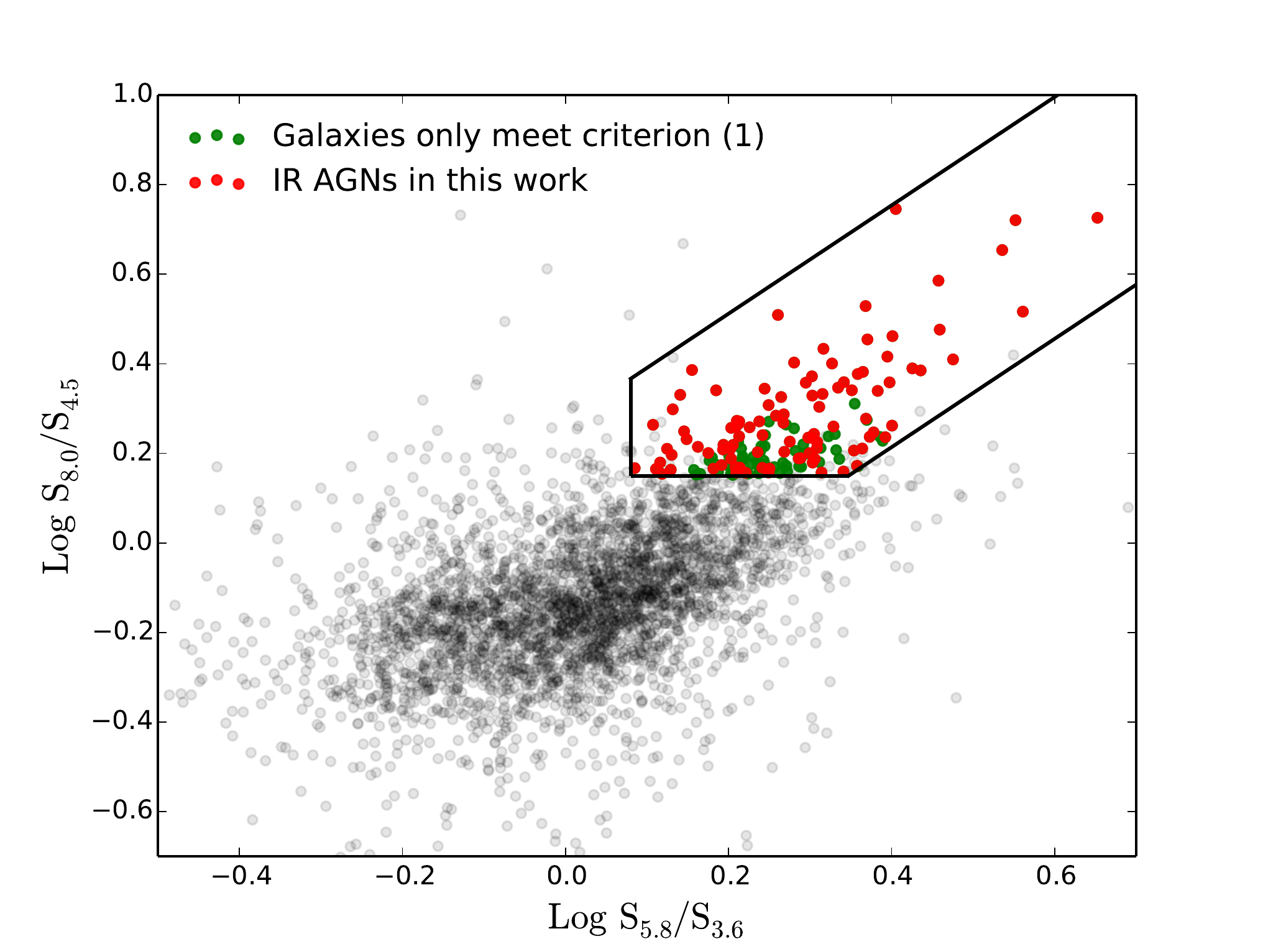}{0.5\textwidth}{}
        }
\caption{IR AGNs selected by the IRAC color-color diagram of \citet{Donley2012}. The red dots are the final IR AGNs, while the green dots are galaxies which only pass criterion (\ref{eqn:C1}). The black solid lines mark the region enclosed by the first three equations of criterion (\ref{eqn:C1}). Note that there are still a small number of grey points in this region, those are galaxies whose fluxes in IRAC four bands are {\it not} monotonically increasing (i.e. do not pass the fourth equation of criterion (\ref{eqn:C1})). The details of the selection method can be found in Section \ref{sec:IR AGN}.} \label{fig:IR_Select}
\end{figure}

Black solid lines in Figure \ref{fig:IR_Select} form the boxy region defined by the first three equations of criterion (\ref{eqn:C1}). Galaxies within it have AGN-like SEDs (see Figure 2 of \citealt{Donley2012}), which has further been confirmed by \citet{Kirkpatrick2013} for a sample of 24 $\rm{\mu m}$-selected 0.5$<z<$4 galaxies with deep {\it Spitzer}/IRS spectroscopy. The stellar bump ($\approx 1.6\mu m$) of normal galaxies at $z>2$ are redshifted into the IRAC 4.5, 5.8 and 8 $\rm{\mu m}$ bands, which can effectively contaminate the IR AGN selection. To overcome this, the fourth equation of criterion (\ref{eqn:C1}) therefore is required to exclude galaxies in the boxy region with non-monotonically rising SEDs. For galaxies at $z>2.7$, the additional criterion is required given that the contamination becomes even worse because the stellar light might dominate all IRAC bands. With the additional criterion (\ref{eqn:C2}), \citet{Donley2012} showed that it can effective exclude (1) galaxies whose spectral shapes in IRAC four bands are consistent with the rest-frame 1.6 $\rm{\mu m}$ stellar bump (the 2nd equation) and (2) galaxies which can be possibly fit by the reddest LIRG/ULIRG templates of \citet{Rieke2009} (the 3rd equation).

With the selection described above, we find 138 IR AGNs, among which 45 ($\approx33\%$, similar percentage (27\%) as found by \citealt{Delvecchio2017}) are also identified as X-ray AGNs. As demonstrated by \citet{Kirkpatrick2013}, the IRAC color selection can miss a fraction of MIR spectroscopically confirmed AGNs which can be better recovered by adding {\it Spitzer}/MIPS 24 $\rm{\mu m}$ and far-infrared (FIR) {\it Herschel}/PACS 100 $\rm{\mu m}$ and SPIRE 250 $\rm{\mu m}$ photometry to the selection. Due to the sensitivity and angular resolution of MIR and FIR observations, unfortunately, only $\rm{\approx}$ 14\% of galaxies in the parent sample simultaneously have 24 $\rm{\mu m}$ and 100/250 $\rm{\mu m}$ photometry. We have checked, among the 46 galaxies which are selected as IR AGNs using the selection criteria of \citet{Kirkpatrick2013}, 39 of them have already been picked up by our IRAC color selection. 

\section{Measurements \& Data Analysis}

\subsection{SED fitting}\label{sec:SED}
Physical parameters, including M$_*$, star formation rate (SFR) and rest-frame colors, are derived via SED fitting. In the following, we detail the fitting procedure and outline the systematics of the measurements. 

Throughout this work, we adopt the SED fitting results of \citet{Lee2018} (hereafter, Lee2018), which uses the \citet{Bruzual2003} stellar population synthesis code, assumes a \citet{Chabrier2003} initial mass function (IMF), fixed solar metallicity and the \citet{Calzetti2000} dust attenuation law. Lee2018 takes advantage of the deep CANDELS multi-wavelength photometry that covers from the rest-frame UV to FIR and the official CANDELS photometric redshift catalog (see \citealt{Dahlen2013, Hsu2014}) where full probability density functions are used in the determination of photometric redshift. A key feature of the Lee2018 SED modeling approach is that the fitting procedure applies an advanced Monte Carlo Markov Chain algorithm to treat star formation history (SFH) as a free parameter during the fits. In Lee2018, using mock observations derived from semi-analytical models of galaxy evolution, it has been demonstrated that their measurements of $M_*$, SFR and luminosity-weighted stellar age are much more robust than those derived by setting the functional form of SFH to a pre-assigned type.

 A concern of using the Lee2018 measurements, in particular for the AGN hosts, is the ignorance of the AGN contribution during the SED fitting. To check these systematics, we have run another set of SED fitting using {\sc Sed3fit} \citep{Berta2013} where the AGN component is included to the modeling. We refer readers to Appendix \ref{app:sed_test} for a detailed analysis for the uncertainty of individual parameters derived in this way. In short, the comparisons between Lee2018 and {\sc Sed3fit} results suggest that, when averaged on the galaxy mix of our sample, neglecting the AGN component in SED modeling
 \begin{itemize}
     \item statistically does not affect the M$_*$ measurement in a significant way, although we do find that properly including AGN contribution is crucial for the M$_*$ measurement of broad line AGNs (BL AGNs). BL AGNs however are a very small fraction ($\approx$5\%) of the entire AGN sample and we have checked that our results are insensitive to including/excluding them.
    \item can lead to an $\approx 0.1$ dex overestimation of SFRs for the AGN hosts. This systematics will be taken into account in the following discussions with regard to the star formation properties of the AGN hosts.
    \item statistically does not significantly affect the measurement of rest-frame apparent (dust-attenuated) colors U$-$V and V$-$J, and dust-corrected colors \UVc\ and \VJc. We also notice that the scatter of the \VJc\ measurement is slightly larger in IR AGNs than X-ray AGNs and non-AGNs, which is likely due to the generally larger AGN contribution to the J band in IR AGN hosts.
 \end{itemize}
 
 While the {\sc Sed3fit} tests reveal some tensions of using Lee2018 measurements for AGN hosts (BL AGNs in particular), fortunately, the rather tight correlations between the parameters derived from the two SED fittings (see Figures in Appendix \ref{app:sed_test}) suggests that the overall determination of the parameters that we have considered is {\it insensitive} to the inclusion of the AGN component. Quantifying systematic differences among different SED fitting procedures to a finer degree of accuracy is beyond the scope of this work. We decide to use Lee2018 measurements because parts of the following discussions rely on the measurement of properties of galaxies on the star-forming main sequence, which has been carefully done for the parent sample of Lee2018. Using different SED fitting algorithms and assumptions for AGNs and non-AGNs might introduce systematic bias owing to the systematic shifts in the measurements of M$_*$ and SFR (see Appendix \ref{app:sed_test} and also other works like \citealt{Leja2019}) that, as small or rare as they are, we prefer to avoid.
 
\begin{figure}
\gridline{\fig{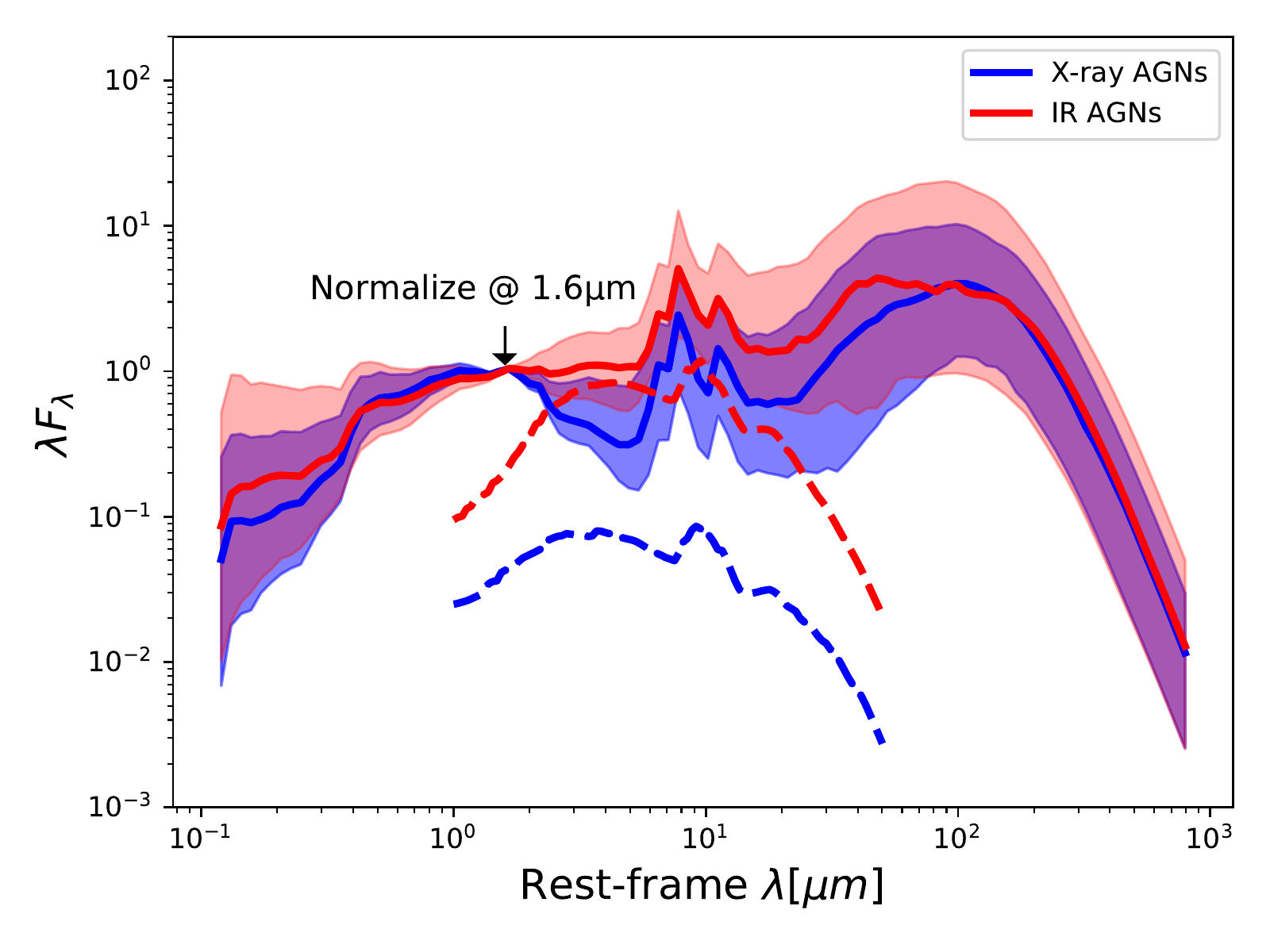}{0.47\textwidth}{}
    }
\caption{{\sc Sed3fit}-derived SEDs of X-ray (blue) and IR (red) AGN hosts. Each best-fit model is normalized to the rest-frame 1.6$\micron$. Solid lines show the medians and shaded regions show the 16th-84th percentile ranges. The dashed lines show the median AGN contributions.} \label{fig:SEDs}
\end{figure}

\begin{figure}
\gridline{\fig{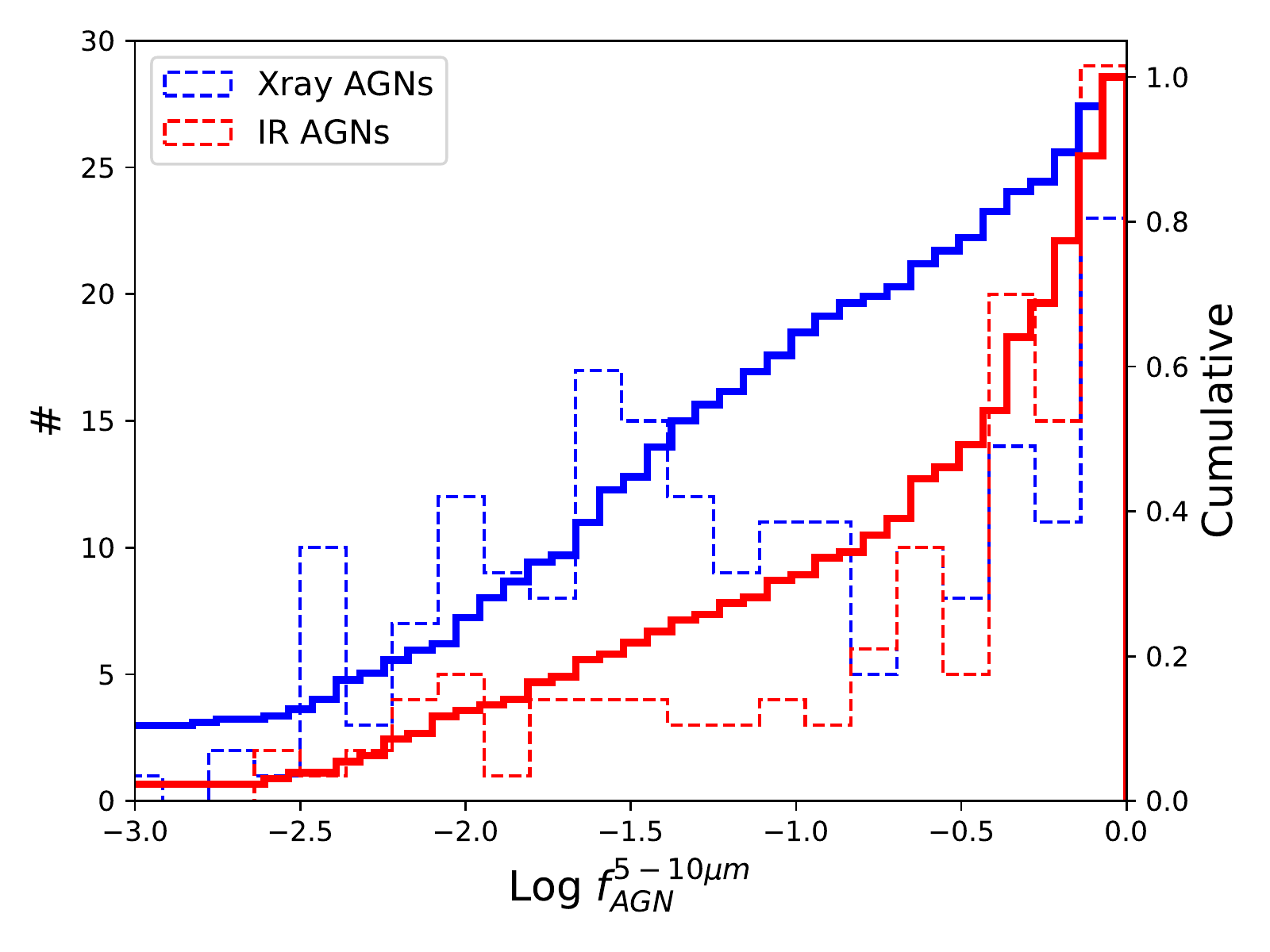}{0.47\textwidth}{}
    }
\caption{{\sc Sed3fit}-derived AGN contributions to rest $5-10\mu m$ IR luminosity ($L_{AGN}^{5-10\mu m}/L_{Total}^{5-10\mu m}=f_{AGN}^{5-10\mu m}$). Blue and red dashed histograms show the distribution of X-ray and IR AGNs respectively. Corresponding cumulative distributions are shown as solid lines.} \label{fig:agn_fraction}
\end{figure}

In addition to comparing with Lee2018 measurements, running {\sc Sed3fit} also helps us validate our IRAC color selection method, as well as quantify AGN luminosity for the IR AGNs (see Section \ref{sec:SF_Lbol} for details). Figure \ref{fig:SEDs} shows the best-fit SEDs of X-ray and IR AGNs derived by {\sc Sed3fit}. Significant AGN contribution to MIR flux is seen in IR AGNs, illustrating the good agreement between the results from the SED decomposition and the adopted IRAC color selection (Section \ref{sec:IR AGN}). The Figure also shows that identifying AGNs at MIR wavelength can sometimes be hard when galaxy stellar SED dominates the total light in the optical/IR part of the spectrum despite the clear presence of the AGN at X-ray wavelengths, which again highlights the importance of selecting AGNs in more than one wavelength range, as we have already discussed in Section \ref{sec:intro}. Figure \ref{fig:agn_fraction} further shows the distribution of the ratio of AGN IR luminosity divided by total rest $5-10\micron$ IR luminosity ($f_{AGN}^{5-10\micron}$). Like those seen in the best-fit SEDs, the IR AGNs have much higher contribution to MIR flux than the X-ray AGNs. 

\subsection{Morphological measurements} \label{sec:mor}
Morphological properties of the AGN hosts and normal galaxies are derived by fitting the CANDELS H$_{160}$ images with 2-dimensional (2D) light profiles using {\sc Galfit} \citep{Peng2010}. Key morphological parameters that we are interested in are: effective radius (R$_e$, a.k.a half light radius), S\'{e}rsic index (n), stellar mass surface density within effective radius (\Se), stellar mass surface density within central 1 kpc (\Sone) and fractional mass within central 1 kpc (\Mone). In the following, we will first introduce the basic setup of {\sc Galfit} and then describe in details on how we measure the aforementioned morphological parameters and their uncertainties. We will also test the validity of the assumed 2D light profile model and discuss the relevant systematics. We will finally describe in details on our purposes and advantages of using \Mone\ to quantify galaxy compactness. 

\subsubsection{{\sc Galfit} fittings and parameter uncertainties}\label{sec:galfit}

Before running {\sc Galfit}, we center on each sample galaxy to make a 6''$\times$6'' cutout. We adopt the H$_{160}$ point spread function (PSF) from the CANDELS team \citep{vanderWel2012}. To get rid of the isophotes contamination from the neighboring galaxies, we first find all galaxies in the cutout image with the aid of the CANDELS H$_{160}$ segmentation map. Then, rather than fitting the neighboring galaxies, we fix and model their light profiles using the best-fit 2D S\'{e}rsic profiles obtained by \citet{vanderWel2012}. For the background sky level of each cutout, we have modelled it in two different ways, namely to (1) set the sky as a free parameters and let {\sc Galfit} find the best-fit value and (2) fix the sky level to be the median pixel value derived from a 3$\sigma$ clipping of the pixel values in the cutout image after masking out all H$_{160}$ detected sources. It turns out that our results are insensitive to the method chosen so we decide to fix the sky level as the median pixel value of each cutout. We fit each target galaxy with a single 2-D S\'{e}rsic profile, from which we can directly obtain n and R$_e$ (= R$_{e,maj}\times\sqrt{b/a}$), as well as $\Sigma_e = M_*/(2\pi R_e^2)$. With the best-fit S\'{e}rsic profile in hand, following the derivation of \citet{Graham2005}, we can get the fractional stellar mass within central 1 kpc through
\begin{equation} \label{eqn:m1m}
\frac{M_{1\rm{kpc}}}{M_*}=\frac{\gamma(2n,\,x)}{\Gamma(2n)},\,x=b_n(\frac{1\rm{kpc}}{R_e})^{1/n}
\end{equation} 
where $\gamma/\Gamma$ is the ratio of incomplete gamma function divided by complete gamma function. When n$>$0.36, $b_n$ is calculated using the approximate expression proposed by (\citealt{Ciotti1999}, their Equation 18, accurate to better than $\rm{10^{-4}}$), otherwise $\rm{b_n}$ is calculated by numerically solving $\Gamma(2n)=2\gamma(2n,b_n)$. Finally, we can obtain the stellar mass surface density within central 1 kpc through 
\begin{equation}
\Sigma_{\rm{1kpc}} =\frac{M_{\rm{1kpc}}}{\pi \cdot1\rm{kpc}^2} = \frac{M_*}{\pi \cdot1\rm{kpc}^2}\frac{\gamma(2n,\,x)}{\Gamma(2n)}
\end{equation}

\begin{figure}
\gridline{\fig{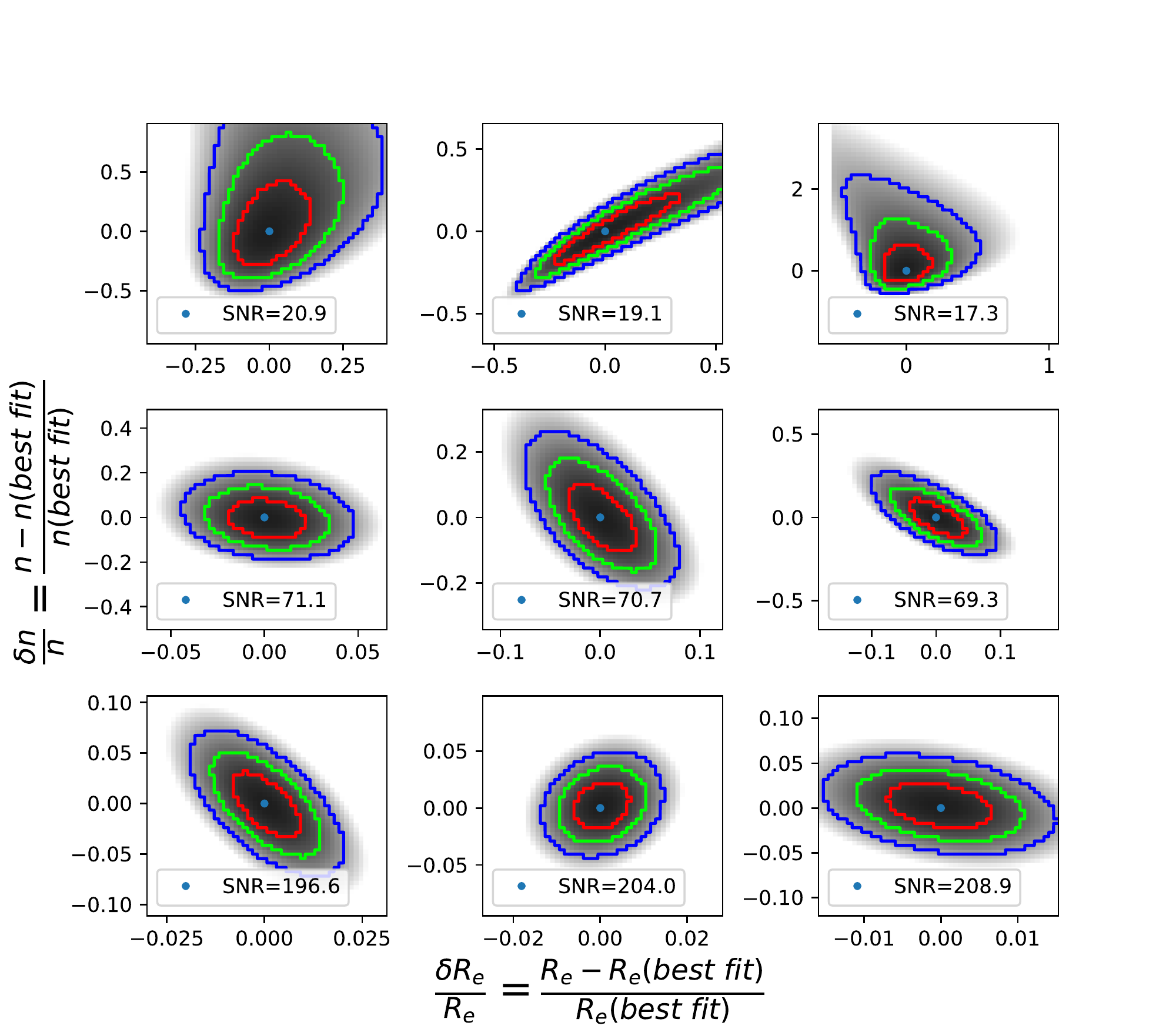}{0.5\textwidth}{}
}
\caption{The R$_e$-n covariances of randomly selected 9 examples. The first, second and third rows show the cases with H$_{160}$ SNR $\sim$ 20, 70 and 200 respectively. Red, green and blue lines show the corresponding 1$\sigma$, 2$\sigma$ and 3$\sigma$ confidence contours.}\label{fig:examples}
\end{figure}

\begin{figure}
\gridline{\fig{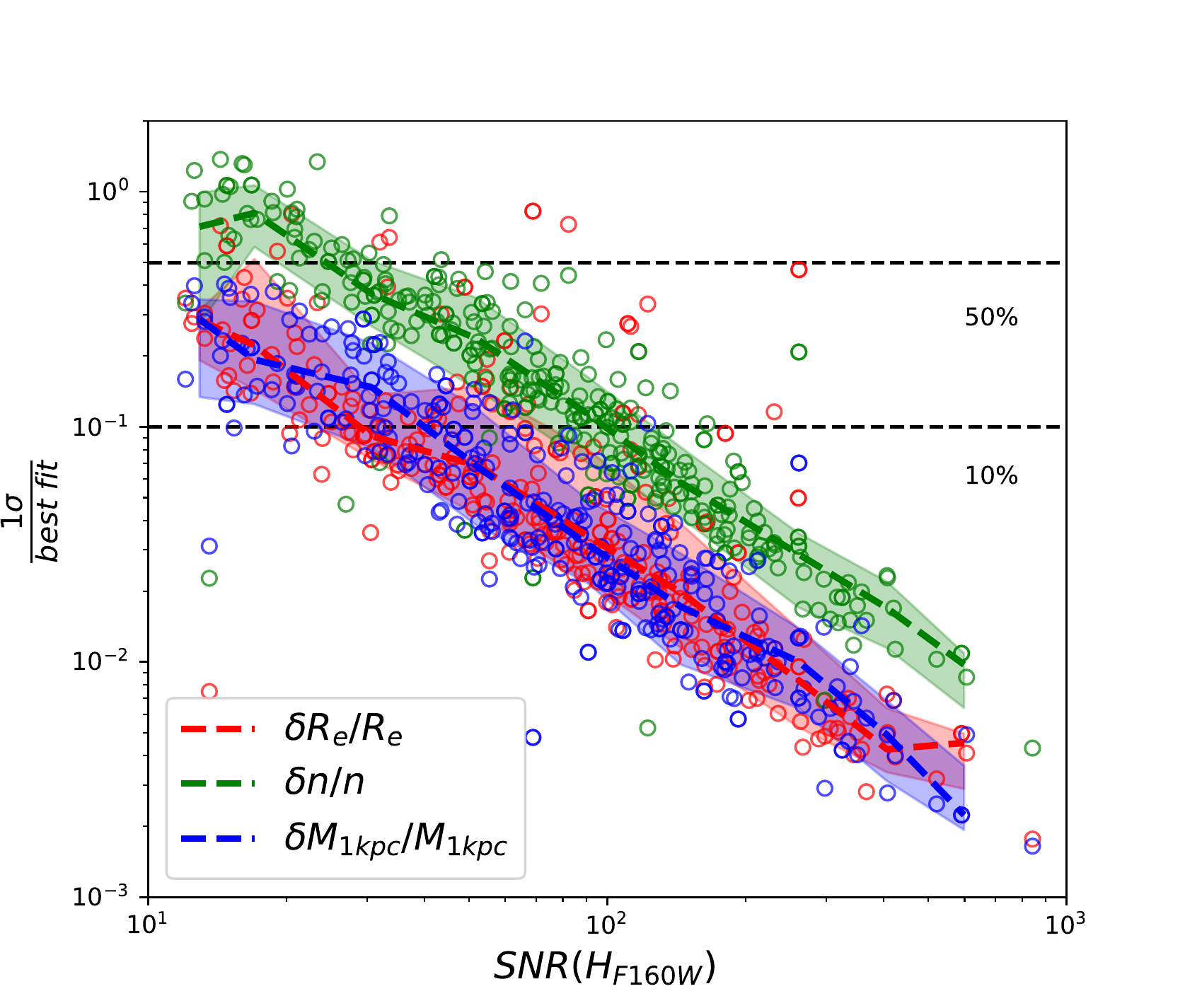}{0.5\textwidth}{}
}
\caption{The uncertainty of R$_e$ (red), n (green) and \Mone\ (blue) as a function of H$_{160}$ SNR. Y-axis shows the ratio of 1$\sigma$ uncertainty divided by the best-fit value, where the 1$\sigma$ range is from the covariance analysis (see Section \ref{sec:mor} for details). Each circle shows the measurement of an AGN in our sample. The dashed lines show the median and the shadow regions show the 16th-84th percentile range. Also marked as the horizontal dashed lines are the 50\% and 10\% accuracy lines.}\label{fig:SNR}
\end{figure}

Quantifying the uncertainty of these morphological parameters is non-trivial due to the covariance between parameters \citep[e.g.][]{Ji2020}. In this work, we have conducted the covariance analysis by measuring the covariance between R$_e$ and n for the entire AGN sample, aiming to estimate error bars of each aforementioned morphological parameter. To do so, we first run {\sc Galfit} to get the best-fit values of all free parameters and then use {\sc Galfit} to generate a number of models by changing n and R$_e$ while fixing any other parameters to the best-fit values. We then calculate the $\chi^2$ distribution of these new models to get the n-R$_e$ covariance. Figure \ref{fig:examples} shows covariances of the randomly-selected 9 AGNs with different H$_{160}$ SNRs. Diverse shapes of the covariances are clearly seen even when sources have similar SNRs, demonstrating that the individual determination of n and R$_e$ is non-trivial. For each AGN, we derive the 1$\sigma$ uncertainty ranges of R$_e$ and n using the covariance. We then plug all possible R$_e$-n combinations along the 1$\sigma$-$\chi^2$ contour into Equation \ref{eqn:m1m} to get the corresponding 1$\sigma$ uncertainty of \Mone. Figure \ref{fig:SNR} shows the derived uncertainty as a function of H$_{160}$ SNR. While the uncertainty overall decreases as SNR increasing, the uncertainty of different morphological parameters is different. For a detection with a descent SNR ($\ge 20$), while R$_e$ is reasonably well-constrained with a typical $\lesssim10\%$ 1$\sigma$ uncertainty, the uncertainty of n can be as large as $\approx 50\%$. Even for a SNR $\approx100$ detection, the uncertainty of n can still be $\approx10\%$. Importantly, although n itself is usually not well-constrained, the measurement of \Mone\ (the combination of R$_e$ and n) is about as good as R$_e$.

\subsubsection{Validity of the single S\'{e}rsic profile assumption}\label{sec:valid_single_sersic}

\begin{figure*}
\gridline{\fig{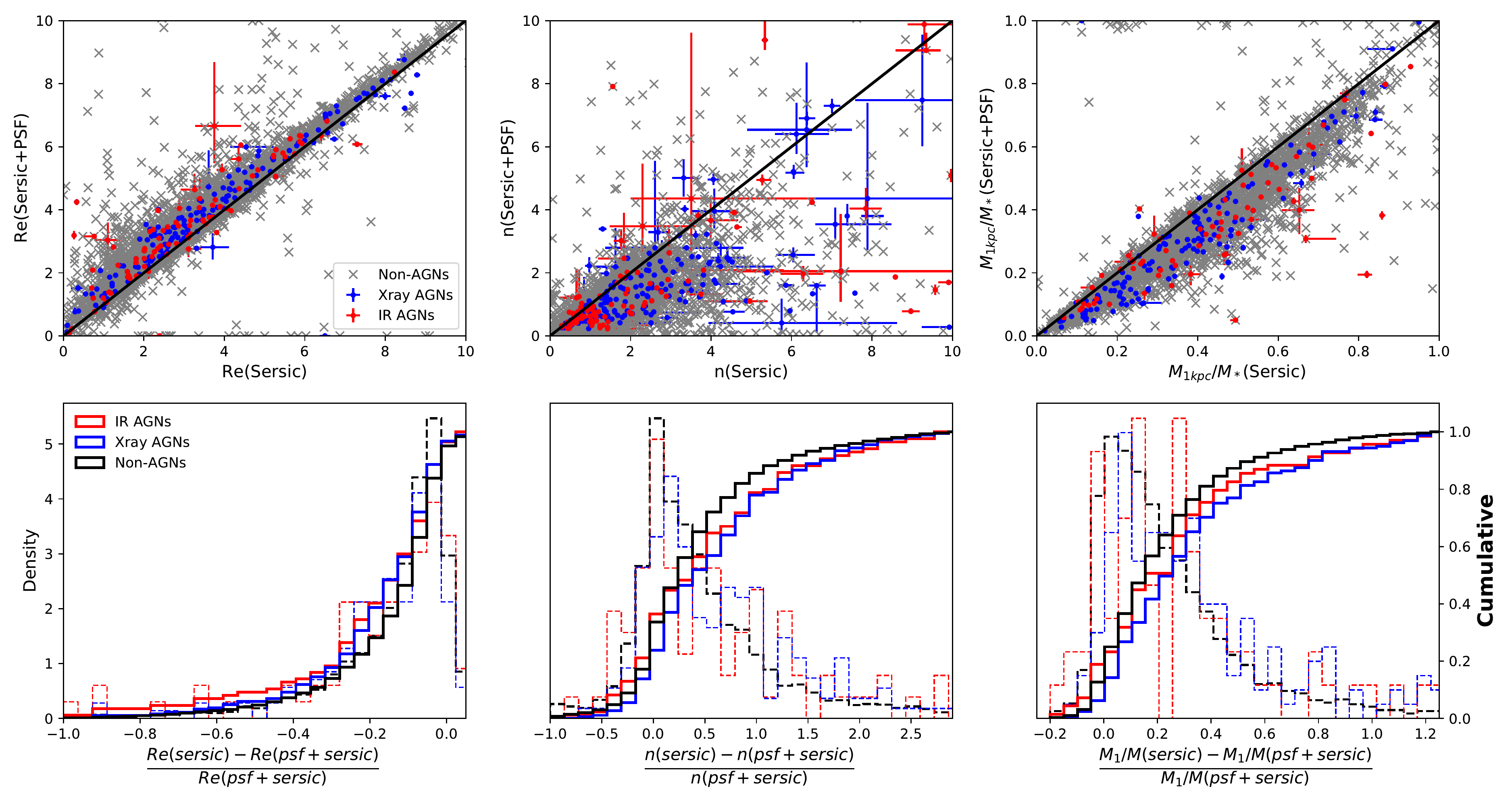}{0.9\textwidth}{}
}
\caption{The comparisons of morphological parameters from the two different assumed light profiles, i.e. S\'{e}rsic-only and S\'{e}rsic+PSF models (see Section \ref{sec:valid_single_sersic} for details). X-ray, IR AGNs and non-AGNs are color-coded with blue, red and grey respectively. The black solid line marks the one-to-one relation. The bottom panels show distributions and cumulative distributions of the relative changes of each individual parameter.}\label{fig:Compare}
\end{figure*}

We now test the validity of the single S\'{e}rsic profile assumption that we made so far for the morphological measurements. The non-stellar AGN radiation can ``pollute'' the stellar light distribution and hence introduce a systematic bias in the morphological measurements of host galaxies. To test for this systematic, we have re-done the morphological measurements assuming a different model, i.e. a 2D S\'{e}rsic profile plus a nuclear point source (S\'{e}rsic+PSF). Similarly as we did for the single 2D S\'{e}rsic profile fittings, we have also measured the covariances between R$_e$ and n for the AGN hosts and derived the corresponding 1$\sigma$ errors. In addition, S\'{e}rsic+PSF fittings have also been done among the non-AGNs in the GOODS-S, which we did in order to compare with the AGNs. Figure \ref{fig:Compare} shows the comparisons of R$_e$, n and \Mone\ between the assumed two different light profiles (S\'{e}rsic-only and S\'{e}rsic+PSF). We see clear correlations of R$_e$ and \Mone\, while a big scatter of S\'{e}rsic index n between the two measurements, suggesting a qualitatively {\it insensitive} dependence of R$_e$ and \Mone, but a much more sensitive dependence of n on the assumed light distribution. This further supports what we have already found in Figure \ref{fig:SNR} that n is not as well-constrained as R$_e$ and \Mone. 

Compared with the S\'{e}rsic-only results, PSF+S\'{e}rsic leads to an increase of R$_e$ and a decrease of \Mone, which is expected since adding a nuclear point source is equivalent to fit a single S\'{e}rsic profile to an image with some fraction of central light removed. In other words, if the nucleated component really has the non-stellar origin like an AGN, stellar morphology of the host galaxy should be more extended (larger R$_e$ and smaller \Mone) than it is seen from the image. Owing to the limited image depth and spatial resolution at high redshift, however, it is hard to conclusively say if adding the central component to the fitting is physically necessary. For example, we notice that the fitting $\chi^2$ generally improves after adding the nuclear point-like component. In particular, the reduced-$\chi^2$ improves by 10\% for the PSF+S\'{e}rsic model. But, we do not know if the improvement of $\chi^2$ indicates the physical requirement of the central component, or simply because the PSF+S\'{e}rsic model has more free parameters than the S\'{e}rsic-only model and (of course) can fit the data ``better''. We can in principle compare the $\chi^2$ change with the expected change that can be theoretically calculated if all the parameters are {\it independent} (which unfortunately is not the case, see Figure \ref{fig:examples}). Even if one can prove that the nucleated point-like source is a physically necessary component, it remains difficult to definitely disentangle its origin, which could be the non-stellar light from an AGN, or the stellar light from galaxy central structures like bulge, or both. It is worth mentioning here that, based on the PSF+S\'{e}rsic fitting results, we find a significant positive correlation, with a Pearson correlation test p-value of $7\times10^{-5}$, between AGN luminosity and $\rm{F_{PSF}/F_{Sersic}}$, i.e. the flux ratio of the PSF component divided by the S\'{e}rsic component. While such correlation can be simply explained in terms of the AGN contamination being more severe to the rest optical stellar morphology as the AGN becomes more luminous, we do find evidence that the real cause(s) behind {\it cannot} merely be the AGN contamination. We defer detailed analysis and discussions of this issue to an upcoming paper.

Distributions of the relative changes of R$_e$, n and \Mone\ are shown in the bottom panels of Figure \ref{fig:Compare}, where relative changes are larger for the AGN hosts than normal galaxies. Interestingly, compared with IR AGNs, the relative changes also seem to be larger for X-ray AGNs, which is consistent with the scenario that X-ray AGNs are less (relative to IR AGNs) obscured such that the central AGN light ``contanminates'' the optical stellar morphology more for X-ray AGNs.  The findings above seem to suggest that AGNs either require an extra nuclear non-stellar component for the morphological fitting, or to be preferentially embedded in galaxies that have developed a central compact structure, or both. Regardless of the actual physical reasons, which we will investigate in a future work, our findings suggest that the two component fitting for AGN hosts very likely is required and removing the nuclear light can reduce the correlation between AGN presence and galaxy compactness that has been found in previous works. Given the magnitudes of relative changes of R$_e$ and \Mone, however, we have checked that this will not change our conclusions that AGN prevalence is fundamentally tied to mass more so than compactness (see Section \ref{sec:color-mor} and \ref{sec:fagn_morp}). In the subsequent analysis, we will use the morphological parameters measured from the S\'{e}rsic-only fittings.

\subsubsection{Quantify galaxy compactness with \Mone}\label{sec:m1m}
\begin{figure}
\gridline{\fig{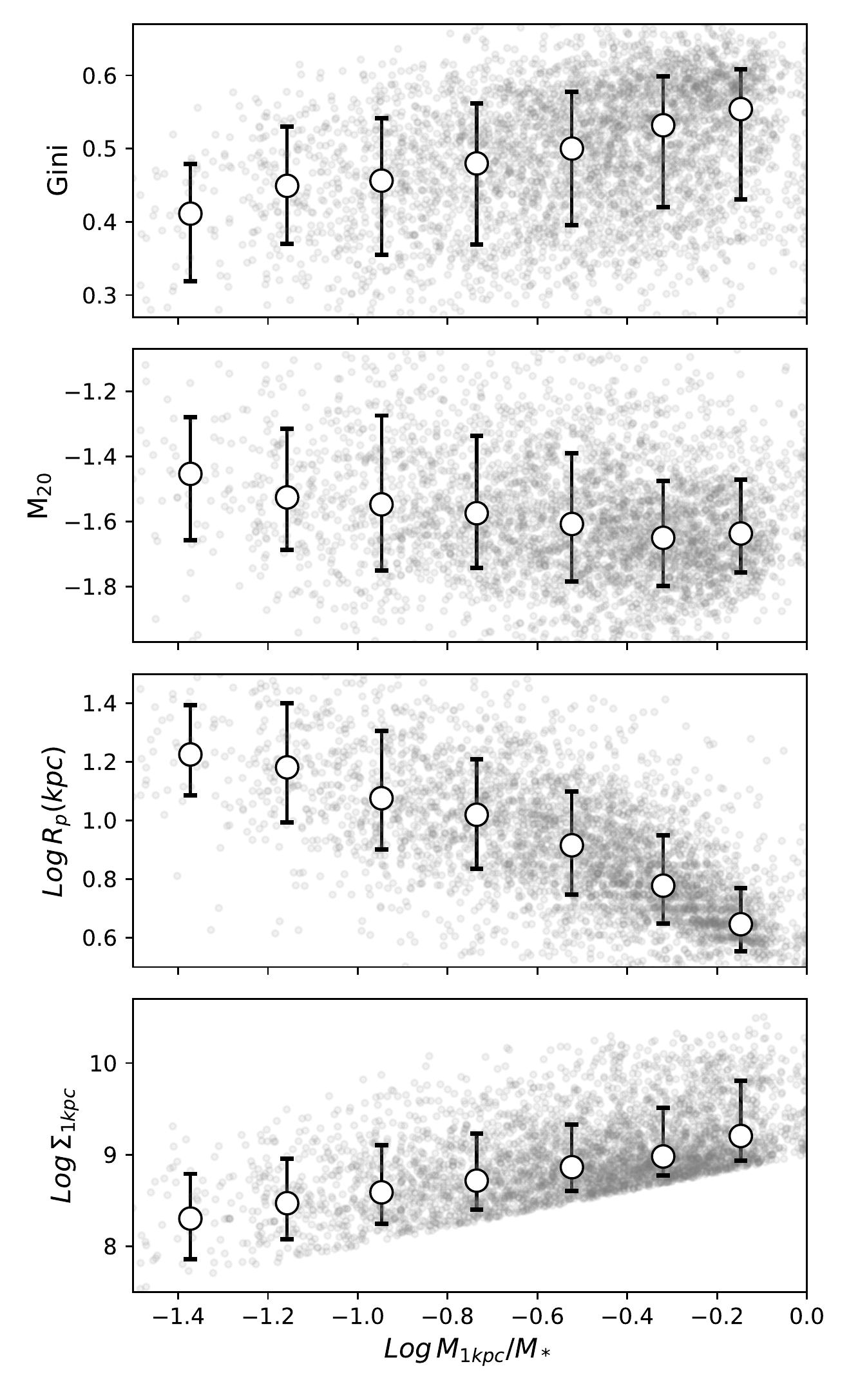}{0.47\textwidth}{}
}
\caption{\Mone\ vs Gini, M$_{20}$, R$_{\rm{p}}$ and \Sone. Individual galaxies are shown as grey dots. The circles with error bars show the median and 16th to 84th percentile range.}\label{fig:M1M_Other}
\end{figure}

We now detail our motivations and the advantages of using \Mone. This parameter measures the fractional stellar mass within the central 1 kpc and is a metric that quantifies the compactness of a galaxy. To check the effectiveness of this metric, in Figure \ref{fig:M1M_Other}, we compare \Mone\ with other commonly-used morphological metrics, namely Petrosian radius $\rm{R_p}$ \citep{Petrosian1976}, Gini, $\rm{M_{20}}$ \citep{Lotz2004} and \Sone. We see that \Mone\ does contain information on galaxy compactness in the sense that galaxies with large \Mone\ statistically are also compact according to other metrics, i.e. large Gini, small M$_{20}$, small R$_{\rm{p}}$ and larger \Sone.

\begin{figure}
\gridline{\fig{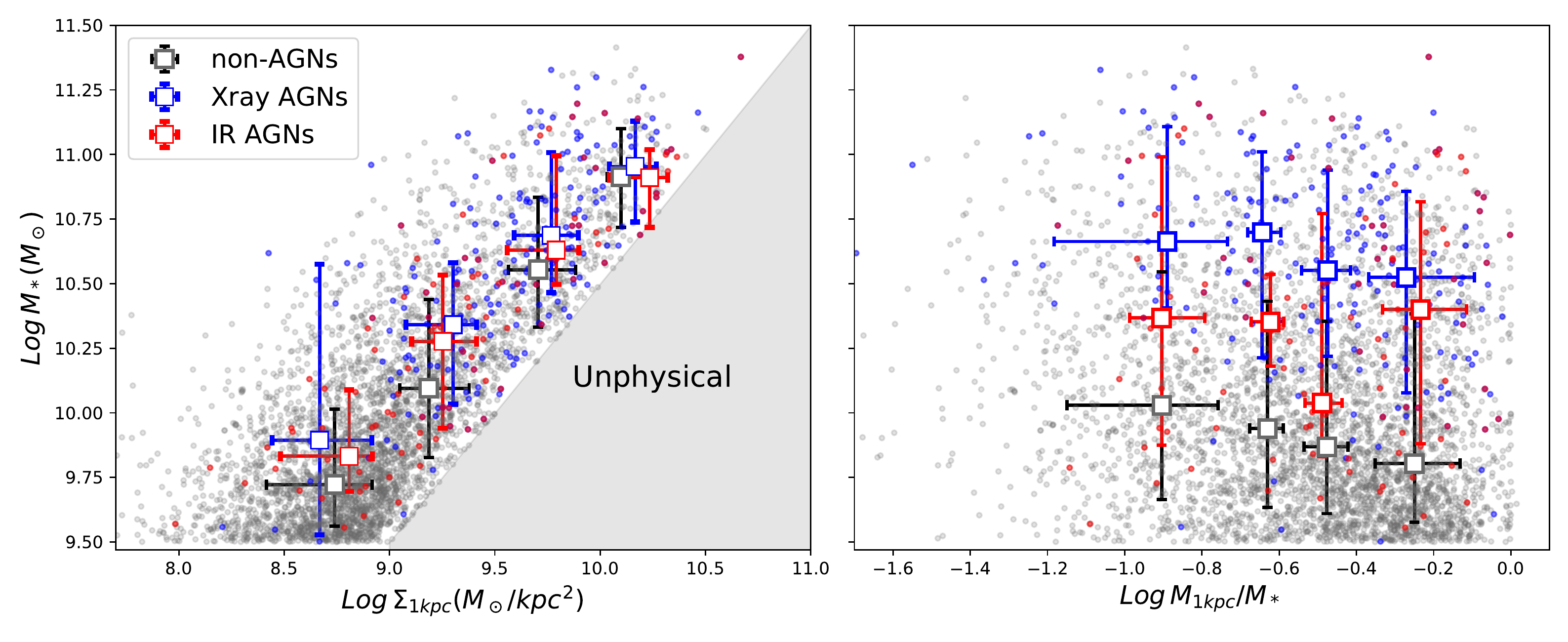}{0.5\textwidth}{}
}
\caption{Scatter plots of M$_*$ vs. $\Sigma_{\rm{1kpc}}$ (left) and M$_*$ vs. \Mone\ (right). It is obvious that the dependence on M$_*$ is much weaker for \Mone\ than for \Sone.}\label{fig:M-S1-M1M}
\end{figure}

As we already discussed in Section \ref{sec:galfit}, \Mone\ can be measured with reasonably small uncertainty. Much more importantly, unlike the commonly-used compactness metrics like \Sone\ and \Se\ which are biased toward more massive galaxies (see details in the next paragraph), the dependence of \Mone\ on M$_*$ is much weaker, which can be explicitly seen from Equation (5). This can also be shown using the existing measurement of the Log\Sone-LogM$_*$ correlation. For example, in CANDELS/GOODS-S, for this correlation \citet{Barro2017} reported a strong but sub-linear relationship with slope of $\beta\approx 0.9$ and 0.7 for SFGs and QGs, respectively. The slopes do not change across the redshift range $0.5<z<3$. If we assume these slopes, we can then get the slope for the Log\Mone-LogM$_*$ correlation, which should be -0.1 for SFGs and -0.3 for QGs. In both cases, \Mone\ have much weaker dependence on M$_*$. Using our sample, Figure \ref{fig:M-S1-M1M} further demonstrates that the strong M$_*$-dependence of \Sone\ is largely  eliminated when using \Mone, and only a slightly decreasing trend with \Mone\ still persists for non-AGNs. This is from the low-mass (LogM$_*<10$) galaxies in our sample, because the 1 kpc scale (compared with galaxy sizes \footnote{If we assume the \citet{vanderWel2014} mass-size relation for SFGs, the median R$_e$ of a 10$^{11}M_\sun$ galaxy is $\approx4$ kpc at z$\sim$2, while it is 2.6 (1.6) $\times$ smaller for a 10$^9$ (10$^{10}$) $M_\sun$ galaxy.}) probes a relatively larger area for a low-mass galaxy than for a high-mass galaxy, which naturally results in generally larger \Mone\ for low-mass galaxies. The trend is much less obvious (it even disappears) for AGN hosts because AGNs are preferentially embedded in more massive galaxies (also see Section \ref{sec:fagn_morp}). 

The criterion commonly used to select compact galaxies in literature are essentially a threshold cut on stellar mass surface density $\Sigma$, which can be formularized as 
\begin{equation}\label{eqn:compact}
\rm{Log\,\Sigma\,>\,\alpha Log\,M_*+\beta}
\end{equation}
If we select compact galaxies using a fixed threshold of $\rm{\Sigma}$, i.e. $\rm{\alpha=0}$, then Equation \ref{eqn:compact} becomes $\rm{Log \Sigma>\beta}$. Given that galaxies follow the well-defined size-mass relation with the form $\rm{LogR\propto \eta LogM_*}$ (e.g. \citealt{vanderWel2014}), the selection criterion then becomes $\rm{(1-2\eta)LogM_*>constant}$. We can now explicitly see that more massive galaxies are more likely to be selected as compact unless $\rm{\eta=0.5}$, which however is not the case (e.g. use the R$_e$-\Se\ relation of \citealt{Barro2017}, $\rm{\eta}$ is $\rm{\approx0.2}$ for SFGs and $\rm{\approx0.8}$ for QGs). To reduce this M$_*$ bias, one can then use a M$_*$-dependent threshold cut on $\Sigma$, i.e. $\rm{\alpha\neq 0}$ (e.g. \citealt{Barro2013,Kocevski2017,Wang2018}). Now, Equation \ref{eqn:compact} becomes to $\rm{(1-2\eta)LogM_*>\alpha LogM_*+constant}$. The bias in principle can be fully removed by choosing $\rm{\alpha=1-2\eta}$. However, the size-mass relation depends on galaxy properties. For example, observations have shown that SFGs and QGs follow different relations (e.g. \citealt{Newman2012,Law2012,Barro2017}). This means that, even with the M$_*$-dependent threshold cut on $\Sigma$, the bias still cannot be fully removed. The bias remains in at least one galaxy population (SFGs or QGs). This selection bias becomes particularly important for data interpretation when trying to identify the driven factor (e.g. mass vs morphology) of some observed correlations. For example, as will be discussed later in Section \ref{sec:fagn_morp}, we find that the prevalence of AGNs positively correlates with \Sone. However, since \Sone\ positively correlates with M$_*$ and the prevalence of AGNs also increases with M$_*$, we do not know if the observed AGN prevalence-\Sone\ correlation is due to M$_*$, or actually infers the causation between the prevalence of AGNs and galaxy compactness. 

To this end, we highlight the advantage of using \Mone. Because of its weak dependence on M$_*$, any relation observed with \Mone\ should be primarily caused by galaxy morphology.

\section{Results}

 In this Section, we aim to investigate the observational evidence of the effects of AGN presences on host galaxies. In the following, we will first compare the star formation properties of AGNs with non-AGNs (Section \ref{sec:SFMS} and Section \ref{sec:SF_Lbol}), and then investigate if the AGN prevalence changes with the star formation properties of their hosts (Section \ref{sec:fagn_SF}). We will then compare the morphological properties of AGNs and non-AGNs (Section \ref{sec:color-mor}), and then investigate if the AGN prevalence changes with the morphological properties of their hosts (Section \ref{sec:fagn_morp}).

\subsection{Star formation properties}

\subsubsection{Distributions of AGNs on the star forming main sequence} \label{sec:SFMS}

\begin{figure*}
\gridline{\fig{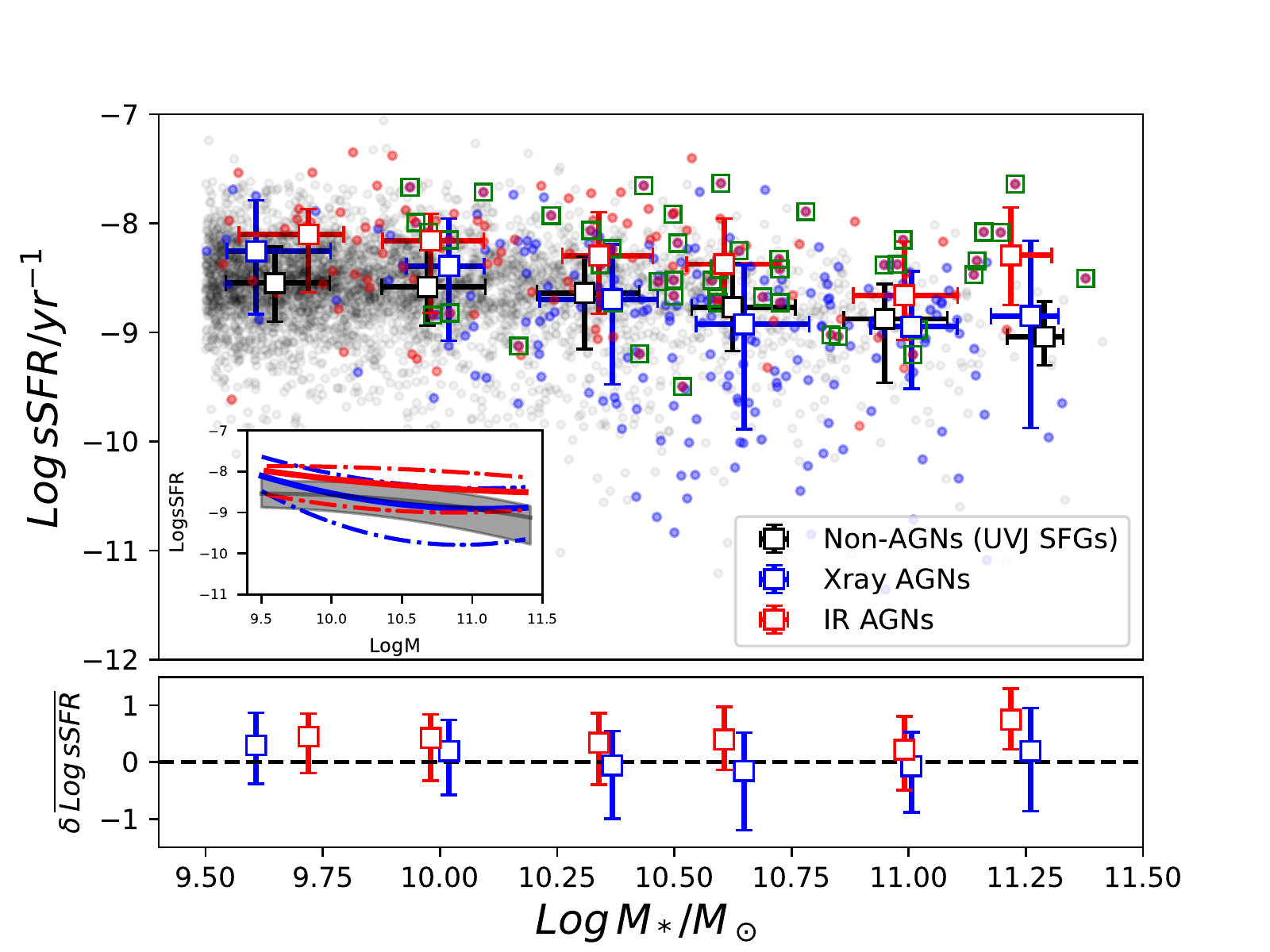}{0.9\textwidth}{}
}
\caption{AGNs and normal SFGs on the sSFR-M$_*$ diagram. X-ray, IR AGNs and non-AGNs are shown as blue, red and black dots respectively. The corresponding color-coded circles with error bars show the median and 16th--84th percentiles of sSFR in each M$_*$ bin. AGNs that are identified as both X-ray and IR AGNs are labelled with green squares. The inset on the bottom left of the main panel shows the medians and 16th--84th ranges for individual populations derived by {\sc cobs} in R using the Constrained B-splines interpolations. The bottom panel shows differences of the mean sSFR between AGNs and non-AGNs in each M$_*$ bin. }\label{fig:SFMS}
\end{figure*}

In Figure \ref{fig:SFMS}, we compare the distributions of AGNs with normal SFGs on the star-forming main sequence (SFMS), i.e. specific star formation rate (sSFR) vs M$_*$. The medians and 1$\sigma$ (16th--84th) ranges for individual populations are derived in two ways. A common way is to compute median and inter-quartile sSFR in arbitrarily defined M$_*$ bins, which are shown as squares with error bars in the main panel of the Figure. The other way of calculating the percentiles is to use the non-parametric quantile regression, in which case no arbitrarily defined bins are required. Here, we adopt the COnstrained B-Splines ({\sc cobs}, see \citealt{Ng2007,Ng2020} for details) package in R to carry out the quantile regressions, where the total number of knots required for the regression B-spline method is determined using the Akaike-type information criterion. The results from {\sc cobs} are inserted to the bottom left of the main panel. Regardless of the way to calculate the median relation, we find that, while the median sSFRs of X-ray AGNs are indistinguishable to normal SFGs, enhanced sSFR is observed in IR AGN hosting galaxies.

Two tests have been done in order to check the robustness of the conclusions above. First, the SFR comparisons in Section \ref{sec:SED} have shown that our SED fittings can on average over-estimate SFRs for AGN hosting galaxies by $\approx$ 0.1 dex due to the ignorance of AGN components. However, the magnitude of this systematics is small compared with the scatter of sSFR distribution of X-ray AGNs and, it is also smaller than the strength of sSFR enhancement ($\sim0.4$ dex) as seen for the whole sample of IR AGNs. We therefore do not expect such over-estimation can significantly affect our sSFR comparisons. Second, different M$_*$ and redshift distributions of AGNs and non-AGNs can potentially affect our sSFR comparisons because of the evolution of the SFMS \citep[e.g.][]{Whitaker2014,Lee2018}. We test this by building the M$_*$-z-matched subsample of non-AGN SFGs, whose sSFR distribution is then used to compare with that of AGNs. We do this in three M$_*$ bins, and for X-ray and IR AGNs {\it separately,} since their M$_*$ and redshift distributions are also different from each other. For each AGN, we select the two non-AGN SFGs which are the closest to the AGN in the M$_*$-z space to build the M$_*$-z-matched subsample. We have checked that our conclusions below do not depend on how the M$_*$-z-matched subsample is built. For instance, we have tried building the subsample by randomly selecting two/three non-AGN SFGs whose redshifts are within $\delta z<0.2$ and M$_*$ are within $\delta LogM<0.3$, and the results remain unchanged.

\begin{figure*}
\gridline{\fig{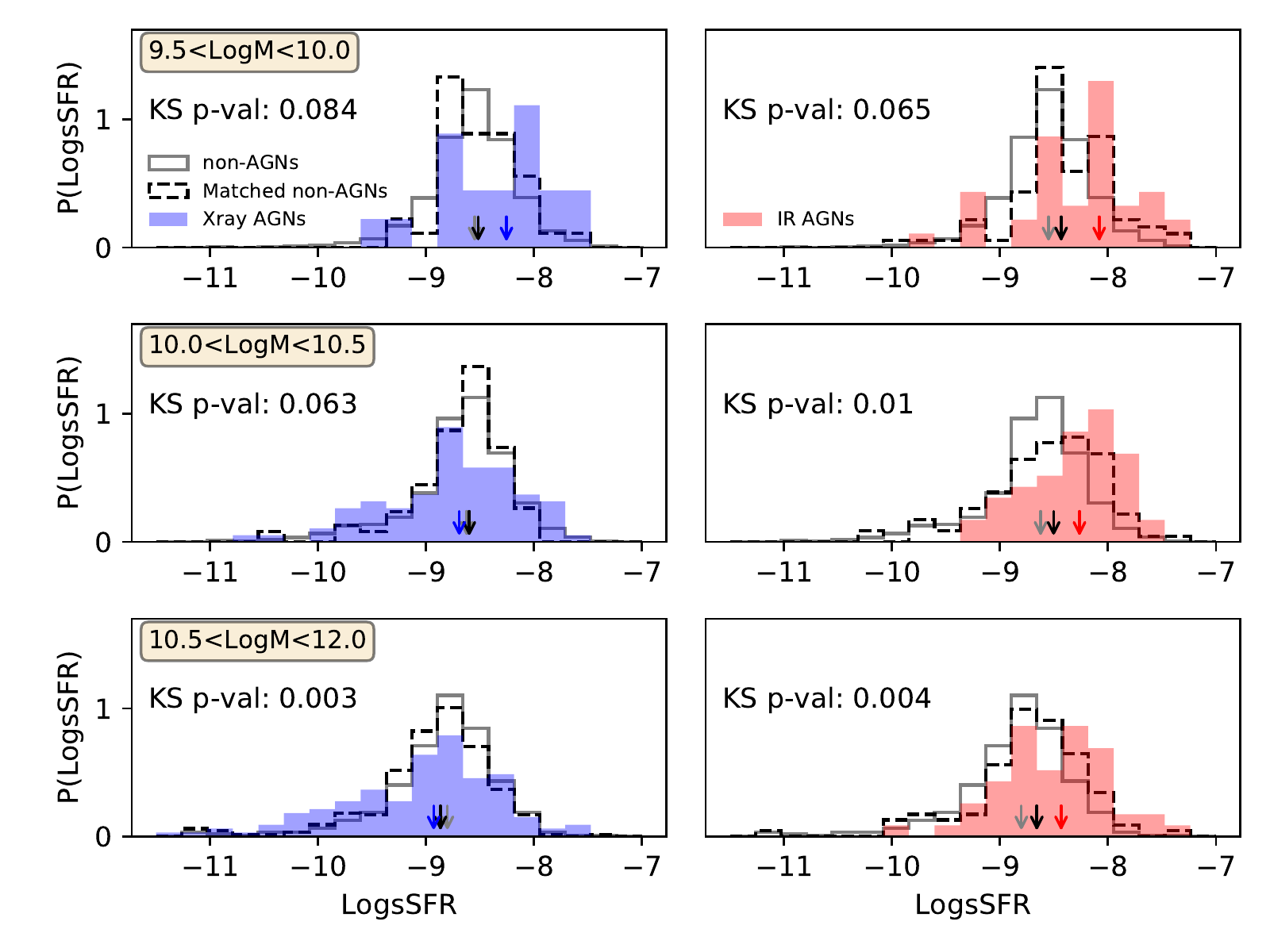}{0.87\textwidth}{}
}
\caption{Comparisons of distributions of sSFR in the three M$_*$ bins, namely 9.5$<$ LogM$_*<$10 (1st row), 10$<$ LogM$_*<$10.5 (2nd row) and 10.5$<$ LogM$_*<$12 (3rd row). The comparisons are between AGNs, where X-ray AGNs are shown in blue in the left panels and IR AGNs are shown in red in the right panels, and non-AGN SFGs, where the entire non-AGN SFGs are shown in grey and M$_*$-z-matched non-AGN SFGs are shown in black. Down arrows in each panel show the medians of individual distributions. Also labelled in each panel is the p-value of the two sample Kolmogorov–Smirnov test  for the null hypothesis that the sSFR distribution of AGN (either X-ray or IR selected) hosting galaxies is identical to that of M$_*$-z-matched non-AGN SFGs.}\label{fig:sf_diff_sig}
\end{figure*}

Figure \ref{fig:sf_diff_sig} shows the detailed comparisons of sSFR distributions for AGNs and non-AGN SFGs. The median sSFR for X-ray AGNs is similar to that of the M$_*$-z-matched, non-AGN SFGs, except in the smallest M$_*$ bin (i.e. 9.5$<$ LogM$_*<$10), where the X-ray AGN sample suffers from small number statistics. In spite of the similar medians, the two sample Kolmogorov–Smirnov tests indicate that we can reject the null hypothesis that the two (matched non-AGN SFGs and X-ray AGNs) sSFR distributions are identical with a $91.6\sim99.7\%$ (i.e. 1.7-3$\sigma$, depending on the M$_*$ bins, see the Figure for details) confidence level. Compared with the M$_*$-z matched non-AGN SFGs, an enhanced sSFR in IR AGNs is still observed, although the magnitude shrinks from $\approx0.4$ to 0.3 dex. A similar enhancement strength is seen in all three M$_*$ bins. The two sample Kolmogorov–Smirnov tests indicate that we can reject the null hypothesis that the two (matched non-AGN SFGs and IR AGNs) sSFR distributions are identical with a $93.5-99.6\%$ (i.e. 1.8-3$\sigma$) confidence level, which, as can be seen in the Figure, is likely driven by the shift toward high sSFR for IR AGNs. To this end, we conclude that, rather than measurement uncertainty or different M$_*$ and z distributions between AGNs and non-AGNs, our results do suggest the median sSFR of (1) IR AGNs is enhanced and (2) that of X-ray AGNs is indistinguishable relative to normal SFGs. In addition, our Kolmogorov–Smirnov tests indicate the entire sSFR distribution for AGNs, either X-ray or IR selected, are different from normal SFGs with a $\approx 2-3\sigma$ confidence level.

For IR AGNs, the enhancement of star formation has also been reported by other works (e.g. \citealt{Cowley2016,Ellison2016,Azadi2017}). The widely-accepted interpretation of it is galaxy merger, a violent process that naturally can both ignite starbursts and fuel luminous AGNs (\citealt{Sanders1988}, also see Figure 6 in \citealt{Alexander2012} for a schematic view). The observational supports on this scenario primarily come from the morphological studies of host galaxies of IR AGNs. \citet{Satyapal2014} studied a sample of WISE-selected AGNs in SDSS, from which they showed the probability to find IR AGNs in post-merger systems is $\approx10-20$ times higher than the control sample. Similar conclusions have also been made by using different MIR selections and at higher redshifts. For instance, \citet{Donley2018} adopted IRAC-color selection criteria (the same as used in this work) to study IR AGN populations at z $\lesssim$ 3 in the CANDELS/COSMOS, from which they concluded that IR AGNs are significantly more likely to be found in interacting/merging systems compared with Seyfert-like AGNs. These, in turn, can also explain why this IR selected AGN population is missed in X-ray since obscuration correlates with merger stage and SMBHs can grow during highly obscured stages of galaxy mergers. If the IR selection is more efficient in picking up the AGNs triggered by galaxy mergers/interactions, then we would expect to see the host galaxies of IR AGN to have enhanced star formation activities, as being the consequence of galaxy mergers/interactions. Based on the sSFR comparison itself, nothing can be said on whether on-going AGN activities have any casual connection with galaxy-wide star formation or not, as the effects (if any) can be easily ``buried'' beneath the effects produced by mergers/interactions. 

For X-ray AGNs, while their median sSFR is indistinguishable from normal SFGs, the sSFR distribution of the X-ray AGNs hosted by massive galaxies (LogM$_*\gtrsim10.3$) is skew to low sSFRs. Moreover, among the massive X-ray AGN hosts, those with high sSFRs often are also identified in IR. If we look at the AGNs which are merely identified by X-ray, skewness to low sSFRs becomes even more clear. These are consistent with the conclusions of \citet{Mullaney2015}, where they found that the mass and redshift-normalized SFR distributions of their X-ray AGNs are broader and peaked at lower value than normal main sequence SFGs, despite that mean SFRs for the two populations are similar. The interpretation of the results above is non-trivial owing to different timescales involved. While AGN is instantaneous, SFR is not. One would be able to measure instantaneous SFR if a correct SFH were known. As a result, no causal link can be indicated merely based on the SFR comparisons between AGNs and normal SFGs unless AGNs have been ``on'' for the same timescale as SFRs are being traced. Even so, the interpretation of the similar median sSFRs between AGNs and normal SFGs is not unique. If a time lag (longer than the timescale of the current star formation episode) is required to enable AGN feedback effects being observable, not too much can be said by looking at on-going AGNs. Alternatively, although the fine-tuning of AGN feedback is required, the observed similar median sSFRs can also be produced by the equally positive and negative feedback of X-ray AGNs. The latter one, however, seems to be disfavored by the observed independency (although error bars are large) between AGN luminosities and star formation activities (Section \ref{sec:SF_Lbol}).

\subsubsection{AGN luminosity vs. Starburstiness}\label{sec:SF_Lbol}

Phenomenologically speaking, if AGN activities do {\it instantaneously} affect galaxy-wide star formation, a correlation between AGN luminosities and their hosts' star formation properties is expected. We therefore study the relation between AGN bolometric luminosities ($L_{bol}$) and starburstiness ($R_{SB}$), which is defined as the SFMS-normalised sSFR,
\begin{equation}
R_{SB} = \frac{sSFR}{sSFR(z,M_*)}
\end{equation}
where sSFR(z,M$_*$) is the sSFR for a galaxy on the SFMS with M$_*$ at z. We adopt the SFMS measured by \citet{Lee2018}, as the relation was measured upon the same galaxy sample using the same SED fitting algorithm. 

The details of $L_{bol}$ measurements can be found in Ji et al. 2021 in prep. (to be submitted) and we only briefly outline the key steps here. For X-ray AGNs, we first take intrinsic X-ray 0.5-7 keV luminosities from \citet{Xue2016} and \citet{Luo2017}, which were measured by correcting the observed X-ray flux with the obscuration empirically calibrated by X-ray band ratios. We assume an AGN spectral photon index $\Gamma=1.8$ and convert the intrinsic 0.5-7 keV to intrinsic 2-10 keV luminosities, which are finally converted to $L_{bol}$ using the 2-10 keV bolometric correction from \citet{Hopkins2007}. For IR AGNs, we first obtain the AGN monochromatic luminosities at 15$\micron$ using the best-fit SED decomposition by {\sc Sed3fit} and convert them to $L_{bol}$ using the 15$\micron$ bolometric correction of \citet{Hopkins2007}. The 15$\micron$-derived $L_{bol}$ is consistent with the direct $L_{bol}$ output from {\sc Sed3fit} (the difference between the two is -0.15$\pm$0.2 dex). We have checked that our conclusions are not sensitive to the choice of MIR derived $L_{bol}$, i.e. 15$\micron$-derived one and direct output from {\sc Sed3fit}.

\begin{figure}
    \gridline{\fig{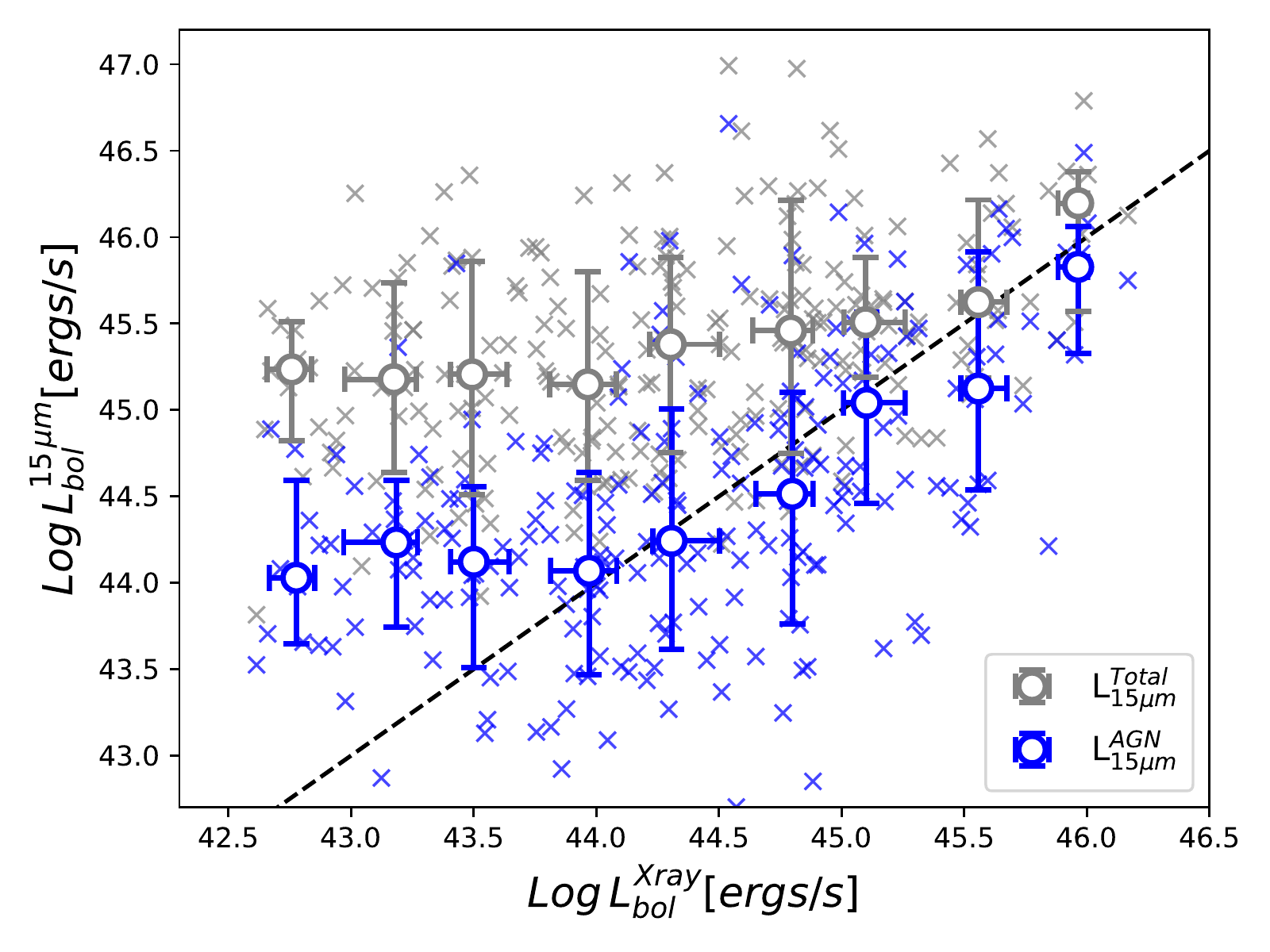}{0.47\textwidth}{}
    }
\caption{The comparison between X-ray- and $15\,\micron$-derived $L_{bol}$ for X-ray AGNs. Two values of $L_{15\,\micron}$ from {\sc Sed3fit} are used to calculate $L_{bol}(15\micron)$. One is the total (AGN+stellar) $L_{15\,\micron}^{Total}$ (grey x), the other is the best-fit AGN {\it only} $L_{15\,\micron}^{AGN}$ (blue x). The circles with error bars show the medians and 16th-84th percentile ranges. The black dashed line marks the one-to-one relation. This comparison shows the essential role that SED decomposition plays in deriving correct $L_{bol}$.}\label{fig:Check_Lbol}
\end{figure}

To check the robustness of the measurements, we first checked that our measurements of the ratio of AGN 2-10 keV luminosity divided by AGN IR luminosity are in good agreement with \citet{Kirkpatrick2017}. We have also further compared the MIR-derived with X-ray-derived $L_{bol}$ for X-ray AGNs and found that the two $L_{bol}$ are consistent with each other when $L_{bol}\gtrsim10^{43.5}\,erg/s$ (see Figure \ref{fig:Check_Lbol} and a detailed discussion in Ji et al. 2021 in prep.), although the scatter between the two measurements is large, with a typical $\pm0.5$ dex which will hopefully be much improved with the coming MIR capability of JWST and future more sensitive X-ray telescopes. Since only a small fraction of the AGNs are fainter, we have checked that including/excluding those faint AGNs cannot affect our conclusions.

\begin{figure*}
\gridline{\fig{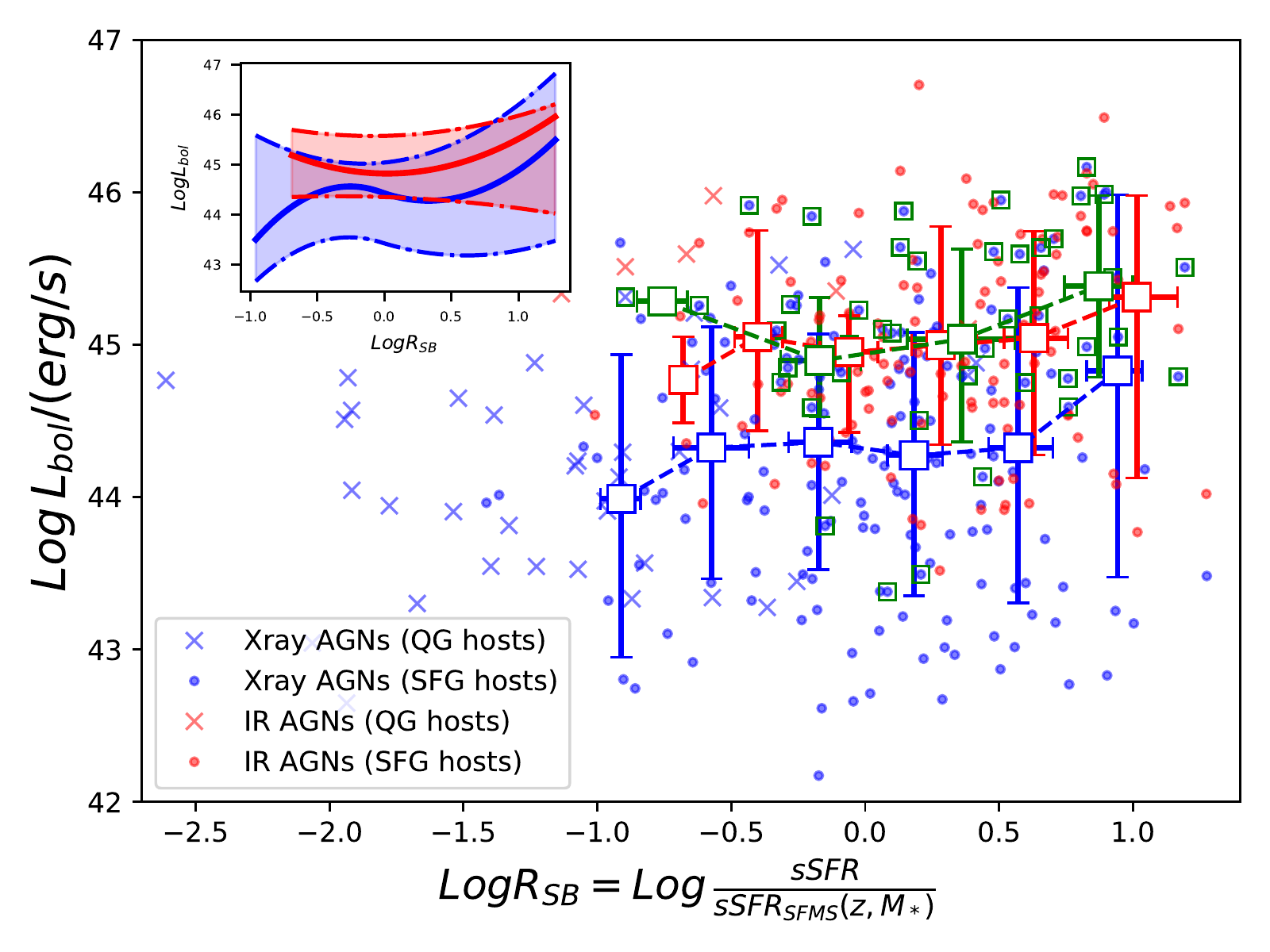}{0.77\textwidth}{}
}
\caption{The $R_{SB}$-$L_{bol}$ scatter plot of X-ray AGNs (blue), IR AGNs (red) and AGNs that are identified as both X-ray and IR AGNs (green). AGNs hosted by QGs are shown as `X's. AGNs hosted by SFGs are shown as dots, with the mean and 16th-84th percentiles over-plotted as circles with error bars. The inset on the top left shows the medians and 16th--84th ranges for SFG hosting X-ray AGNs and IR AGNs derived by {\sc cobs} in R using the Constrained B-splines interpolations. }\label{fig:SB_Lbol}
\end{figure*}

In Figure \ref{fig:SB_Lbol}, $R_{SB}$ is plotted against $L_{bol}$. IR AGNs in our sample are in general brighter than X-ray AGNs by $\approx0.5$ dex, indicating that the IRAC-color selection adopted by us is less sensitive, hence detects only most powerful AGNs. The $L_{bol}-R_{SB}$ correlation is neither seen for X-ray AGNs nor seen for IR AGNs, which seemingly suggests that {\it instantaneous} AGN activities do not affect galaxy-wide star formation. We point out, however, that the measurement uncertainty of the relation, particularly along the $L_{bol}$ axis, is large which may potentially wash out an existing trend. Moreover, stochastic AGN variability can easily weaken the correlations between the observed AGN activities and the star formation properties of AGN hosts \citep{Hickox2014}. 

While the overall trend between $L_{bol}$ and $R_{SB}$ is unclear, we do notice that the galaxies with the most intense star formation activities (i.e. the highest $R_{SB}$) seem to also have the most powerful AGNs. In addition, we also see very tentative evidence that, for X-ray AGNs, the median $L_{bol}$ is smaller at the low-end of $R_{SB}$ although the scatter is very large. Like we did in Section \ref{sec:SFMS}, we also use the constrained B-splines regressions (i.e. {\sc cobs}) to get the $L_{bol}$--$R_{SB}$ quantile curves (top-left inset of Figure \ref{fig:SB_Lbol}), according to which we reach the similar conclusions. These findings are consistent with the X-ray stacking results obtained by \citet{Rodighiero2015}, where they found an enhancement (deficit) of X-ray luminosity in their stacked starburst (green valley) galaxies. Possible interpretations of the enhanced X-ray flux in starburst systems are (1) starbursts are more X-ray active just as they are more star forming (\citealt{Rodighiero2015} reported a factor of 2 larger BH accretion rate per star formation rate (BHAR/SFR) for starbursts than galaxies on the SFMS) and (2) the increasing fraction of AGNs driven by mergers as $L_{bol}$ increasing, which has been observationally demonstrated by \citet{Treister2012}.

\begin{figure}
\gridline{\fig{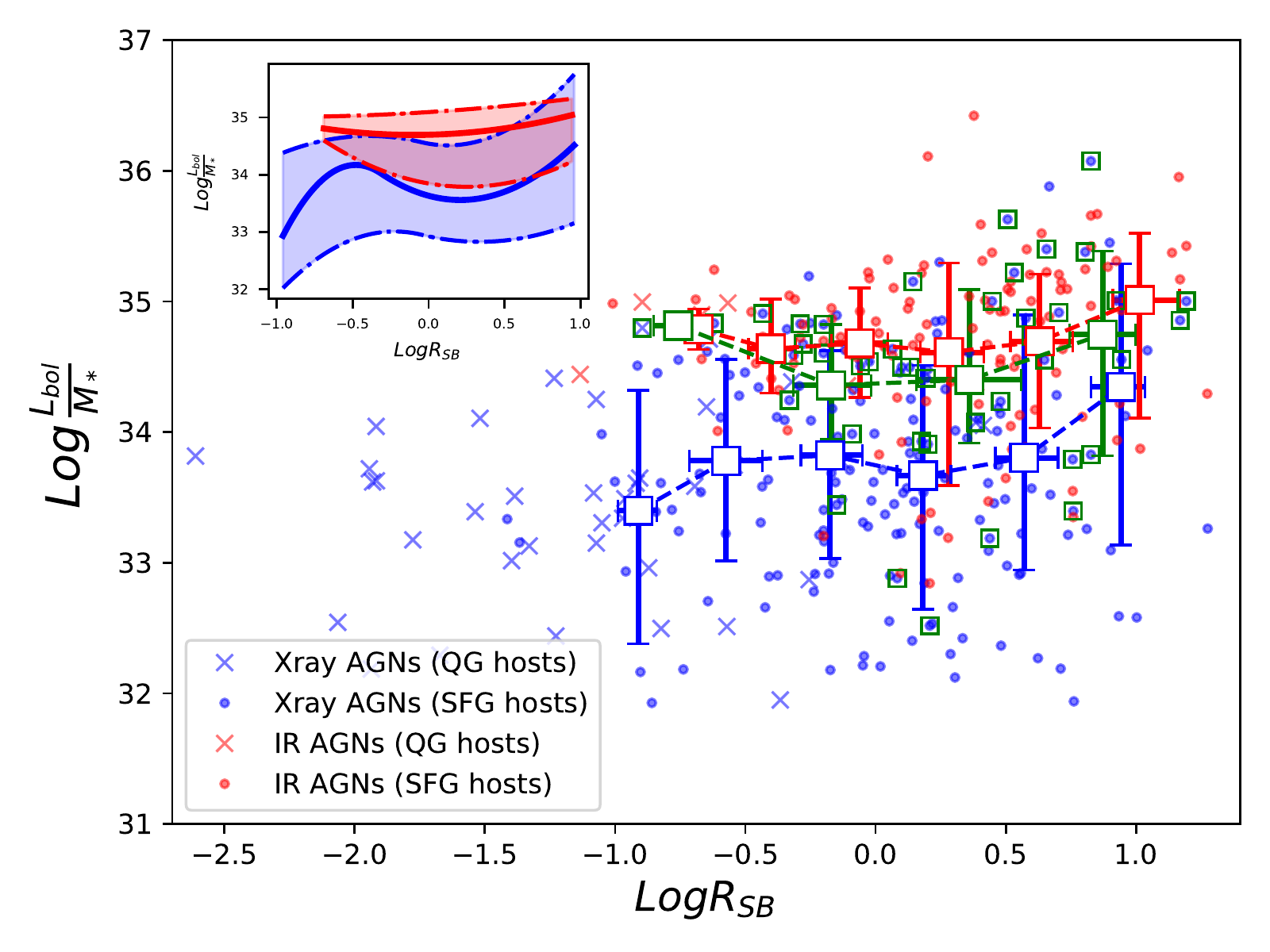}{0.47\textwidth}{}
}
\caption{Similar as Figure \ref{fig:SB_Lbol}, but y-axis is changed to L$_{bol}/M_*$.}\label{fig:SB_LbolperM}
\end{figure}

Finally, we look into the relation of $R_{SB}$ with AGN bolometric luminosity per stellar mass ($L_{bol}/M_*$). Similar as what have been found for $L_{bol}$, Figure \ref{fig:SB_LbolperM} shows that (1) $L_{bol}/M_*$ is larger for our IR-selected AGNs and (2) no clear correlation is seen between $L_{bol}/M_*$ and $R_{SB}$. Unlike that $L_{bol}$ measures the total radiative energy released from a SMBH, $L_{bol}/M_*$ measures its accretion efficiency \footnote{Note that $L_{bol}/M_*$ can be easily converted to the {\it Eddington Ratio} by assuming a M$_{BH}$-M$_*$ relation}. The larger $L_{bol}/M_*$ suggests a higher accretion efficiency for IR AGNs than X-ray AGNs, which possibly indicates different fueling mechanisms of SMBHs. While X-ray AGNs are more likely powered by the stochastic fueling processes like secular evolution of galaxies themselves or galactic disk instabilities, IR AGNs are likely triggered by the violent events like galaxy mergers, which are consistent with the findings of morphological studies of AGN hosts (e.g. \citealt{Kartaltepe2010,Cisternas2011,Kocevski2012,Villforth2014,Ellison2016,Donley2018}). Consistent results have also been found recently by \citet{Delvecchio2020}, where they empirically modelled AGN luminosity functions for galaxies on and above the SFMS. They found that higher {\it Eddington Ratios} are required to reproduce the luminosity function for starburst galaxies.

\subsubsection{AGN prevalence vs. star formation properties}\label{sec:fagn_SF}

\begin{figure*}
\gridline{\fig{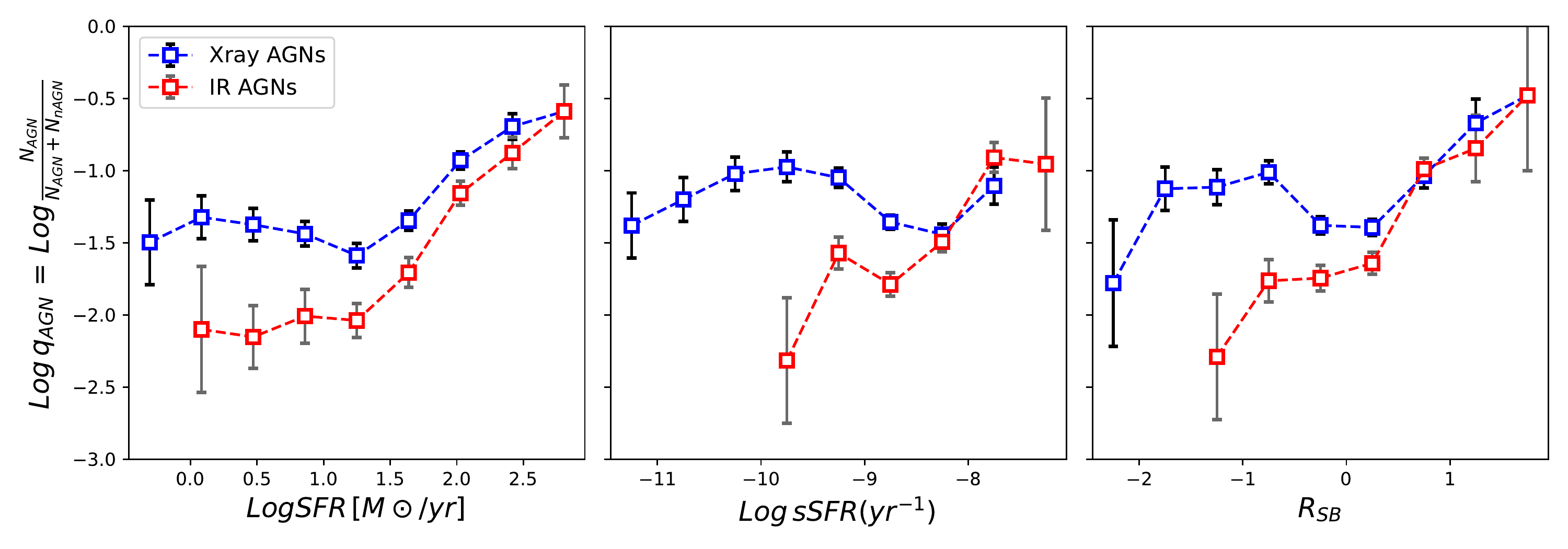}{0.9\textwidth}{}
}
\gridline{\fig{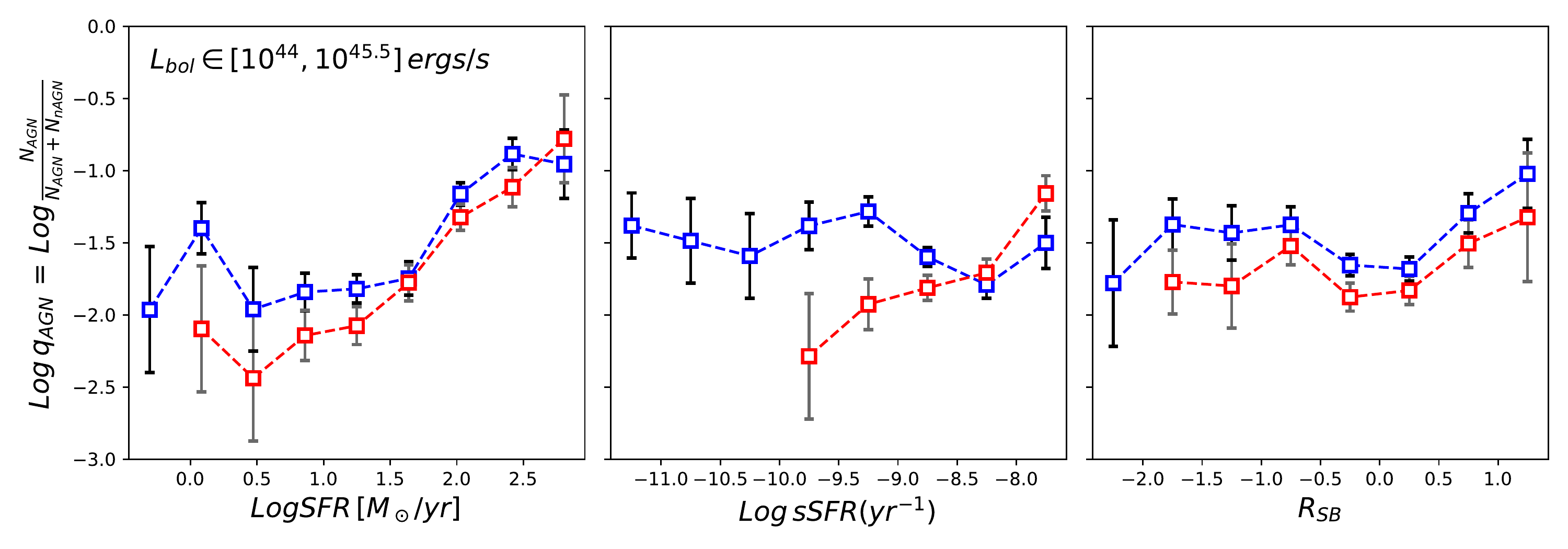}{0.9\textwidth}{}
}
\caption{{\bf Upper:} AGN prevalence vs. star formation properties. From left to right: the dependence of $\rm{q_{AGN}}$ on SFR, sSFR and $R_{SB}$. We refer readers to Figure \ref{fig:fAGN} for the relation of AGN prevalence with M$_*$. {\bf Bottom:} Similar to the upper panels, but a L$_{bol}$ range cut ($10^{44}-10^{45.5}$ erg/s) is posted on both AGN populations.}\label{fig:fAGN_SF}
\end{figure*}

We now investigate the dependence of the AGN prevalence on star formation properties of host galaxies. The prevalence of AGNs is quantified by AGN fraction (q$_{\rm{AGN}}$) which is defined as the ratio of the number of AGNs (N$_{\rm{AGN}}$) divided by the total number of galaxies (N = N$_{\rm{AGN}}$+N$_{\rm{nAGN}}$), i.e. q$_{\rm{AGN}}$ = N$_{\rm{AGN}}$/N. The three panels of Figure \ref{fig:fAGN_SF} show the changes of q$_{\rm{AGN}}$ with SFR, sSFR and $R_{SB}$ respectively. Recall that both sSFR and $R_{SB}$ are essentially normalized SFR, with the former being normalized by M$_*$ and the latter being normalized by both M$_*$ and z.

We start with the prevalence of X-ray AGN. First, regardless of the adopted metric of star formation intensity (SFR, sSFR or $R_{SB}$), q$_{\rm{AGN}}$ is high in galaxies with intense star formation activities. Second, for galaxies with normal/suppressed star formation rates LogSFR$\le$1.5, q$_{\rm{AGN}}$ stays approximately flat with SFR. Because both SFR and q$_{\rm{AGN}}$ increase with M$_*$ (the q$_{\rm{AGN}}$-M$_*$ relation will be studied in Section \ref{sec:fagn_morp}), normalizing SFR with M$_*$ (i.e. sSFR) can effectively mitigate the M$_*$ dependence to allow a more direct view on the link between q$_{\rm{AGN}}$ and star formation activities. Compared with galaxies with moderate sSFR ($\sim$ 1 Gyr$^{-1}$), a higher incidence of X-ray AGNs is observed in galaxies with suppressed sSFR (also have green colors, which will be shown in Section \ref{sec:color-mor}). A similar trend is also seen when using $R_{SB}$ which mitigates not only the M$_*$ but also redshift dependence by normalizing each galaxy with the SFMS. The findings above are consistent with what have been reported by \citet{Aird2019} (see their Figure 10 and 11 in particular), where they showed that the X-ray AGN prevalence is larger both for galaxies with suppressed star formation and for starbust galaxies, although it should be pointed out that, apart from galaxies with X-ray detections, they adopted a Bayesian methodology to also include the X-ray information for galaxies lacking direct flux detection into their analysis while we do not follow such an approach here.

Unlike X-ray AGN prevalence, q$_{\rm{AGN}}$ of IR AGNs generally increases with SFR, sSFR and $R_{SB}$. The increasing q$_{\rm{AGN}}$ towards galaxies with intense star formation is consistent with the picture of merger-driven scenario. Compared with X-ray AGNs, the unseen over-abundant IR AGNs hosting by galaxies with suppressed star formation show the differences between the two AGN populations, highlights the importance of the AGN selection effect (e.g. X-ray vs IR) in altering the distribution of host galaxy properties and as a result in building up a comprehensive picture of AGN effects on host galaxies.

Finally, as we already discussed in Section \ref{sec:SF_Lbol}, because the sensitivities of the two AGN selection methods are different, namely that the IR selection is less sensitive at fixed bolometric luminosity (Figure \ref{fig:SB_Lbol}), the L$_{bol}$ difference, in principle, can lead to the distinct q$_{\rm{AGN}}$ trends seen between X-ray and IR AGNs, if there is a strong dependence of star formation properties with AGN luminosity, which however is not seen (Section \ref{sec:SF_Lbol}) despite of the still large uncertainty in the L$_{bol}$ measurements. Nevertheless, we do test this possibility by setting a cut in L$_{bol}$, i.e. $10^{44}\,erg/s\le L_{bol}\le 10^{45.5}\,erg/s$, on both AGN populations. The cut at the low end of the L$_{bol}$ distribution aims to exclude the faint AGNs that currently are not picked up by our IR selection. The high-end cut, on the other side, aims to exclude the brightest and highly obscured AGNs missed by the X-ray selection. As Figure \ref{fig:SB_Lbol} shows, both selection methods are similarly sensitive with the adopted L$_{bol}$ range. As the bottom panels of Figure \ref{fig:fAGN_SF} show, our conclusions do not change after doing the L$_{bol}$ cut.

\subsection{Morphological properties}
\subsubsection{Distributions of AGNs on color-morphology diagrams} \label{sec:color-mor}
\begin{figure*}
\gridline{\fig{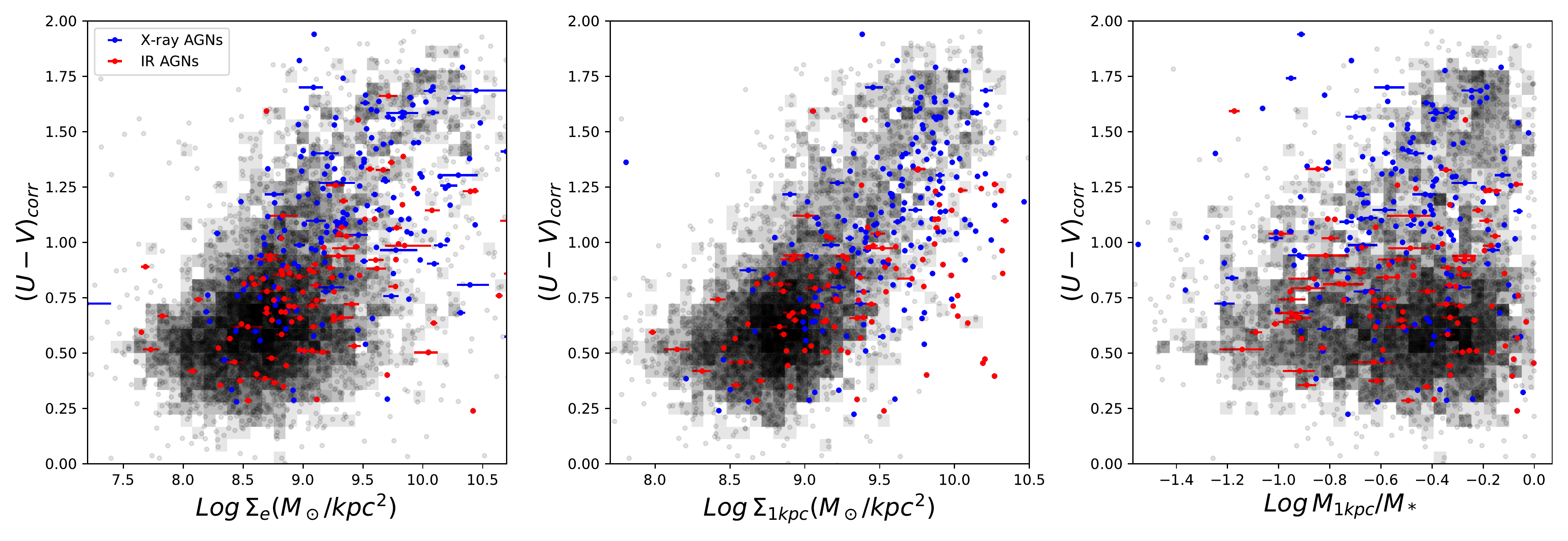}{0.9\textwidth}{}
}
\gridline{\fig{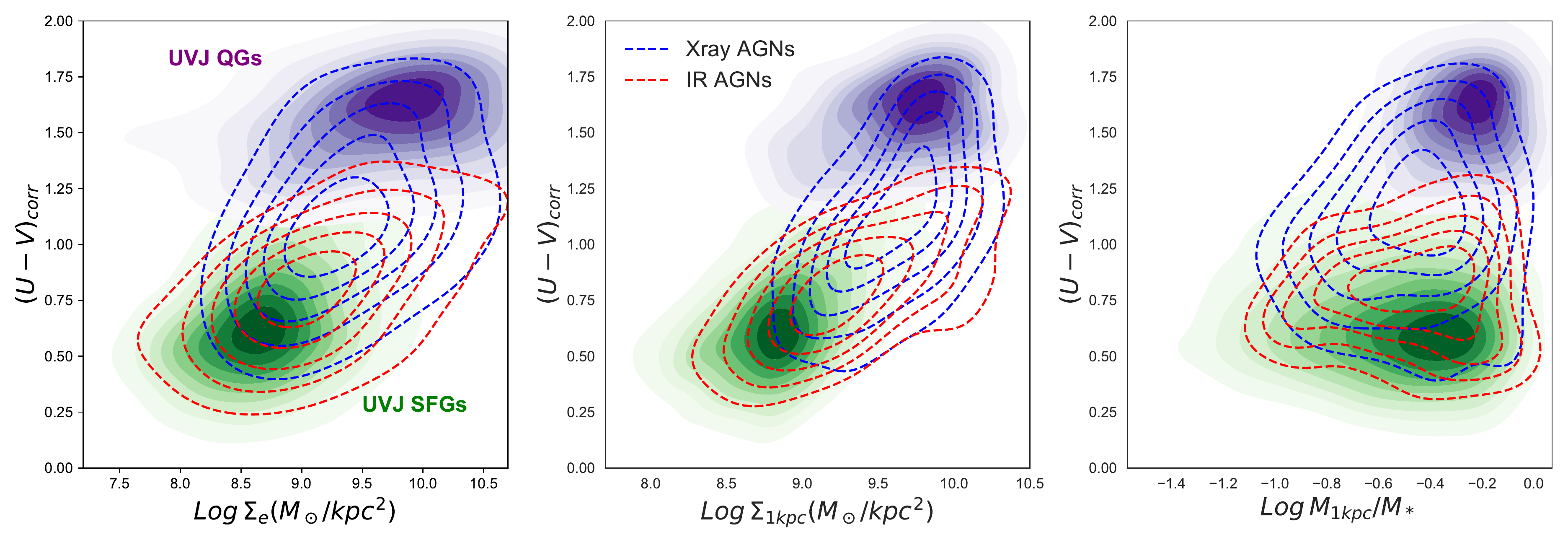}{0.9\textwidth}{}
}
\caption{Dust-corrected rest-frame color (U $-$ V)$_{corr}$ vs. morphological properties. {\bf Upper:} Scatter plots for X-ray AGNs (blue dots), IR AGNs (red dots) and non-AGNs  (black dots). From left to right, each panel shows (U $-$ V)$_{corr}$ vs. \Se, \Sone\ and \Mone\ respectively. The 2-D histograms of non-AGNs are over-plotted in grey scales. {\bf Bottom:} Number density distributions for AGNs and  non-AGNs. Non-AGNs are further divided into star-forming (green) and quiescent (purple) galaxies according to the UVJ color-color diagram. The number density distributions are estimated using the Gaussian kernel.}\label{fig:UV_morp}
\end{figure*}

In this Section, we study how AGNs and non-AGNs distribute in the color-morphology parameter space. In particular, we study their distributions in the diagrams of dust-corrected rest-frame color \UVc\ vs. \Se, \Sone\ and \Mone\ respectively. The reasons of using \UVc, rather than \VJc, are that \UVc\ (1) better probes star formation properties and (2) is less sensitive to the assumption of dust attenuation (see Figure \ref{fig:check_color}). We notice that, after doing the dust correction, \UVc\ itself can effectively separate SFGs and QGs (see Figure \ref{fig:UV_morp}). The separation boundary is \UVc $\approx$ 1.1 mag, fully consistent with \citet{Kocevski2017}.

As Figure \ref{fig:UV_morp} shows, compared with non-AGNs, X-ray AGN hosts are over-abundantly seen to be hosted by galaxies with green \UVc\ colors, which is consistent with Section \ref{sec:fagn_SF} where the relations between q$_{\rm{AGN}}$ and star formation properties were investigated. Consistent conclusions also have been obtained by many other studies on X-ray AGNs, both in the local Universe \citep[e.g.][]{Martin2007,Salim2007,Schawinski2010} and at high redshifts \citep[e.g.][]{Nandra2007,Coil2009}. With regard to morphological properties, compared with non-AGNs, X-ray AGNs share the similar locus of parameter space with QGs, which also have larger stellar mass surface density (\Se\ and \Sone) than SFGs. This fully aligns with the finding of \citet{Kocevski2017}, where they reported a large fraction of compact SFGs hosting X-ray AGNs at $z\sim2$.

Figure \ref{fig:UV_morp} also clearly shows that IR AGNs distribute differently in the color-morphology space when compared with X-ray AGNs. Specifically, unlike X-ray AGN hosts peaked in the region with green colors, IR AGNs are bluer and have similar (but slightly redder) colors as normal SFGs. Meanwhile, while IR AGNs seem to have larger surface stellar mass density than SFGs, they are not as compact as X-ray AGNs, immediately showing the importance of AGN selection on the distributions of physical properties of AGN hosting galaxies.

\begin{figure*}
\gridline{\fig{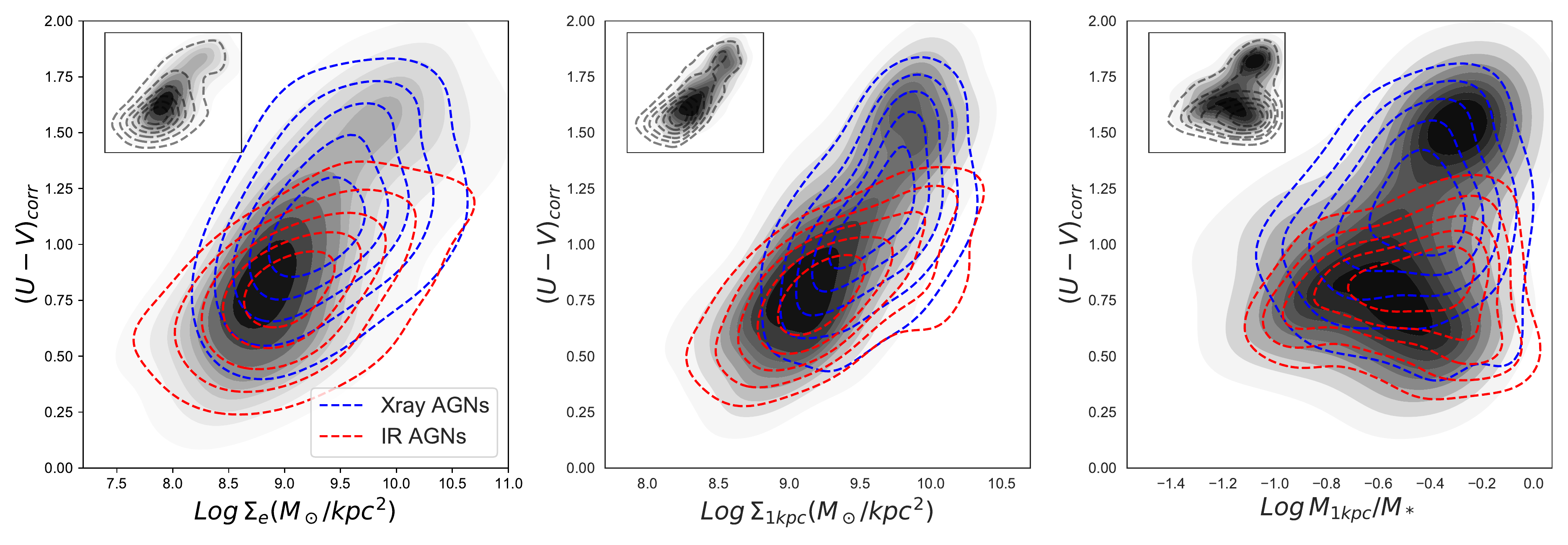}{0.9\textwidth}{}
}
\gridline{\fig{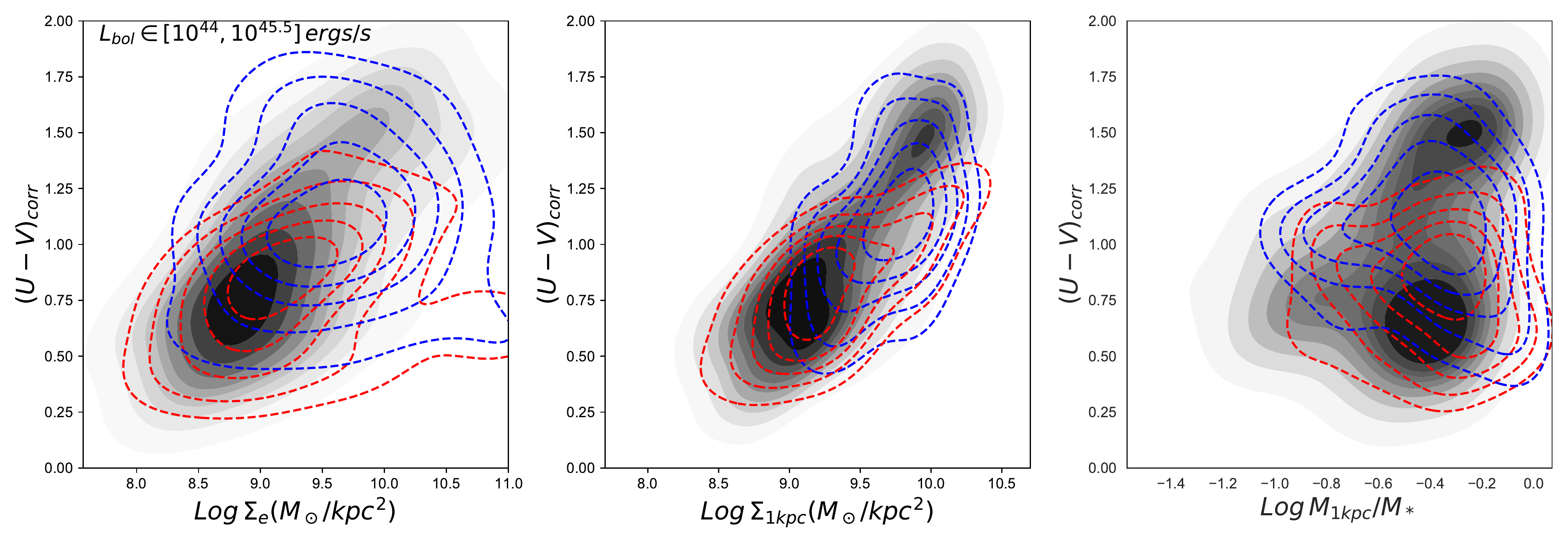}{0.9\textwidth}{}
}
\caption{{\bf Upper:} Distributions of AGNs (X-ray: blue, IR: red) and M$_*$-z matched non-AGNs (grey shades) in color-morphology diagrams. Number density distributions are estimated using the Gaussian kernel. For comparison, also plotted in the inserted sub-figure are the distributions of the entire sample of non-AGNs (grey lines) and the M$_*$-z matched subsample (grey shades). {\bf Bottom:} Similar to the upper panels, but a L$_{bol}$ range cut ($10^{44}-10^{45.5}\,erg/s$) is posted on both AGN populations. Again, background grey contours show the distribution of non-AGNs whose M$_*$ and z are matched to the L$_{bol}$ cut AGN sample.}\label{fig:UV_morp_match}
\end{figure*}

Because both \Se\ and \Sone\ strongly and positively correlate with M$_*$ (see Section \ref{sec:m1m}), the observed larger \Se\ and \Sone\ of AGN hosts (both X-ray and IR) than SFGs can possibly be explained by the fact that AGN hosts are systematically more massive than non-AGNs (Section \ref{sec:fagn_morp}), rather than the intrinsic relation between galaxy compactness and AGN activities. To check this, in the right-most panel of Figure \ref{fig:UV_morp}, \UVc\ is plotted against \Mone, our compactness metric that only weakly depends on M$_*$ (Section \ref{sec:m1m}). Unlike using \Se\ and \Sone, the \Mone\ distribution of AGNs is very similar to that of SFGs, suggesting {\it no} clear link between galaxy compactness and AGN activities. In addition, served as an alternative test, we have compared the \Se, \Sone\ and \Mone\ distributions of AGNs with a sub-sample of M$_*$-matched non-AGNs (the upper panels of Figure \ref{fig:UV_morp_match}). Similarly to what we did in Section \ref{sec:SED}, for each AGN, we selected the closest two non-AGNs in the M$_*$-z space. We have checked, by choosing the closest three/four non-AGNs, that our results do not change. After doing the M$_*$-z matching, the distributions of both \Se\ and \Sone\ of non-AGNs move toward larger values, making the tendency of AGNs being more compact {\it less obvious}. Also noticed in the Figure is that the \Mone\ distribution does not significantly change after matching M$_*$, again showing the only weak M$_*$-dependence nature of \Mone\ that has already been discussed in details in Section \ref{sec:m1m}.  

Not only M$_*$, because of the different sensitivities of the two AGN selection methods, the L$_{bol}$ difference could also result in the distinct color-morphology distributions seen between X-ray and IR AGNs, if L$_{bol}$ somehow plays a crucial role in determining host galaxies' colors and morphology. To test this, like we did in Section \ref{sec:fagn_SF}, we post a L$_{bol}$ range cut on both AGN samples. Our conclusions do not change after doing the L$_{bol}$ cut (the bottom panels of Figure \ref{fig:UV_morp_match}). We have also checked that our conclusions will not change, if we do the faint end cut only, i.e. L$_{bol}\ge10^{44}$ erg/s. Nevertheless, we do notice that the distribution of X-ray AGNs seem to shift {\it slightly} towards bluer \UVc\ after excluding the faint X-ray AGNs, because the X-ray AGNs hosted by QGs are seemingly fainter than those hosted by SFGs (as already discussed in Section \ref{sec:SF_Lbol}, also see Figure \ref{fig:rsb_m1m_lbol} below).

\begin{figure}
\gridline{\fig{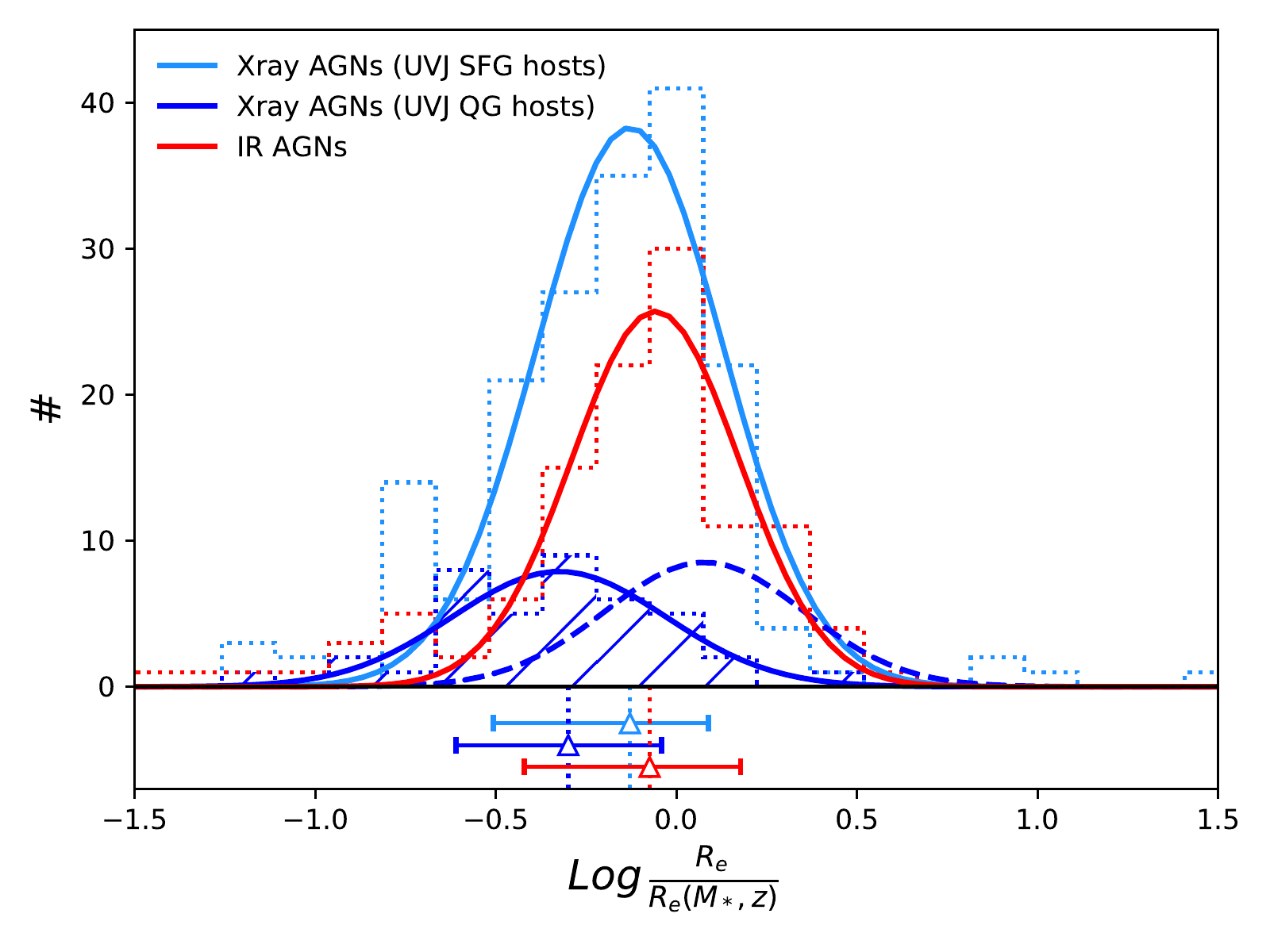}{0.49\textwidth}{}
}
\caption{Distributions of normalized R$_e$ of X-ray AGNs hosted by SFGs (light blue), X-ray AGNs hosted by QGs (blue) and IR AGNs (red). Each R$_e$ is normalized with the best-fit M$_*$-R$_e$ relation for normal {\it late-type} galaxies (taken from \citealt{vanderWel2014}). Solid lines are best-fit Gaussian distributions. Bottom panel shows medians and 16th-84th ranges of individual distributions. Blue dashed line shows the distribution of X-ray AGNs hosted QGs if we instead normalize R$_e$ with the best-fit M$_*$-R$_e$ relation for {\it early-type} galaxies.}\label{fig:normed_re}
\end{figure}

Finally, we compare normalized R$_e$ of AGNs with non-AGNs. In order to remove the M$_*$ and z dependence, each R$_e$ is divided by the median R$_e$ of a galaxy with the same M$_*$ and at z. To do so, we adopt the galaxy mass-size relation measured by \citet{vanderWel2014}, which was done for all 3D-HST$+$CANDELS galaxies at $z<3$. In particular, we normalize R$_e$ of individual galaxies in our sample with the best-fit M$_*$-R$_e$ relation for {\it late-type} galaxies at the closest redshift bin of \citealt{vanderWel2014} (see their Table 1). Because R$_e$ in \citet{vanderWel2014} is the size of rest-frame 5000\AA, we convert it to the size of H$_{160}$ using the Equation (1) and (2) in \citet{vanderWel2014}. Figure \ref{fig:normed_re} shows distributions of normalized R$_e$ for X-ray and IR AGNs, where X-ray AGNs are further divided into two subsamples according to star formation properties of host galaxies (Note that almost all IR AGNs are hosted by SFGs so we decide not to divide them into subsamples). The distribution of normalized R$_e$ of IR AGNs shows that the sizes are in general consistent with sizes of normal SFGs, with a $-0.07$ dex median. Normalized R$_e$ of X-ray AGNs hosted by SFGs are also consistent with normal SFGs, with a median and 16-84 percentile range of $-0.13^{+0.22}_{-0.37}$ dex, although it seems to be smaller than IR AGNs and skews to low normalized R$_e$. X-ray AGNs hosted by QGs have smaller normalized R$_e$, with a median and 16-84 percentile range of $-0.30^{+0.26}_{-0.31}$ dex. This is expected as QGs are in general more compact than SFGs at fixed M$_*$. If, instead, R$_e$ of X-ray AGNs hosted by QGs is normalized with the best-fit M$_*$-R$_e$ relation of \citet{vanderWel2014} for {\it early-type} galaxies (blue dashed curve in the Figure), normalized R$_e$ changes to $+0.09^{+0.30}_{-0.27}$ dex, indicating that the sizes of X-ray AGN hosed by QGs are consistent with normal QGs. 

\subsubsection{AGN prevalence vs. morphological properties}\label{sec:fagn_morp}
We now investigate the dependence of $\rm{q_{AGN}}$ on M$_*$, \Se, \Sone\ and \Mone. 

\begin{figure*}
\gridline{\fig{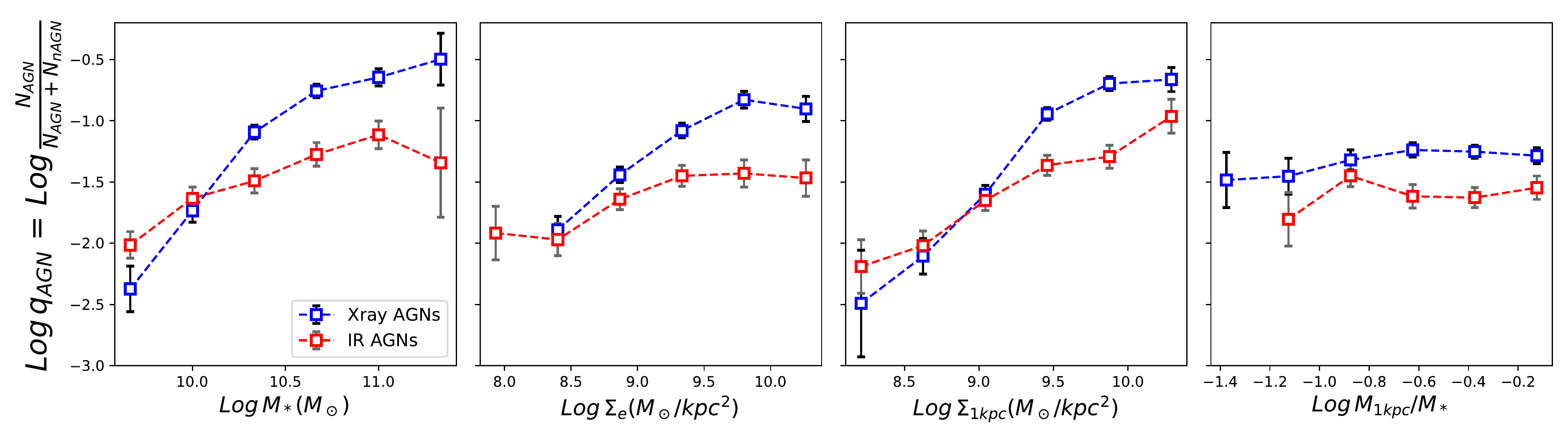}{0.9\textwidth}{}
}
\gridline{\fig{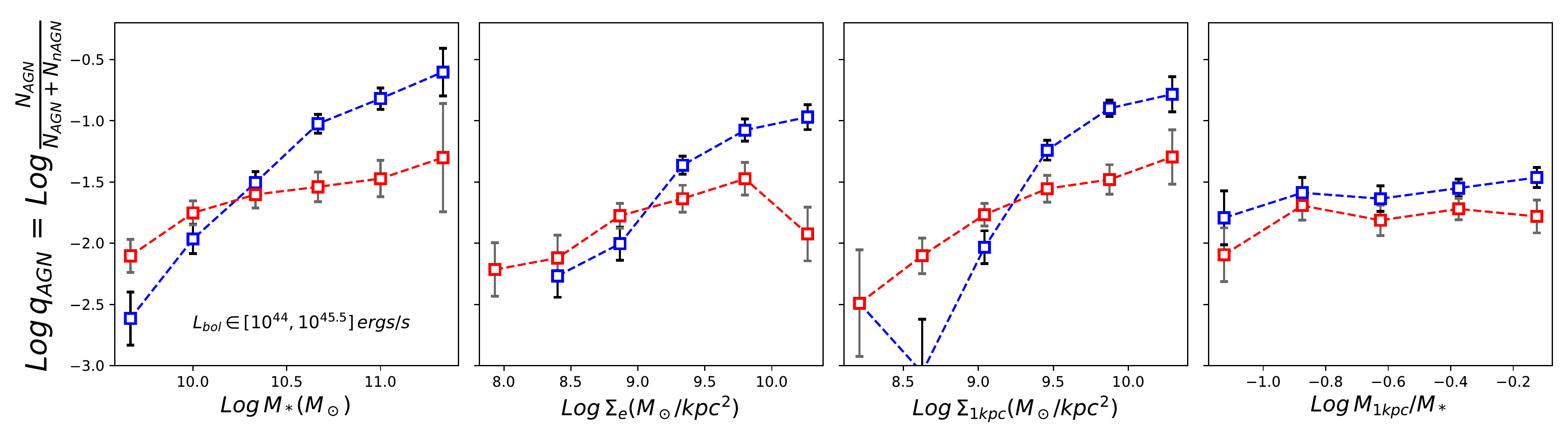}{0.9\textwidth}{}
}
\caption{Similar to Figure \ref{fig:fAGN_SF}. From left to right respectively: the dependence of $\rm{q_{AGN}}$ on $M_*$, \Se\, \Sone\ and \Mone. }\label{fig:fAGN}
\end{figure*}

To begin, $\rm{q_{AGN}}$ increases with M$_*$ (the left-most panel of Figure \ref{fig:fAGN}), the similar conclusion has also been made by many other authors (e.g. SDSS emission line selected AGNs: \citealt{Kauffmann2003}; X-ray AGNs: \citealt{Xue2010,Aird2012}). Given the well-known correlations among M$\rm{_{BH}}$, bulge mass and M$_*$, the positive dependence of q$_{\rm{AGN}}$ on M$_*$ is not surprising. Specifically speaking, an AGN is fueled by accretion onto a central SMBH, the rate and radiative efficiency of which together determine its luminosity. M$\rm{_{BH}}$ is positively and tightly correlated with bulge mass (see \citealt{Kormendy2013} and references therein) and, it is also positively (and likely superlinearly, \citealt{Delvecchio2019}) correlated with $\rm{M_*}$ although the correlation is not as tight as seen with bulge mass, (\citealt{Reines2015,Volonteri2016,Savorgnan2016,Bentz2018}). Therefore, galaxies with larger M$_*$ also statistically have larger M$\rm{_{BH}}$, and hence tend to have higher {\it absolute} accretion rates \citep[e.g.][]{Mullaney2012,Yang2018}. Given that AGNs are essentially selected according to some luminosity threshold, they are naturally expected to be more likely found in more massive galaxies. The positive trend between q$_{\rm{AGN}}$ and M$_*$ is seen both for X-ray and IR AGNs, but with evidence that host galaxies of IR AGNs are less massive than those of X-ray AGNs. Combined with that IR AGNs are also more star-forming than X-ray AGNs (Figure \ref{fig:uvj} and Section \ref{sec:color-mor}), our findings for IR AGNs are consistent with \citet{Hickox2009} where they reported that IR AGN hosts are bluer and less massive than X-ray AGN hosts.  

Also learnt from Figure \ref{fig:fAGN} (middle two panels) is that $\rm{q_{AGN}}$ increases with stellar surface density, which is observed both for X-ray and IR AGNs. The q$_{\rm{AGN}}$-\Sone\ trend could indicate that AGNs prevalently embed in galaxies with high \Sone, i.e. central compactness. Alternatively, the trend could be the ``by-product'' of the positive relations between q$_{\rm{AGN}}$ and M$_*$ and, between $\rm{M_*}$ and \Sone. Since $\rm{q_{AGN}}$ increases with $\rm{M_*}$, even without any intrinsic relation between q$_{\rm{AGN}}$ and galaxy central compactness, we still expect to see the increasing trend of $\rm{q_{AGN}}$ with \Sone\ (similar argument above can be used for \Se). In order to check if there is a causation between $\rm{q_{AGN}}$ and compactness of galaxies, instead of using the morphological metrics which are correlated with $\rm{M_*}$ like \Sone\ and \Se, we therefore look at the relation between $\rm{q_{AGN}}$ with \Mone\ which has much weaker dependence on M$_*$.

As shown in the right-most panel of Figure \ref{fig:fAGN}, unlike increasing with $\rm{M_*}$ and \Sone, $\rm{q_{AGN}}$ stays more or less as a constant with \Mone, which again is observed for both X-ray and IR AGNs. The flat trend suggests that the probability of the presence of AGNs does not depend on galaxy compactness, i.e. no clear evidence on the prevalence of AGNs in compact galaxies. This, in return, indicates that the observed increasing trend between $\rm{q_{AGN}}$ and \Sone\ is {\it primarily} caused by the dependence of $\rm{q_{AGN}}$ on M$_*$, while the intrinsic connection (if any) between AGN and \Sone\ can only be the secondary. Similar conclusion has also been reached by \citet{Ni2019}, where they found the sample-averaged BH accretion rate does not significantly depend on \Sone\ and \Se\ once SFR and M$_*$ among galaxies are controlled. 

Like what we did before (Section \ref{sec:fagn_SF} and \ref{sec:color-mor}), we have checked, by posting a L$_{bol}$ range cut on both AGN populations (the bottom panels of Figure \ref{fig:fAGN}) to ensure the X-ray and IR selections probe the similarly powerful AGNs, our conclusions above do not change.

\begin{figure*}
\gridline{\fig{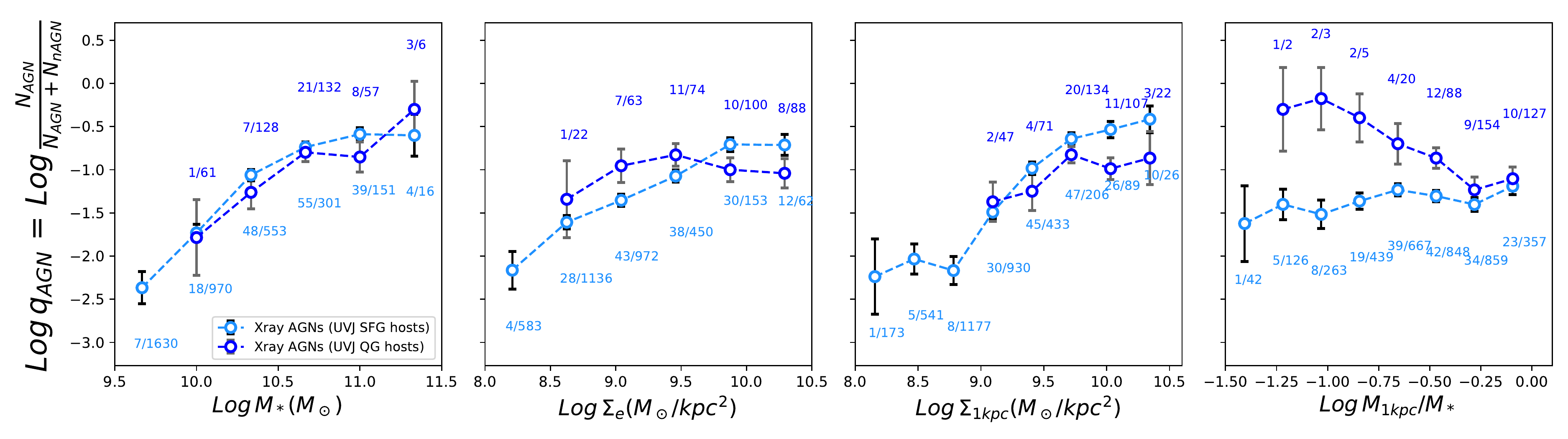}{0.9\textwidth}{}
}
\caption{Similar to Figure \ref{fig:fAGN}. X-ray AGNs are divided into two sub groups based on star formation properties of their hosts, namely SFGs and QGs. The seemingly decreasing trend of $\rm{q_{AGN}}$ with \Mone\ (the right-most panel) for X-ray AGNs hosted by QGs suffers from the small number statistics at the low-end of \Mone.}\label{fig:fAGN_X_SF_Q}
\end{figure*}

Finally, we investigate if the relations seen above depend on star formation properties of host galaxies. This investigation is {\it only} conducted for X-ray AGNs because almost all IR AGN hosts are SFGs (see Figure \ref{fig:uvj}). We first use UVJ-diagram to separate X-ray AGNs into two sub groups, namely SFGs and QGs. Figure \ref{fig:fAGN_X_SF_Q} shows the dependence of $\rm{q_{AGN}}$ on each morphological parameter for X-ray AGNs hosted by UVJ-selected SFGs and QGs. While increasing trends between $\rm{q_{AGN}}$ and M$_*$ are observed among both SFGs and QGs hosting X-ray AGNs, relations between $\rm{q_{AGN}}$ and \Sone, \Se\ depend on types of host galaxies. In particular, while SFGs hosting X-ray AGNs have the similar increasing trends between $\rm{q_{AGN}}$ and \Sone, \Se\ as observed for the entire sample of X-ray AGNs, flatter trends are observed for the ones hosted by QGs, which are consistent with the findings of \citet{Kocevski2017}. For $\rm{q_{AGN}}$-\Mone\ relation, SFGs hosting X-ray AGNs show a flat trend, while QGs hosting X-ray AGNs seem to have a decreasing trend which however is far from conclusive at this point owing to the small sample size. We further test the findings by sub-grouping X-ray AGN hosts using $R_{SB}$. In this case, we divide the entire sample into four sub groups, namely starburst (SB, $R_{SB}>3$), main sequence (MS, $1/3<R_{SB}<3$), green valley (GV, $1/30<R_{SB}<1/3$) and QG ($R_{SB}<1/30$). Like what we have seen when separating the sample with UVJ-diagram, except QG-hosting X-ray AGNs which show almost flat trend of $\rm{q_{AGN}}$ with \Se\ and \Sone, all other X-ray AGN hosts show the similar trends as seen for the entire X-ray AGN sample. 

\begin{figure*}
\gridline{\fig{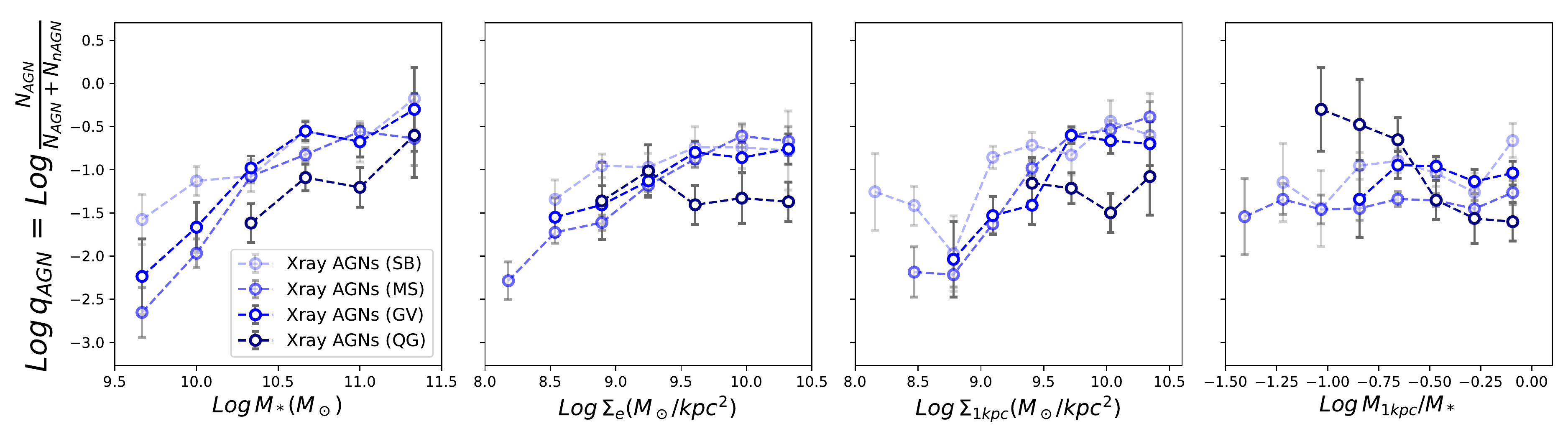}{0.9\textwidth}{}
}
\caption{Similar to Figure \ref{fig:fAGN_X_SF_Q}, but X-ray AGN hosts are now divided based on their distances from SFMS $R_{SB}$ (Section \ref{sec:SF_Lbol}). In particular, X-ray AGNs are divided into 4 sub groups, namely starburst (SB), main sequence (MS), green valley (GV) and QG.}\label{fig:fAGN_X_SB}
\end{figure*}

\section{Discussions}

\begin{figure*}
\gridline{\fig{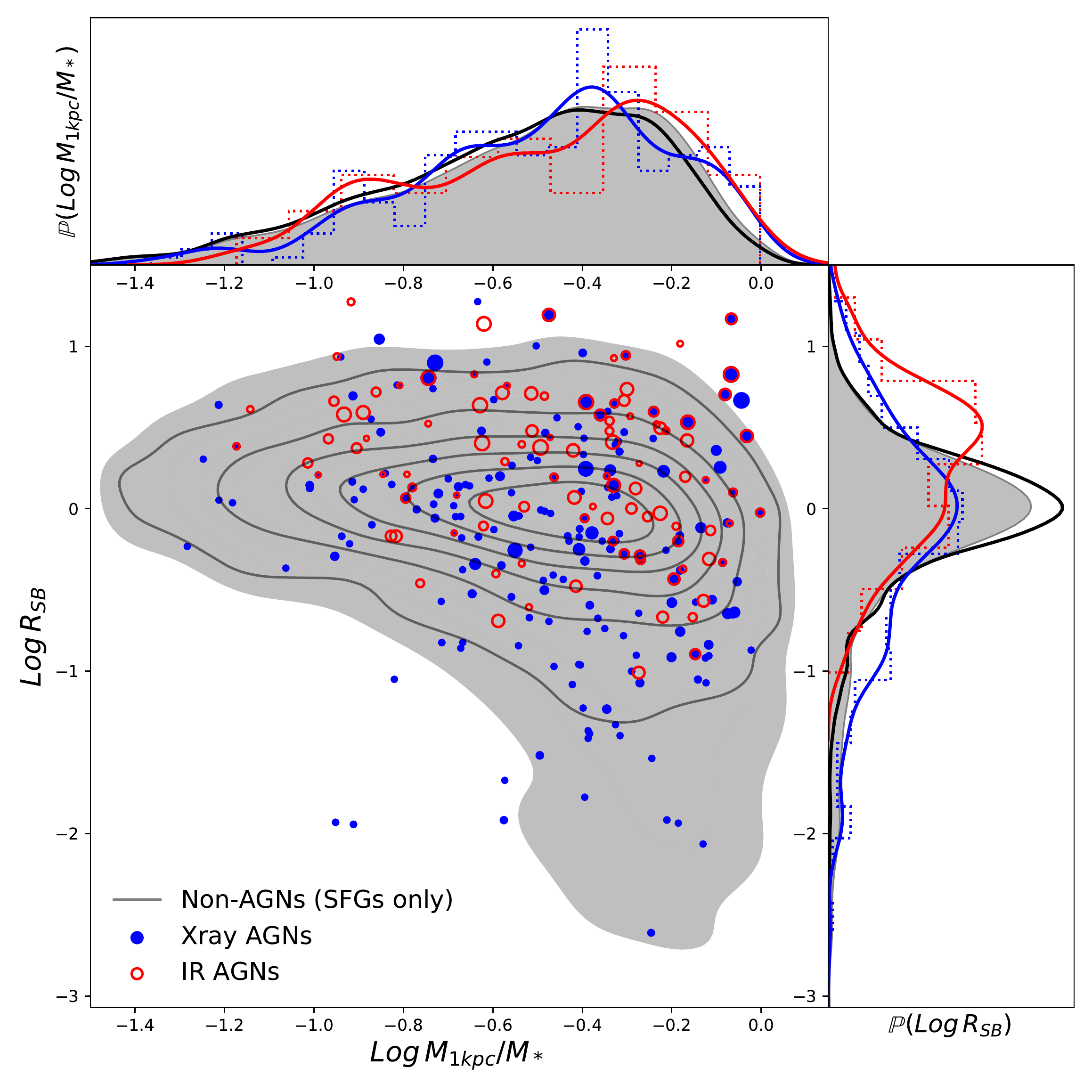}{0.87\textwidth}{}
}
\caption{Distributions of AGNs and non-AGNs on the $R_{SB}$-\Mone\ diagram. Black solid contours show the distribution of normal UVJ-selected SFGs, while the grey shaded contour shows the distribution for all non-AGNs (i.e. SFGs+QGs). X-ray and IR AGNs are shown as blue filled and red empty circles, sizes of which are scaled with L$_{bol}/M_*$. Also shown in the two sub-panels are the distributions of \Mone\ and $R_{SB}$ for different galaxy populations.}\label{fig:rsb_m1m_lbol}
\end{figure*}

The main findings of this work can effectively be summarized by the scatter plot of Figure \ref{fig:rsb_m1m_lbol}, where the M$_*$-dependence has been more or less removed for all shown parameters, including R$_{SB}$ (y-axis), \Mone\ (x-axis) and L$_{bol}$/M$_*$ (point size). Despite the still relative large uncertainty in the L$_{bol}$ measurement (Section \ref{sec:SF_Lbol}), some general conclusions can be drawn. While there is no clear trend of L$_{bol}$/M$_*$ with R$_{SB}$ in the sample of SFGs hosting AGNs, the QGs hosting AGNs, which almost all come from the X-ray selection, appear to have overall lower L$_{bol}$/M$_*$ than the SFGs hosting ones. Both X-ray and IR AGNs share similar \Mone\ with normal SFGs, suggesting no clear link between galaxy compactness and the presences of AGNs. At the same time, although the median R$_{SB}$ of X-ray AGNs is consistent with normal SFGs, its distribution is skewed to low R$_{SB}$. A different distribution of R$_{SB}$ is observed for IR AGNs which generally have larger R$_{SB}$ than normal SFGs. These show that the high incidence of AGNs being hosted by galaxies in the SFG-to-QG transitional region is {\it only} observed for X-ray AGNs, rather than for IR AGNs. In the following, we detail our discussions on how our findings can help constrain the effects of the AGN presences on galaxy quenching.  

\subsection{Towards the general picture of AGN quenching}\label{sec:agn_qenching}

While current cosmological simulations (e.g. Illustris \citep{Vogelsberger2014}, EAGLE \citep{Schaller2015}, IllustrisTNG \citep{Pillepich2018}, SIMBA \citep{Dave2019}) can reproduce the observed statistics of massive galaxies by implementing AGN quenching of star formation, no consensus has yet emerged from the observations that such mechanism is effective in real galaxies. 

Comparing star formation properties between AGNs and non-AGNs seems to be among the most straightforward tests. Our finding, that the median sSFR/R$_{SB}$ of AGNs is either similar to (X-ray AGNs) or larger than (IR AGNs) non-AGNs (Section \ref{sec:SFMS}), shows little evidence that the presence of an AGN suppresses the galaxy-wide star formation. Merely comparing the median/mean star formation properties of AGNs with non-AGNs may bias our view \citep{Mullaney2015}. A more detailed look shows that the distributions for X-ray AGN hosts are skewed to low sSFR/R$_{SB}$, which seemingly suggests a negative effect of AGNs on their hosts' star formation. However, because the similar distributions are {\it not} seen for IR AGNs, this calls into question whether the skewed distributions of star formation properties of X-ray AGNs are a manifestation of AGNs and quenching or simply an AGN selection effect. Similar results have also been obtained by \citet{Ellison2016}, where they found that, compared with the M$_*$-z-environment matched non-AGNs, the SFR distributions of AGNs are different among different AGN selections. In particular, they found that their optical-selected AGNs have wide and skewed to low SFR distribution, while the distribution for their MIR-selected AGNs is skewed to high SFR (their Figure 3).

Similarly to what we found by comparing the distributions of star formation properties, we further study the relations of q$_{\rm{AGN}}$ with SFR, sSFR and $R_{SB}$ (Section \ref{sec:fagn_SF}). While both X-ray AGNs and IR AGNs show higher incidence in galaxies with enhanced star formation relative to the main sequence, a higher incidence of X-ray AGNs is also seen in galaxies with suppressed star formation, which, however, {\it is not} seen in IR AGNs. Empirically speaking, any physical process that is observed to be preferentially taken place in the SFG-to-QG transitional phase may contain critical information of galaxy quenching \citep[e.g.][]{Strateva2001,Bell2004,Faber2007}. While the over-abundance of X-ray AGNs is observed in galaxies with suppressed star formation, such conclusion certainly cannot be extrapolated to all AGNs, since we know that it is invalid for IR AGNs. We therefore conclude that the direct comparisons of the star formation properties between AGNs and non-AGNs show no clear evidence of a causal link between the presences of AGNs and galaxy quenching.

Next, if on-going AGN activities really were to play an observable role in affecting galaxy star formation properties, a correlation between AGN luminosities and star formation properties would be expected. In fact, a number of theoretical works predict the existence of a strong link between star formation and BH growth because both processes require cold gas supply (e.g. \citealt{DiMatteo2005,Hopkins2010,Angles-Alcazar2013}). Regardless of the still relatively large measurement uncertainty, analysis in Section \ref{sec:SF_Lbol} shows null correlation between L$_{bol}$ and $R_{SB}$ in the SFG-hosting AGNs (both X-ray and IR), except that we see evidence that the brightest AGNs have the most intense star formation activities. These findings are fully aligned with other works (e.g. \citealt{Lutz2008,Mullaney2012,Harrison2012}), suggesting a rather weak/no link between star formation and BH growth. However, because stochastic AGN variabilities can diminish the underlying strong star formation-BH correlation \citep{Hickox2014} and, unfortunately, little is known about the AGN duty cycle, it is impossible to conclusively say the real cause(s) of the L$_{bol}$-$R_{SB}$ null correlation. 

While our understanding of the detailed physics driving galaxy quenching is still incomplete, theories suggest one possible evolutionary path, which has been shown by high resolution zoom-in simulations to be particularly effective in the early Universe when dissipative gas inflowing rate is high, namely that a galaxy undergoes a process of compaction as it transforming from a SFG to a QG (e.g. \citealt{Zolotov2015,Tacchella2016}). Some evidence supporting such mechanism has been reported based on recent observations, including the similar number densities, masses and sizes between compact SFGs and compact QGs \citep{Barro2013}, as well as the ALMA observed compact distribution of molecular gas and highly intense nucleated star formation activities in galaxies at $z\approx2$ \citep[e.g.][]{Barro2016,Tadaki2017,Kaasinen2020}. Similar to what has been reported by \citet{Kocevski2017}, our analysis in Section \ref{sec:color-mor} also finds a comparatively higher incidence of X-ray AGN hosts occupying a similar morphological parameter space as compact SFGs. This evidence seemingly suggests a causal link among the presence of an AGN, galaxy compaction and quenching. After adding IR AGNs to the same diagrams (Figure \ref{fig:UV_morp}), however, we immediately realize that IR AGNs occupy the different parts of color-morphology space, namely to distribute more like normal SFGs, with similar \UVc, \Se\ and \Sone. These findings again question the claimed physical association between AGNs and galaxy quenching in the sense that the high incidence of AGNs being hosted by the SFG-to-QG transitional galaxies is only observed for X-ray AGNs and nor for IR AGNs. This significantly weakens the argument of that (X-ray) AGNs preferentially being hosted in compact SFGs is evidence of AGN quenching, since it depends on how the AGNs are selected and similar morphological characteristic of the host galaxies, i.e. frequent high compactness, is not observed for IR AGN hosts.

A further issue about the causal link between the presences of AGNs and galaxy compactness is noticed when we look at \Mone, the morphological parameter that we introduced as an alternative compactness metric (see Section \ref{sec:m1m} for details), which has the distinct advantage of the weak dependence on M$_*$. If \Mone\ is used to define galaxy compactness, we see that not only IR AGNs but also X-ray AGNs have similar \Mone\ distributions to normal SFGs' (Section \ref{sec:color-mor} and the right most panels of Figure \ref{fig:UV_morp} and \ref{fig:UV_morp_match}). This is different from the conclusions made upon the \Se\ and \Sone\ comparisons where X-ray AGNs seem to be more compact (larger \Se\ and \Sone) than normal SFGs. We remind that, however, our purpose here is not to argue which parameter is better in quantifying galaxy compactness. In fact, there is no universal definition of galaxy compactness and the physical meanings of \Se, \Sone\ and \Mone\ obviously are all closely related. Depending on the specific analysis, we view the strong M$_*$-dependence of \Se\ and \Sone\ as a significant drawback when investigating the link between AGN activities and galaxy compactness, because the combination of the M$_*$-M$_{BH}$ and M$_*$-\Se(\Sone) correlations can mimic a null correlation between AGN and galaxy compactness as a real one. This can be clearly seen in Section \ref{sec:fagn_morp} where the relations of q$_{\rm{AGN}}$ with M$_*$, \Se, \Sone\ and \Mone\ are studied. While q$_{\rm{AGN}}$ increases with M$_*$. \Se\ and \Sone, a flat trend is observed with \Mone\ for both X-ray and IR AGNs, indicating that the higher incidence of AGN with larger \Sone\ is primarily due to M$_*$ rather than a morphological reason. It is also possible that the $\rm{q_{AGN}}$-$\Sigma$ trends are driven by some other M$_*$ surrogates, such as bulge fraction (B/T). As already mentioned in Section \ref{sec:valid_single_sersic}, we leave the relevant discussions of this possibility to a separate work.

\begin{figure}
\gridline{\fig{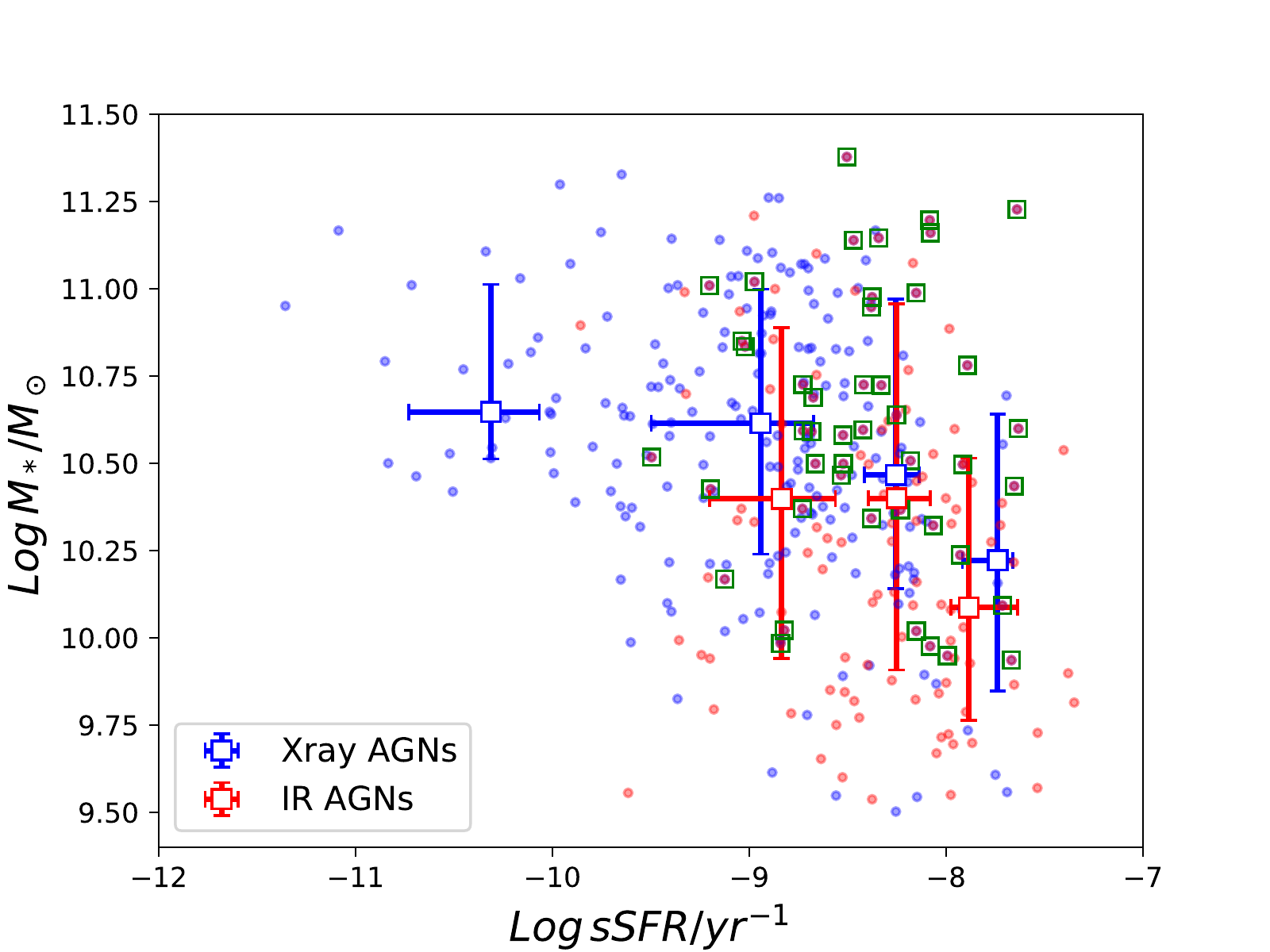}{0.47\textwidth}{}
}
\caption{Median M$_*$ at fixed SSFR of X-ray (blue) and IR (red) AGNs. Green squares mark the AGNs both selected by X-ray and IR.}\label{fig:m_fixed_sfr}
\end{figure}

Our combined study of X-ray and IR AGNs highlights the essential importance of AGN selection effects on the distributions of host galaxy properties. It is likely that different AGN selection methods are sensitive to different galaxy evolutionary states. In a simple BH-galaxy co-evolutionary model, one would expect a dusty BH growth, which tends to be picked up by IR observations, to occur in the early state of normal SFGs when both bulge and BH mass are built-up and to precede the less/un- obscured phase of BH growth that X-ray observations tend to pick up. Observational studies based on the host morphology of IR and X-ray selected AGNs find consistent results with this scenario. For example, \citet{Kocevski2015} argued that their IR selection preferentially selects obscured sources in SFGs before quenching has started, while X-ray selection preferentially finds unobscured sources after the central bulge has built-up and quenching has begun (see their Figure 10). The fact that we see IR AGNs live in galaxies like normal SFGs and X-ray AGNs live in transitional galaxies is in agreement with this simple evolutionary picture. A direct way to test this scenario would be to obtain high quality measurements of ages and BH accretion histories in the AGN hosts and then compare the differences between different AGN populations, which is the subject of our currently investigation. Since SFGs evolve by growing their stellar mass along the SFMS, however, a simplified version of this test would be  to fix the SSFR and use the average M$_*$ as a crude proxy for the age of the stellar populations. Figure \ref{fig:m_fixed_sfr} shows that, at fixed SSFR, IR AGN hosts are systematically less massive than X-ray AGN ones, supporting the idea that IR AGNs are observed at a younger stage, which is consistent with the BH-galaxy co-evolutionary model above. Therefore, the findings of this work certainly are not in direct conflict with AGN quenching, i.e. the scenario that AGNs drive the quenching of star formation in massive galaxies. On the other hand, however, the incidence of AGNs hosted by transitional galaxies, namely those with suppressed star formation and larger central surface mass density, depends on how the AGNs are selected. Thus, the fact that X-ray AGNs are preferentially hosted in transitional galaxies cannot be used as evidence of AGN quenching either, because this could also be simply a reflection of unobscured AGNs being more likely to show up during the later, less obscured evolutionary stages of the host without them having anything to do with the quenching of the host.  

\subsection{Outlook for future studies -- Combining all
radiative AGNs selected by different methods}

Our discussions so far have treated X-ray and IR AGNs as two separate populations. Ideally, they should be combined as a single AGN population to compare with their non-AGN counterparts to investigate if the presences of AGNs in general, no matter how they are selected, correlate with host properties. However, the task of combining different AGN populations is non-trivial. 

Regarding AGN luminosities, different selection methods are not homogeneous, which, for example, can be seen in Figure \ref{fig:SB_Lbol} that our IR selection is not as sensitive as the X-ray selection for faint AGNs. While it remains difficult to use the existing IR data to push the IR selection to the similar faint limit as of X-ray selection, this issue of luminosity inhomogeneity among different AGN selections, hopefully, can be resolved, or at least greatly mitigated, once the MIR capabilities of JWST will be online \citep{Rieke2019}. 

Since the timescale for a galaxy shining like an AGN is extremely short compared with the lifetime of the galaxy, the AGN is, in practice, an almost instantaneous event and thus the total number of AGNs expected in a given field and at a given epoch of the Universe (e.g. redshift intervals) should be statistically proportional to the timescale of AGN presences. As a result, if the timescales change with the phases (e.g. obscured and unobscured) of BH growths, simply adding different AGNs together means to give more weights to the phase with longer timescale. Because different AGN selection methods are sensitive to different BH growth phases, e.g. IR/X-ray selection for the obscured/unobscured AGNs, the  distributions of the host properties for a simple combined AGN population are biased towards the AGN selection that corresponds to the AGN phase with the longer timescale. Instead of simply adding up all AGNs, if we could add different types of AGNs with weights of the inverse of the corresponding timescales, the bias, in principle, would be eliminated. This would rely on a comprehensive knowledge of AGN duty cycle, the information of which unfortunately remains dramatically missing both theoretically and observationally. 

Moreover, it is well known that radiative AGNs can effectively, but are not limited to, be identified by MIR colors and X-ray. AGNs selected through other methods, e.g. optical emission lines \citep[e.g.][]{Agostino2019} or MIR SED decompositions, also occupy a significant fraction of AGN populations (e.g. \citealt{Delvecchio2017}). Our work immediately shows that potentially biased conclusions can be made based on a  specific AGN selection method. To draw a comprehensive picture of AGN effects on host galaxies, it is therefore crucial for future studies to include all AGNs selected by different methods.

\section{Caveats}
Finally, we mention the caveats of this work. First, we emphasize that  conclusions above should only apply to luminous AGNs. The current MIR data only allows for identifying very bright IR AGNs. While the deep X-ray data push the detection limit to $\approx1$ dex fainter for X-ray AGNs (Figure \ref{fig:SB_Lbol}), the identified AGNs are still relatively bright ones. It is unclear that how the fainter and unidentified AGNs can affect our conclusions. Second, the effects of dust obscuration on our conclusions remain to be tested. Dust gradient can affect the H$_{160}$ ($\approx$ rest-frame V-band at z$\sim$2) light profiles. If, statistically, X-ray and IR AGN hosting galaxies share similar dust gradient, then the results of this work should not be greatly affected. However, if there exists intrinsic difference of dust obscuration between the two AGN populations, our conclusions might be affected. Since the IR selection is more sensitive to highly obscured AGNs, we would expect IR AGNs to more likely be hosted by galaxies with more nucleated dust obscuration, which might be able to even completely bury the central light from starbursts/AGNs. This actually is consistent with what we see in Figure \ref{fig:Compare} (see Section \ref{sec:valid_single_sersic} for details). However, it is unclear to what extent the potentially different dust distributions can affect our conclusions, particularly for the distinct color-morphology distributions seen between X-ray and IR AGNs. In particular, if a high column density of dust only exists along  lines of sight with a very small opening angle such that little stellar light is ``blocked'', our results should stand. On the other hand, if the opening angle of the high column density dust is large that significant amount of central stellar light is ``blocked'', then the IR AGNs could be more (although we do not know how much more) compact than they are seen in the H$_{160}$ images. It will soon be possible to use JWST high angular resolution imaging in MIR, which is much less affected by dust obscuration, to finally investigate this issue.

\section{Summary}

In this work, we carry out a combined study of X-ray and IR AGNs at $z\approx2$ and compare the star formation and morphological properties of AGN and non-AGN host galaxies. We show that the criteria used to select AGNs have profound impacts on the distributions of host galaxy properties. 

With regard to star formation properties,
\begin{itemize}[noitemsep,topsep=0pt]
    \item  while the distributions of star formation properties (sSFR and R$_{SB}$) for X-ray AGN hosts is skewed to low values, the medians are similar to normal (i.e. non-AGN) SFGs on the SFMS. A similar distribution is {\it not} seen for IR AGNs, which show enhanced star formation relative to galaxies on the SFMS (Section \ref{sec:SFMS}). 
    \item large measurement uncertainty of L$_{bol}$ notwithstanding, no clear trends, neither for X-ray AGNs nor for IR AGNs, are seen between L$_{bol}$ and R$_{SB}$ (Section \ref{sec:SF_Lbol})
    \item the trends of q$_{\rm{AGN}}$ with SFR, sSFR and R$_{SB}$ show that, despite high incidence is seen for both X-ray and IR AGNs in galaxies with intense star formation, the incidence of X-ray AGNs is also high in galaxies with suppressed star formation (sSFR and R$_{SB}$), which however is {\it not} seen for IR AGNs (Section \ref{sec:fagn_SF}).  
\end{itemize}

With regard to morphological properties,
\begin{itemize}[noitemsep,topsep=0pt]
    \item distributions of morphological properties of X-ray and IR AGN hosts are very different in the color-morphology space (Section \ref{sec:color-mor}). In particular, while X-ray AGN hosts tend to have green colors and large stellar surface mass densities (both \Se\ and \Sone), IR AGN hosts show distributions that are much more similar to those of normal SFGs. Because both \Se\ and \Sone\ are strongly correlated with M$_*$, we introduce a new diagnostic of compactness \Mone, that significantly eliminates the dependence on M$_*$. We show that the distributions of \Mone\ for both X-ray and IR AGNs are similar to normal SFGs'. Consistent results are also obtained by comparing the distribution of {\it normalized} R$_e$ between AGNs and non-AGNs.
    \item  increasing trends of q$_{\rm{AGN}}$ with \Se\ and \Sone\ are seen for both X-ray and IR AGN hosts. The trends with \Mone, however, remain more or less flat, indicating that the correlation with \Se\ and \Sone\ are primarily driven by M$_*$ (Section \ref{sec:fagn_morp}).
\end{itemize}

While the findings presented above are not in direct conflict with the scenario of AGNs driving the quenching of massive galaxies, they do not support it either. Our findings show that the frequency of AGNs hosted by transitional (from SFGs to QGs) galaxies, namely galaxies with suppressed star formation and large surface stellar mass density, depends crucially on how the AGNs are selected. Thus, this calls into question the notion that there is a causal relationship between the presences of AGNs and the quenching of star formation. In fact, interpreting the different physical properties between the two AGN population hosts as evidence of different evolutionary phases of their ISM obscuration, for example, could imply another, yet unidentified, mechanism responsible for both quenching and the apparent evolution of the AGN properties.

\begin{acknowledgments}
We thank the anonymous referee for useful comments. AK and CH gratefully acknowledge support from the NASA/FINESST award (PID:19-ASTRO20-0078).
\end{acknowledgments}

\bibliography{ji_2021_AGN}

\appendix
\section{Tests of systematic uncertainty when using Lee2018 SED fitting measurements} \label{app:sed_test}
We use {\sc Sed3fit} to test that if neglecting the AGN component when carrying out SED fitting with the methodology of Lee2018 can introduce significant systematic bias into the physical parameters of the AGN sample. {\sc Sed3fit} is built upon {\sc Magphys} \citep{daCunha2008} and includes an AGN component into the modeling. During the {\sc Sed3fit} fitting, a galaxy's SED is modelled as the combination of stellar emission, dust emission (PAHs, hot dust and cold dust) and an user-defined AGN spectral library. The basic setup of {\sc Sed3fit} is the same as {\sc Magphys}, namely using the BC03 stellar population synthesis code, assuming \citet{Chabrier2003} IMF and \citet{Charlot2000} dust attenuation model. {\sc Sed3fit} uses a parametric SFH, which is assumed to be the summation of two components -- an underlying exponential decline SFR$(t)\propto exp(-\gamma t)$ with random bursts superimposed. Given that the data coverage of each sample galaxy is limited (typical number of photometric bands is $\approx$ 15, $\approx60\%$ of AGN samples also have MIPS/24$\micron$ data), we followed the same procedure as \citet{Delvecchio2014} to only adopt a subset of the AGN spectral library of \citet{Fritz2006} and \citet{Feltre2012}. By testing with the mock galaxies hosting different types of AGNs (Type I, Type II or intermediate), \citet{Ciesla2015} showed that the derived parameters such as M$_*$ and SFR are insensitive to the adopted AGN library. 

\subsection{Stellar mass M$_*$}
Figure \ref{fig:SED_Test} (a) shows comparisons of the M$_*$ measurements between Lee2018 and {\sc Sed3fit}. A clear correlation between the two M$_*$ is seen for the AGN sample. An $\approx0.2$ dex offset between the two measurements is also seen, with M$_*$ derived by {\sc Sed3fit} being larger than that derived by Lee2018. This offset can be attributed to the different setups between the two SED-fitting procedures, either due to including the AGN component to the modelling, or due to other different assumptions on things like SFH and dust attenuation law which are not related to  the presences of AGNs. To check this, we ran {\sc Magphys} over a subsample of normal galaxies whose M$_*$ and redshift distributions are matched to the AGN sample. For each AGN, we select two normal galaxies who are closest to the AGN in the M$_*$-z space.

Like what is seen for the AGNs, an offset of the M$_*$ measurements between Lee2018 and {\sc Sed3fit} is also observed for the M$_*$-z-matched normal galaxies. Moreover, as the right panel of Figure \ref{fig:SED_Test} (a) shows, the magnitude of the offset seen for the non-AGNs ($\approx 0.15$ dex) is also similar as that seen for the AGNs ($\approx 0.17$ dex). These suggest that the offset between the two M$_*$ measurements of our AGNs is primarily from the different assumptions {\it unrelated} to the presences of AGNs, very likely due to different assumptions on SFHs and dust reddening laws. This conclusion actually is not surprising given that the rest-frame UV to NIR SEDs are usually dominated by stellar light even when AGNs exist (see Section 4 in \citealt{Brandt2015} and references therein). We have further tested the M$_*$ measurements by looking at the relation between the M$_*$ difference $\delta$LogM$_*$ and AGN X-ray intrinsic luminosity taken from \citet{Xue2016} and \citet{Luo2017}. If the AGN component plays a vital role in the determination of M$_*$, then we would expect to see a dependence of $\delta$LogM$_*$ on AGN luminosity, which however is not seen in Figure \ref{fig:deltaM_Lx}. We therefore conclude that our M$_*$ measurement using Lee2018 is robust.  

\begin{figure*}
\gridline{\fig{Check_M.pdf}{0.9\textwidth}{(a)}
}
\gridline{
	\fig{Check_SFR.pdf}{0.9\textwidth}{(b)}
        }
\caption{Comparisons between physical parameters derived from two SED-fitting procedures, namely Lee2018 and {\sc Sed3fit} (see Section \ref{sec:SED} for details). {\bf (a) Left:} The comparison of M$_*$ for the AGNs (green) and M$_*$-redshift matched normal galaxies (orange). BL AGNs are labelled with magenta open circles. The black dashed line marks the one-to-one relation. {\bf Right:} Distributions of $\delta$LogM$_*$ for the AGNs (green) and non-AGNs (orange). The AGN sample has been further divided into X-ray AGNs (blue) and IR AGNs (red). Also tabulated are the 16th, 50th and 84th percentiles of the distributions. {\bf (b):} Similar as (a) but for the comparison of SFR.} \label{fig:SED_Test}
\end{figure*}

\begin{figure}
\gridline{\fig{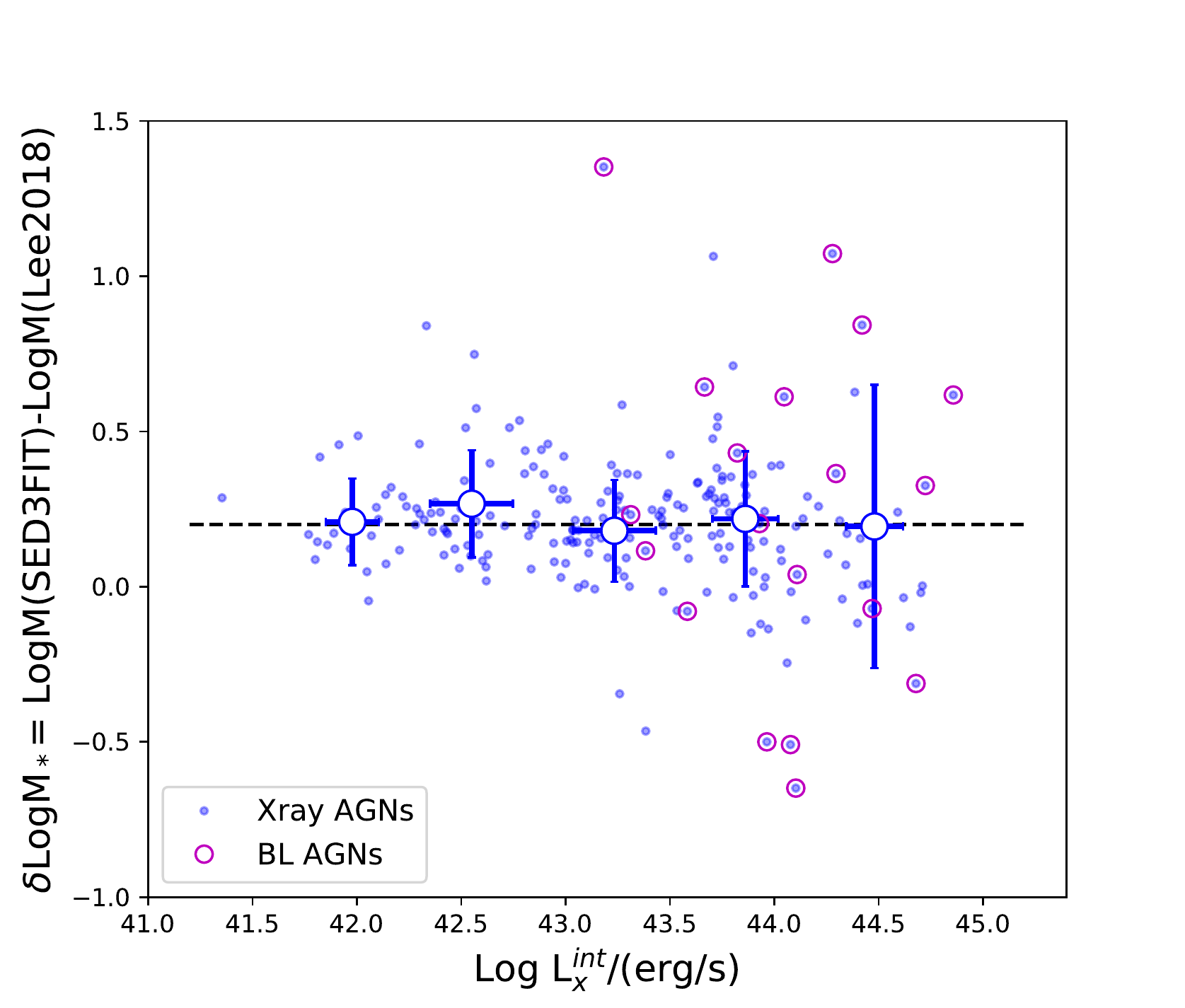}{0.5\textwidth}{}
}
\caption{The relation between $\delta$LogM$_*$ and intrinsic X-ray luminosity. Blue dots are individual X-ray AGNs. In each X-ray luminosity bin, the mean and standard deviation of $\delta$LogM$_*$ is shown as a blue square with error bars. The median $\delta$LogM$_*$ of the entire sample is marked as the black dashed line. BL AGNs are labelled with magenta open circles.} \label{fig:deltaM_Lx}
\end{figure}

Finally, we do notice that some (though a small fraction) AGNs have rather large deviations of M$_*$ measurements between the two SED-fittings (both in Figure \ref{fig:SED_Test} and Figure \ref{fig:deltaM_Lx}). \citet{Santini2012} checked the robustness of their M$_*$ measurements of Type 1 and Type 2 AGNs using two sets of SED fittings, one with the AGN component included while the other without. They found the M$_*$ of Type 2 AGNs can be very well-constrained even without including the AGN component during their SED-fittings, with the mean difference in M$_*$ measurements between the two SED-fittings being zero and only in 1.3\% of objects the difference is larger than a factor of 2. For Type 1 AGNs, although the mean difference in M$_*$ is still consistent with $\approx 0$, the difference in 29\% of objects is larger than a factor of 2. Similar results have also been found by other authors \citep[e.g.][]{Yang2018}. Motivated by the potential different systematics between Type 1 and Type 2 AGNs, we have cross-matched our AGN sample with the broad-line (BL) X-ray AGNs in the GOODS-S \citep{Silverman2010} and GOODS-N \citep{Barger2003}. 19 BL AGNs are found and marked as magenta open circles in Figure \ref{fig:SED_Test}. We do find that a considerable number of AGNs with large differences in M$_*$ measurements are BL AGNs. After comparing the difference of M$_*$ measurements between BL AGNs and other AGNs, we reach the similar conclusion as \citet{Santini2012}, namely that the mean M$_*$ differences are similar between BL and non-BL AGNs. The distribution of M$_*$ difference of BL AGNs, however, is broader than that of non-BL AGNs, with the standard deviation of the former being larger than that of the latter by a factor of 
\begin{equation}
\frac{\sigma(\delta \rm{M}_{*,\rm{BL AGNs}})}{\sigma(\delta \rm{M}_{*,\rm{nonBL AGNs}})} \approx \frac{0.53\, \rm{dex}}{0.28\, \rm{dex}}\approx 1.9.
\end{equation}
It is therefore important to properly model the AGN component to get good measures of M$_*$ for BL AGNs, the detailed methodologies of which are beyond the scope of this work. We decide not to remove BL AGNs from our sample since they are a small fraction of the entire AGN sample and we have checked that our results are insensitive to including/excluding them.

\subsection{Star formation rate SFR}
Figure \ref{fig:SED_Test} (b) shows the comparison of SFRs between the two SED-fitting measurements. We refer readers to Lee2018 for a detailed discussion about the uncertainty of their SFR measurements using mock galaxies from semi-analytical simulations. A clear correlation between the two SFRs is observed both for the AGNs and the M$_*$-z matched non-AGNs, despite that the scatter is larger than in the case of the M$_*$ comparison. The right panel of Figure \ref{fig:SED_Test} (b) compares the distributions of SFR difference between the AGNs and non-AGNs. Generally speaking, the two SED procedures yield consistent estimates of SFRs for the non-AGNs, with a median $\delta$LogSFR $=0.08$ dex. For the AGNs, SFRs from Lee2018 are on average larger than those from {\sc Sed3fit} by 0.22 dex. 

\begin{figure}
\gridline{\fig{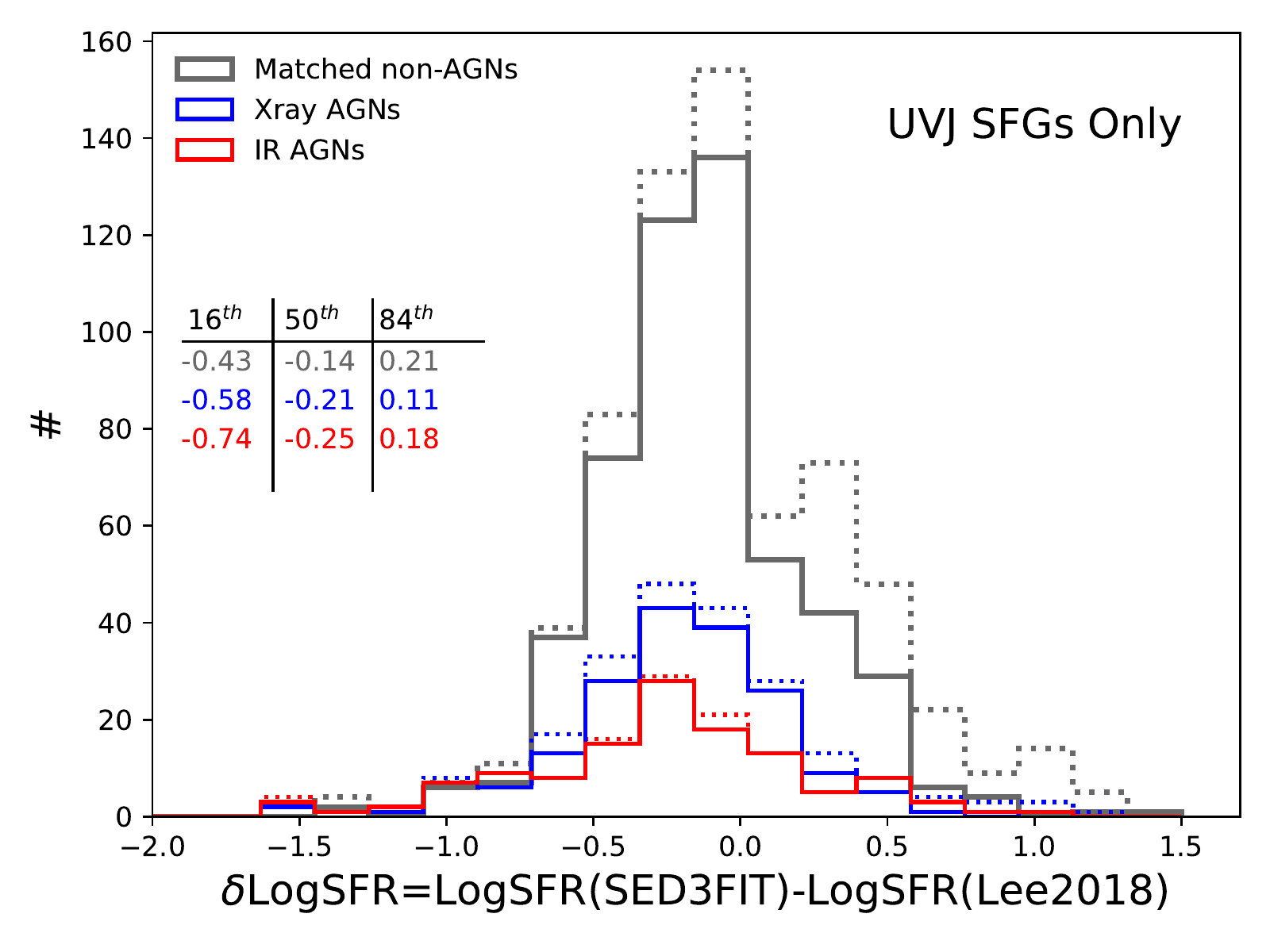}{0.47\textwidth}{}
}
\caption{Similar to the right panel of Figure \ref{fig:SED_Test} (b), but the solid histograms show the results for SFGs only, i.e. AGNs hosted by SFGs and M$_*$-z matched normal SFGs. The histograms of Figure \ref{fig:SED_Test} (b) are also plotted (dotted) for comparison. }  \label{fig:check_sfr_sfg}
\end{figure}

A closer look at the left panel of Figure \ref{fig:SED_Test} (b) shows that, unlike for galaxies with moderate/high SFRs, {\sc Sed3fit} predicts larger SFRs than Lee2018 for galaxies with low SFRs, indicating different systematics of SFR measurements between SFGs and QGs. Such systematics are very likely due to the assumptions of SFHs. In Lee2018, they tested the SED-derived SFRs using mock galaxies (see their Figure 6), where they showed that an incorrect SFH can lead to significantly biased SFR measurements for QGs. Also in \citet{Leja2019}, they showed that treating SFH as a free parameter is essential for the unbiased measurements of SFRs. While assumptions on the SFH can result in strong deviations of the SFR measurements from the intrinsic values for QGs, the situation for SFGs seems to be much better (see Figure 6 in Lee2018 for an example). Therefore, as Figure \ref{fig:check_sfr_sfg} shows, we only compare the SFRs between AGN hosted by SFGs and M$_*$-z matched normal SFGs (note that before we did not put any constrains on the star formation properties when building the M$_*$-z matched normal galaxies). Compared with Figure \ref{fig:SED_Test} (b), we find that the difference of $\delta$LogSFR between AGNs and non-AGNs decreases from $\approx0.2$ dex to $\approx0.1$ dex, which is much smaller than the 1$\sigma$ range of entire distributions ($\gtrsim0.7$ dex). Nevertheless, the remaining 0.1 dex difference between the two SED fittings is very likely due to neglecting the AGN component in Lee2018 fits. 

\begin{figure}
\gridline{\fig{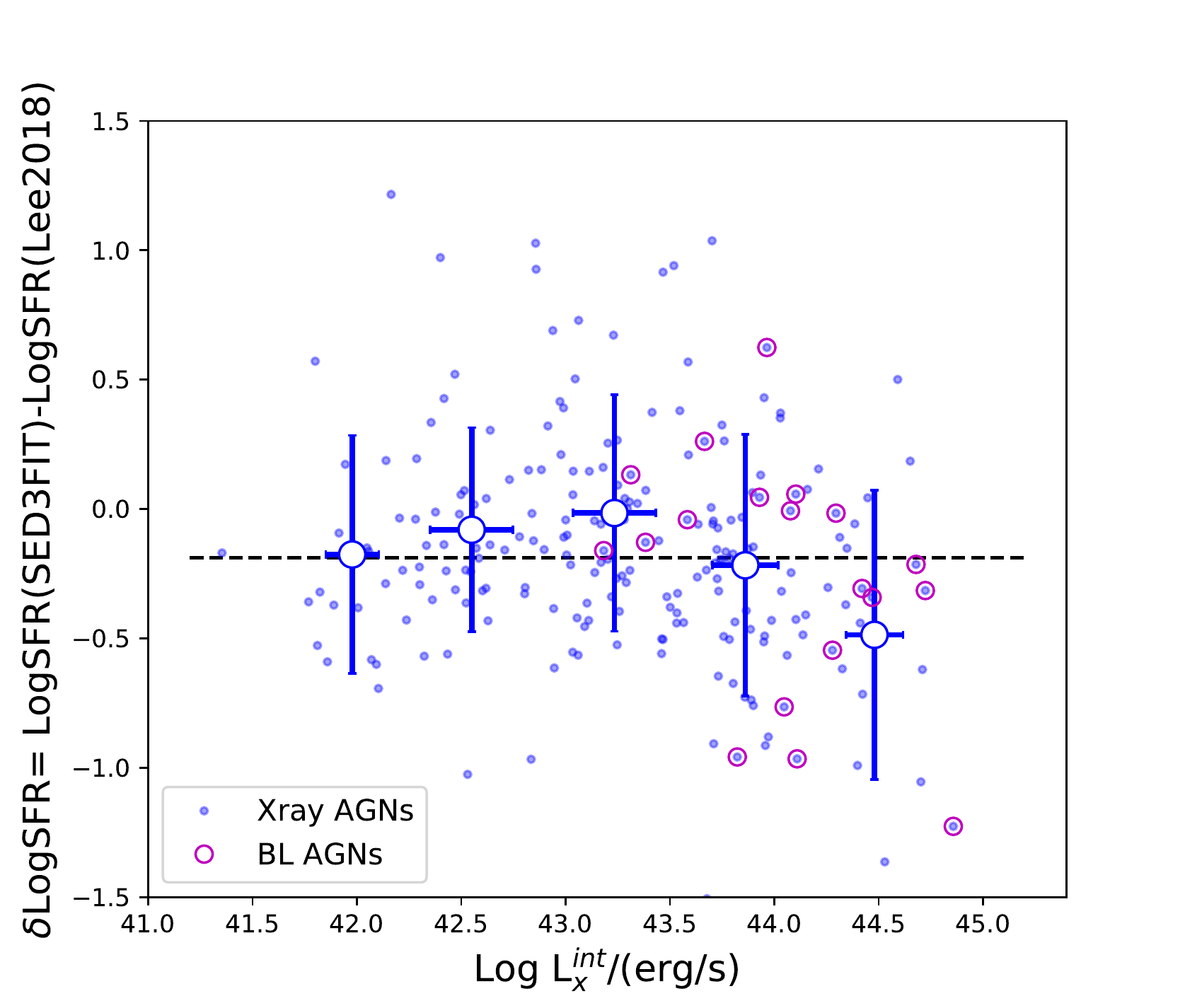}{0.5\textwidth}{}
}
\caption{Similar to Figure \ref{fig:deltaM_Lx} but for $\delta$LogSFR.} \label{fig:deltaSFR_Lx}
\end{figure}

In Figure \ref{fig:deltaSFR_Lx}, we also investigate the relation between $\delta$LogSFR and intrinsic X-ray luminosity. We do not see any clear trend except for the brightest bin ($\rm{L_X \sim 10^{44.5}\, ergs/s}$), where the over-estimation of SFR can be as large as $\approx$ 0.5 dex.

\subsection{Rest-frame colors}
\begin{figure*}
\gridline{\fig{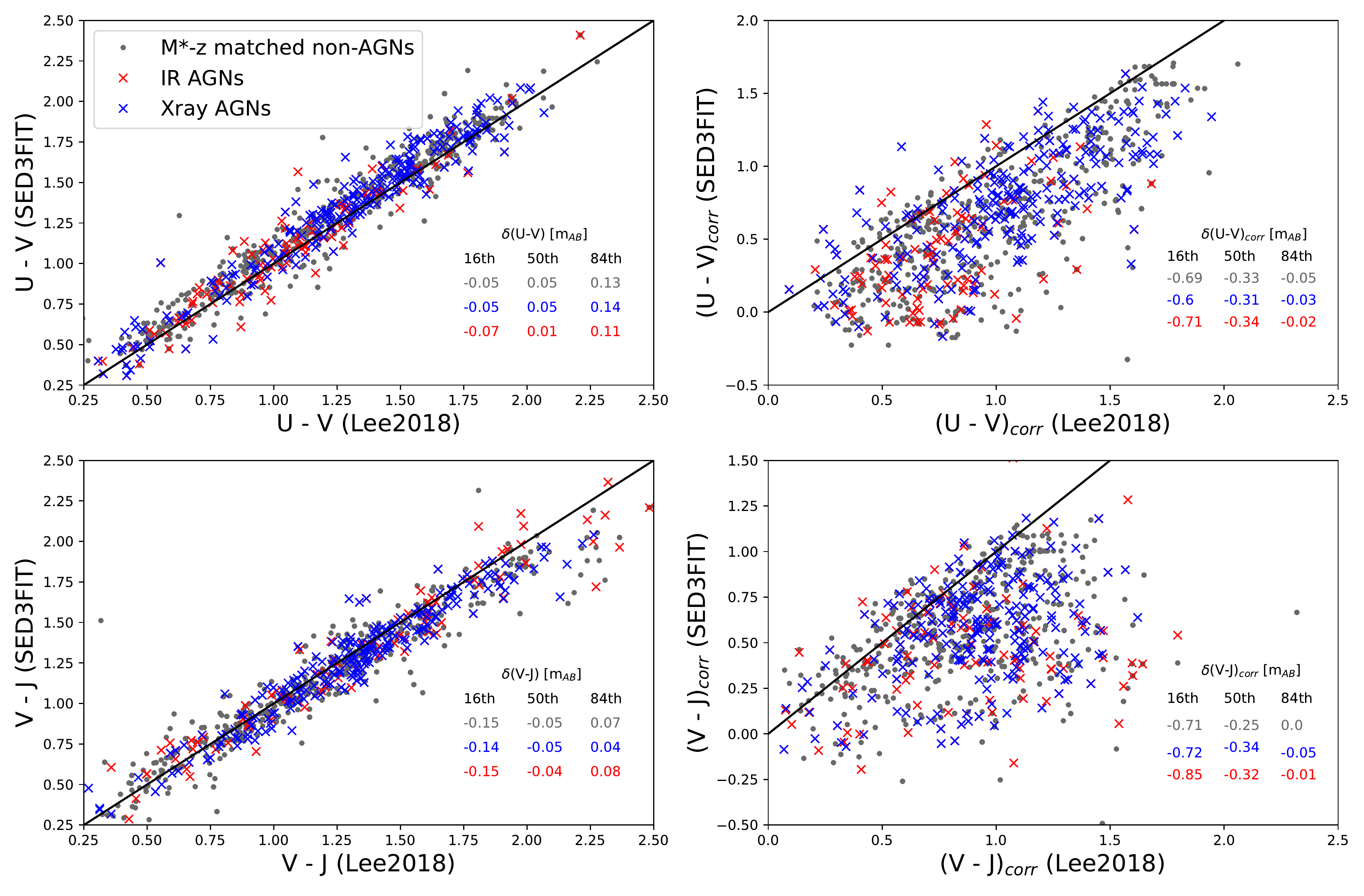}{0.9\textwidth}{}
}
\caption{Comparisons of rest-frame colors derived by the two SED fitting procedures. {\bf Left:} Comparisons between the apparent (i.e. dust-uncorrected) colors U$-$V and V$-$J. {\bf Right:} Comparisons between the dust-corrected colors \UVc\ and \VJc. Also tabulated in each panel are 16th, 50th and 84th percentiles of the corresponding differences between the two measurements for X-ray AGNs (blue), IR AGNs (red) and M$_*$-z matched non-AGNs (grey).}  \label{fig:check_color}
\end{figure*}

The last parameters that we have checked are the rest-frame colors. We first compare the apparent (i.e. dust-attenuated) rest colors $U-V$ and $V-J$ derived from the two SED fittings. To do so, we convolve the best-fit spectra with the Bessel U, V and 2MASS J filters respectively. As the left two panels of Figure \ref{fig:check_color} shows, the apparent rest-frame colors derived from the two SED procedures are in very good agreement with each other. The distributions of AGNs on the UVJ diagram (Figure \ref{fig:uvj}) therefore are not sensitive to the choice of the SED fitting procedure.

A more involved measurement is that of the dust-corrected rest-frame colors, i.e. colors that are corrected for dust attenuation (recall that Lee2018 assumes \citealt{Calzetti2000} dust attenuation law and {\sc Sed3fit} assumes \citealt{Charlot2000} model). The comparisons of the dust-corrected colors \UVc\ and \VJc\ are shown in the right panels of Figure \ref{fig:check_color}. Unlike the dust-uncorrected colors, both systematic offsets and larger scatters between the two measurements are seen for the dust-corrected colors, which illustrate the essential role played by the assumed dust attenuation models when measuring the properties of stellar populations. For \UVc, similar offsets ($\approx0.3m_{AB}$) and scatters are seen for both AGNs and non-AGNs, which indicates that the differences between the two SED measurements are primarily driven by assumptions {\it unrelated} to the presences of AGNs. Similar conclusion can also be made for \VJc, although the scatters of IR AGNs seem to be larger than X-ray AGNs and non-AGNs, very likely because the AGN contribution to the J band is generally larger for IR AGNs than for X-ray AGNs (see Figure \ref{fig:SEDs}). We caution that, however, our conclusions, which are drawn based on the comparisons of rest-frame colors above, depend on correctness of the assumed AGN models. While the SEDs of Type 1 AGNs have been empirically well characterized at UV through NIR wavelengths, the situation for the fainter Type 2 AGN in general, such as the majority considered here, is more uncertain. While significant progress has been made to observationally constrain the SEDs of faint AGNs at MIR wavelengths, comparatively little is known about their SEDs at rest-frame UV/optical. If the adopted AGN templates considerably deviate from the true AGN spectral shapes in the UV/optical window, our tests on the optical colors will be biased.

\end{document}